\documentclass[
prd,
superscriptaddress,
10 pt,
preprintnumbers,
notitlepage,
secnumarabic,
nobibnotes,
nofootinbib,
showpacs
]{revtex4-1}

\usepackage[T1]{fontenc}
\usepackage[pdftex]{color,graphicx}
\usepackage{mathtools}
\usepackage{amssymb}
\usepackage{bm}
\usepackage{dsfont}
\usepackage{mathrsfs}
\usepackage{calligra}
\usepackage{tabularx}
\usepackage[artemisia]{textgreek}
\usepackage{hyperref}
\usepackage{xcolor}
\usepackage{enumerate}
\usepackage{multirow}
\usepackage{url}
\usepackage{caption}
\usepackage[normalem]{ulem}
\usepackage{perpage}
\MakePerPage{footnote}
\renewcommand{\.}{\!\;}		%	2 point space
\renewcommand{\@}{\!\:\!}	%	-2 point space

\DeclareMathAlphabet\mathbfcal{OMS}{cmsy}{b}{n}
\newcommand{\kbar}{\mathchar'26\mkern-9mu k}
\renewcommand{\i}{\!\:\mathrm{i}}
\newcommand{\T}{\!\!\;\mathsf{T}}
\newcommand{\enn}{\mathcall{\Large{n}\!\.}}
\newcommand{\mathcall}[1]{\text{\calligra\footnotesize #1\,}}

\newcommand{\hash}{\text{\raisebox{1pt}{$\scriptstyle\#$}}}
\newcommand{\bighash}{\text{\large\bf\#\@}}
\newcommand{\squarearrowleft}{\llcorner\!\!\!\Lsh}
\newcommand{\squarearrowright}{\Rsh\!\!\!\lrcorner}
\newcommand{\pos}[1]{\ensuremath{\@\langle#1\rangle}}
\usepackage{tikz}
\usetikzlibrary{shapes.geometric}
\tikzset{cb05/.style={draw=black, line width=0.5pt}}
\tikzset{cb15/.style={draw=black, line width=1.5pt}}
\tikzset{cb25/.style={draw=black, line width=2.5pt}}
\tikzset{cg05/.style={draw=gray, line width=0.5pt}}
\tikzset{cg15/.style={draw=gray, line width=1.5pt}}
\tikzset{cg25/.style={draw=gray, line width=2.5pt}}
\tikzset{cl05/.style={draw=lightgray, line width=0.5pt}}
\tikzset{cl15/.style={draw=lightgray, line width=1.5pt}}
\tikzset{cl25/.style={draw=lightgray, line width=2.5pt}}
\tikzset{cl35/.style={draw=lightgray, line width=3.5pt}}
\tikzset{db05/.style={dashed, draw=black, line width=0.5pt}}
\tikzset{db15/.style={dashed, draw=black, line width=1.5pt}}
\tikzset{db25/.style={dashed, draw=black, line width=2.5pt}}
\tikzset{dg05/.style={dashed, draw=gray, line width=0.5pt}}
\tikzset{dg15/.style={dashed, draw=gray, line width=1.5pt}}
\tikzset{dg25/.style={dashed, draw=gray, line width=2.5pt}}
\begin{document}
\title{Relativistic classical theory II.
\\
Holonomy-flux representation of gravitational degrees of freedom}

\author{Jakub Bilski}
\email{bilski@zjut.edu.cn}
\affiliation{Institute for Theoretical Physics and Cosmology, Zhejiang University of Technology, 310023 Hangzhou, China}

	%%%%%%%%%	%%%%%%%%%	%%%%%%%%%	%%%%%%%%%

	%%%%%%%%%	%%%%%%%%%	%%%%%%%%%	%%%%%%%%%
\begin{abstract}
\noindent
This article describes the regularization of the generally relativistic gauge field representation of gravity on a piecewise linear lattice. It is a part of the program concerning the classical relativistic theory of fundamental interactions, represented by minimally coupled gauge vector field densities and half-densities. The correspondence between the local Darboux coordinates on phase space and the local structure of the links of the lattice, embedded in the spatial manifold, is demonstrated. Thus, the canonical coordinates are replaceable by links-related quantities. This idea and the significant part of formalism are directly based on the model of canonical loop quantum gravity (CLQG).

The first stage of this program is formulated regarding the gauge field, which dynamics is independent of other fundamental fields, but contributes to their dynamics. This gauge field, which determines systems equivalence in the actions defining all fundamental interactions, represents Einsteinian gravity. The related links-defined quantities depend on holonomies of gravitational connections and fluxes of densitized dreibeins. This article demonstrates how to determine these quantities, which lead to a nonpertubative formalism that preserves the general postulate of relativity. From this perspective, the formalism presented in this article is analogous to the Ashtekar-Barbero-Holst formulation on which CLQG is based. However, in this project, it is additionally required that the fields' coordinates are quantizable in the standard canonical procedure for a gauge theory and that any approximation in the construction of the model is at least as precisely demonstrated as the gauge invariance. These requirements lead to new relations between holonomies and connections, and the representation of the densitized deibein determinant that is more precise than the volume representation in CLQG.
\end{abstract}
	%%%%%%%%%	%%%%%%%%%	%%%%%%%%%	%%%%%%%%%

\maketitle

	%%%%%%%%%	%%%%%%%%%	%%%%%%%%%	%%%%%%%%% 

	%%%%%%%%%	%%%%%%%%%	%%%%%%%%%	%%%%%%%%%
\section{Introduction}\label{Sec_Introduction}

	%%%%%%%%%	%%%%%%%%%	%%%%%%%%%	%%%%%%%%%
\subsection{Equivalence principle on a lattice}\label{Sec_Introduction_/_Equivalence_principle}

\noindent
This article is the second element in the series presenting a self-consistent and methodologically uniform procedure of the lattice regularization of a relativistic field theory; the first article is forthcoming \cite{Bilski:2021_RCT_I}. This manuscript concerns the gravitational degrees of freedom, which are represented by a unique quantity --- the field that contributes to each element of any action of any relativistic theory. This feature is related to the general postulate of relativity \cite{Einstein:1916vd} that implements covariance by contracting all spacetime coordinates-dependent quantities with the metric tensor. This tensor is assumed to be a propagating field itself, hence any other field is at least minimally coupled with the gravitational field's degrees of freedom. Moreover, the relativistic lattice field theory project assumes that the formulation of each interaction is generally invariant under any change of coordinates. Therefore, the gravitational field itself has to be formulated according to the general postulate of relativity; i.e. the strong equivalence principle is assumed. This refers to the factual meaning of this postulate, thus any conventional use of one of the popular frameworks that are considered as models for gravitational interactions is \textit{a priori} rejected as inconsistent with general relativity. Therefore, as discussed in \cite{Bilski:2021_RCT_I}, the general postulate of relativity restrictions are imposed regarding the gravitational field (and any other fundamental field) theory formulation. These restrictions are known as a systems-equivalent description \cite{Einstein:1907,Einstein:1911vc} and background-independent predictions \cite{Einstein:1907iag,Einstein:1916vd}, shortly, the systems equivalence (SE) and background independence (BI). Consequently, the whole framework, which is constructed in this article, is designed to preserve SE and BI in the standard canonical procedure of a gauge field quantization, without adding any \textit{ad hoc} modifications.

A particular role in the methodology of this project has the relation between time and quantum uncertainty. Both issues have many different notions in physics. The canonical quantization procedure in quantum field theory (QFT) is based on the Hamiltonian formalism applied to a gauge theory. This formalism allows describing the evolution of a physical system in terms of twice as many first-order differential equations as the second-order Euler-Lagrange equations. The latter ones are derived, by definition, from a postulated action. In this equality of descriptions, a fixed relation between the time coordinate and the so-called generalized coordinates is assumed. This relation is given by the map, known as the Legendre transform, from the Euler-Lagrange formalism into the Hamiltonian one. Thus, the Hamiltonian formalism \cite{Hamilton:1834,Hamilton:1835}, which became adjusted to gauge theories 115 years after its construction \cite{Dirac:1950pj}, indicates a particular role of time by imposing the fixed relation between the time coordinate and all the other coordinates. As argued in \cite{Bilski:2021_RCT_I},the Hamiltonian formalism is not in contradiction with the general principle of relativity, neither with the peculiar meaning of time in physics\footnote{The author refers, for instance, to the notion of entropy in the second law of thermodynamics \cite{Caratheodory:1909,Planck:1926}.}. Moreover, in this time determining approach, physically meaningful objects are more likely chosen as fundamental theoretical quantities than purely mathematical objects --- the latter ones are preferred in the Euler-Lagrange formalism. For instance, in the case of electrodynamics, this choice is regarding the electric and magnetic fields over the vector potential and its curvature. The former quantities are measurable, the latter allow to describe the theory in a spacetime diffeomorphisms-invariant formalism. This latter formalism would require the De Donder-Weyl generalized-covariant approach (GCA) to the Hamiltonian field theory \cite{De_Donder:1930,Weyl:1935} (see \cite{Ivanenko:1984vf} for a gravitational application of this formalism). The standard Hamiltonian formalism breaks the spacetime diffeomorphism symmetry into the spatial diffeomorphism invariance and the time reparametrization symmetry. Thus, this formalism is going to be called the time gauge approach (TGA). This breakdown is known as the $3+1$ Arnowitt-Deser-Misner (ADM) decomposition \cite{Arnowitt:1959ah,Arnowitt:1960es,Arnowitt:1962hi}.

What is most important concerning the discussed issue, is the fact that both GCA and TGA satisfy SE and BI (see Sec.~\ref{Sec_Regularization_/_ADM} and the analysis in \cite{Bilski:2021_RCT_I} for more details). However, in the case of TGA, the theory is not explicitly Lorentz-invariant. This mathematical symmetry is replaced by the more general spatial and temporal diffeomorphism invariance in the ADM form. It is worth noting that this form is often considered less elegant because it is expressed by longer mathematical formulas than, for instance, less general Lorentz-invariant perturbative approaches. However, in the author's opinion, this viewpoint is completely wrong. It is in contradiction with the postulate that assumes the universality of laws of nature and with the results of observational verification of these laws, for instance, in tests of general relativity (GR).

In the series of publications that is initiated by this article, TGA is assumed. This assumption allows one to apply many methods that already exist in the literature regarding quantum gravity. In particular, the regularization procedure presented in this article is based mainly on Thiemann's regularization in canonical loop quantum gravity (CLQG) \cite{Thiemann:1996ay,Thiemann:1996aw,Thiemann:1997rt,Ashtekar:2004eh,Thiemann:2007zz}. Moreover, this latter approach is considered as a reference procedure in this manuscript. Furthermore, concerning this approach, the improved method is constructed \cite{Bilski:2020poi}. Although some sections of this article may look like a criticism of Thiemann's method, it should be emphasized that without his results this series of articles would never be created\footnote{The author of this article considers the presented analysis as direct development and improvement of the regularization procedure that has been first described in \cite{Thiemann:1996ay} as a method to construct the so-named quantum spin dynamics.}.

Another advantage of using TGA is that the Hamiltonian formalism applied to a QFT formulation regarding special relativity is easily adjustable to GR. In the case of GCA, only Lagrange formalism-related results of QFT would be directly generalizable to GR (or the results that were derived in the De Donder-Weyl formalism).

The third advantage is the possibility of using the reduced phase space methods \cite{Dirac:1950pj,Henneaux:1992ig}, which allow one to study much simpler quantum models of gauge-reduced systems, for instance, the models with a direct cosmological applicability \cite{Ashtekar:2003hd,Bojowald:2008zzb,Ashtekar:2009vc,Ashtekar:2011ni,Bilski:2021fki}.

Finally, using the ADM decomposition, the non-physical degrees of freedom, known as the ghost fields, are decoupled from the system before quantization. This fourth advantage, which is formally a partial gauge fixing in the rigorous phase space reduction method, could also be recognized as a disadvantage of TGA. In modern academical physics, the mathematical formulas' simplicity and the popular formalism usage are often considered as better or even more precise than more general applicability, formalism uniformity, methodological self-consistency, and even higher symmetry invariance. In the methodology of this series of articles, the opposite, more physically- than mathematically-motivated quality valuation is assumed. However, it should be emphasized that despite many advantages, TGA is a less symmetric formulation than GCA. One could say that, a more symmetric formalism ``has been traded'' in exchange for a more physical theory, or that there is given a priority to the laws of nature manifestation over their mathematical description.

By assuming TGA, many difficulties that could appear in the case of GCA in the further construction of quantum formalism should not be a problem. For instance, one will obtain discrete space but continuous time. This latter quantity can be viewed then as an emerging time in QFT. Moreover, after the derivation of the classical or semiclassical limit, all the issues concerning time in GR will be addressable (with possibly different results due to quantum corrections). Thus, one can ask: how the peculiar points in GR solutions, for instance, the Schw{\"a}rzshild horizon or the so-called Bing Bang, will look like? This and analogous questions can be answered by solving the presented here formalism. Moreover, some of them are simply formulable by imposing appropriate gauge conditions \cite{Dirac:1950pj,Henneaux:1992ig}, which represent a model of physical phenomena. It is worth mentioning that keeping these global conditions locally unmodified is required by the gauge invariance of the related symmetries \cite{Bilski:2019tji,Bilski:2020xfq}.

This article concerns the bosonic field that is used to describe gravitational interactions. Following Bose-Einstein statistics is, without a doubt, a physical (not mathematical) feature. Thus, it may be reflected in some unique qualities of the formalism that distinguishes this feature from following fermionic statistics. However, the fermions-related formalism cannot be neglected because (as it has been already mentioned) gravity is coupled to all the fundamental fields. The fermionic formalism inclusion was formulated in detail in \cite{Kibble:1961ba} by using the so-called Einstein-Cartan theory \cite{Cartan:1922,Cartan:1923zea,Cartan:1924yea,Cartan:1925}. It has been demonstrated that the complete formulation of GR has to include the torsional degrees of freedom, which contribute to the formalism in the presence of a fermionic field \cite{Hehl:1976kj,Hehl:1994ue}. However, these degrees of freedom should not factually contribute to the description of the system's dynamics in the absence of this field. Therefore, the gravitational connection (defined in Sec.~\ref{Sec_Regularization_/_Gauge_bosons}), which specifies how fundamental fields (\textit{cf.} \cite{Bilski:2021_RCT_I}) are transported along affine geodesics in non-Euclidean geometry, is a torsion-full object \cite{Ashtekar:1986yd,Ashtekar:1987gu,Barbero:1994ap,Mercuri:2006um,Bojowald:2007nu}. The torsional part, however, can be neglected in the bosonic systems, as the element not contributing to their dynamics.

The last issue worth mentioning in the introduction is the explanation of the quadrilaterally hexahedral graph choice, the structure of which can represent the spatial manifold geometry. It should be added that the discussed manifold is the one on which the theory of gravity is founded. Thus, one detail should be specified before the kinematical analysis of this theory. Gravity is related to a metric, hence the metrizability determines strong conditions on the manifold, which has to be a perfectly normal Hausdorff space (T$\@_6$ space). In the framework of this project, all the vector fields (including the one that represents gravity) are smeared along graph's edges. The structure of these edges is determined by the spatial diffeomorphism invariance. Their lengths are specified by the smeared gravitational variables. In this way, the graph becomes embedded in the spatial manifold, and the resulting graph-space relation is known as a lattice. It should be emphasized that its structure and scales are not coupled. The invariance of the first quantity represents GR, and it concerns all the fundamental fields in the same manner (not only the gravitational field). The variance of the second quantity represents the dynamics of geometry, and it is described as the gravitational field's evolution. Consequently, the diffeomorphism symmetry is decoupled from geometrical scales that vary, leading to changes in relative distances. This latter dynamical process is going to be identified as gravitation. Its decoupling from the diffeomorphism invariance is determined by the Hamiltonian-Dirac formalism \cite{Hamilton:1834,Hamilton:1835,Dirac:1950pj} that distinguishes the diffeomorphism constraint and the Hamiltonian constraint, where only the latter object contains information about dynamics.

Then, the quadrilaterally hexahedral graph tessellation of space is favored by three arguments. First, it specifies the graphs embedding in the spatial manifold. The lattice obtained by this embedding does not have to be restricted by any fixed pattern; hence it remains diffeomorphism-invariant\footnote{The same property occurs regarding any other regular tilling --- see Sec.~\ref{Sec_Regularization_/_Tessellations}.}. Second, the quadrilaterally hexahedral structure of each lattice' cell coincides with the continuous-to-discrete map of the differential form $d^3x$ (see the remarks in \cite{Bilski:2020poi,Bilski:2021fki}). Third, this structure allows to identify a triple of opposite faces of each cell and then allows to slice each cell on a countably infinite number of layers between the opposite faces. This latter slicing of cells allows to approximate continuous distributions of variables inside each cell arbitrarily well \cite{Bilski:2021ysc} (see also Sec.~\ref{Sec_Regularization_/_Holonomy-flux_algebra}). The last pair of arguments concerns the transition method from the continuous formulation of the classical field theory into the lattice formulation. By this transition, the lattice theory, quantizable in the standard canonical procedure \cite{Dirac:1950pj}, is obtained. This fact allows applying the phase space reduction to derive symmetry-fixed quantum models of physical phenomena, and applying the correspondence principle to link quantum numbers with the values of classical variables \cite{Bohr:1920}. This reduction is directly construable concerning the quadrilaterally hexahedral lattice.
\\

The notation in this article is compatible with the manuscript methodologically first in this series \cite{Bilski:2021_RCT_I}. All the indices in the four-dimensional manifold are raised or lowered by the metric tensor $g_{\mu\nu}:=e^{\alpha}_{\mu}e^{\beta}_{\nu}\eta_{\alpha\beta}$, where $e^{\alpha}_{\mu}$ is the co-vierbien and $\eta_{\alpha\beta}$ is the flat Minkowski metric with the $(-,+,+,+)$ signature. The $3+1$ decomposition introduces the four-dimensional manifold slicing normalized along the proper time direction. Each slice is the Cauchy hypersurface; the related spatial metric tensor is $q_{ab}:=e^i_a e^j_b\delta_{ij}$, where $e^i_a$ denotes the co-dreibein and $\delta_{ij}$ is the Kronecker delta. The indices contracted with the $\mathfrak{su}(2)$ generators $\tau^i:=-\frac{\i}{2}\sigma^i$ ($\sigma^i$ are Pauli matrices) label directions in the internal space corresponding to the non-Abelian gauge symmetry representation of gravitational variables. The Einstein coupling constant related to this interaction is defined by $\kappa:=16\pi G$, where the speed of light is normalized to $c=1$. All the repeated indices except the ones written in brackets are contracted according to the Einstein convention.

This manuscript is organized as follows. The second section of the introduction describes the idea of the lattice formulation of a quantum gauge theory. The main part of this article applies this idea regarding the gravitational field. It starts from the establishment of the gauge vector field's representation of gravity in Sec.~\ref{Sec_Regularization_/_Gauge_bosons}. The regularization of the momentum degrees of freedom concerning the free field's theory is introduced in Sec.~\ref{Sec_Regularization_/_ADM}, and then in Sec.~\ref{Sec_Regularization_/_volume}, it is extended to be applicable to the minimal coupling of gravity with other fundamental fields. The holonomy representation of the gravitational connection is described in Sec.~\ref{Sec_Regularization_/_graph}. The related SU$(2)$ group-$\mathfrak{su(2)}$ representation relation is analyzed in Sec.~\ref{Sec_Regularization_/_Wigner}. This analysis leads to the establishment of the accurately symmetry-preserving lattice representation of the gravitational connection in Sec.~\ref{Sec_Regularization_/_holonomy}. Sec.~\ref{Sec_Regularization_/_revisited} describes the method of increasing the regularization of the momentum variable from Sec.~\ref{Sec_Regularization_/_volume}. In Sec.~\ref{Sec_Regularization_/_Tessellations} the construction of a piecewise linear lattice is introduced by defining its relation with a metrizable manifold. The separability of this manifold allows one to specify its divisibility into elementary cells that are characterized in Sec.~\ref{Sec_Regularization_/_Elementary_cell}. Sec.~\ref{Sec_Regularization_/_Holonomy-flux_algebra} initiates the analysis of the algebra of canonical variables on a lattice. This issue becomes completely elaborated in Sec.~\ref{Sec_Regularization_/_geometry} and in Sec.~\ref{Sec_Regularization_/_rapidity} it is extended on all the variables contributing to the Hamiltonian. Finally, the procedure of the lattice regularization of the gravitational variables is concluded in the last part of this article, namely, in \ref{Sec_Summary}.

	%%%%%%%%%	%%%%%%%%%	%%%%%%%%%	%%%%%%%%%
\subsection{Motivation: Wilson loops}\label{Sec_Introduction_/_Wilson}

\noindent
In this article, gravitational degrees of freedom are represented by torsion-full lattice-smeared $\mathfrak{su}(2)$-invariant vector fields \cite{Ashtekar:1986yd,Ashtekar:1987gu,Barbero:1994ap,Mercuri:2006um,Bojowald:2007nu}. The first complete formulation of a gauge theory on a lattice was proposed by Wilson \cite{Wilson:1974sk}. The relation between vector potential fields and Wilson loops was later improved by Giles \cite{Giles:1981ej}. They elaborated a gauge-invariant quantization method of a vector field in a flat geometry. The flat geometry space allowed them to define a quantum theory on a discrete cubic lattice. In this geometry, the discretization of Cartesian directions can be implemented, for instance, by a compactification of the real orthogonal directions.

%
%%%	FIGURE	%%%
%
\begin{figure}[h]
\vspace{5pt}%
\begin{center}
\begin{tikzpicture}[scale=1]
%holonomy
\draw[cl35,->]{(0.5,0.8) -- (3.5,0.8)};
\draw[cl35,->]{(3.5,0.8) -- (3.5,3.8)};
\draw[cl35,->]{(3.5,3.8) -- (0.5,3.8)};
\draw[cl35,->]{(0.5,3.8) -- (0.5,0.8)};
%secondary horizontal
\draw[cb15]{(-2.5,0.8) -- (3.5,0.8)};
%secondary diagonal
\draw[cb15]{(0.5,0.8) -- (0,0)};
%secondary vertical
\draw[cb15]{(0.5,3.8) -- (0.5,-2.2)};
%tertiary diagonal
\draw[cb05]{(1,1.6) -- (0.5,0.8)};
%%%	labeling		%%%
\node at (1.1,1.08) {$A\big(x\big)$};
\node at (4.3,1.1) {$A\big(x\!+\!l_{(q)}\big)$};
\node at (1.4,4.1) {$A\big(x\!+\!l_{(r)}\big)$};
\node at (4.6,4.1) {$A\big(x\!+\!l_{(q)}\!+\!l_{(r)}\big)$};
%%%	dressing		%%%
\node at (0.7,0.5) {$x$};
\node at (-0.4,-0.2) {$l_{(\@p\@)}\!\big(x\big)$};
\node at (1.6,1.8) {$l_{\@(\@p\@)}\!\big(x\!-\!l_{(p)}\@\big)$};
\node at (2.3,0.45) {$l_{(\@q\@)}\!\big(x\big)$};
\node at (-1.6,0.4) {$l_{\@(\@q\@)}\!\big(x\!-\!l_{(q)}\@\big)$};
\node at (0,2.8) {$l_{(\@r\@)}\!\big(x\big)$};
\node at (-0.3,-1.3) {$l_{\@(\@r\@)}\!\big(x\!-\!l_{(r)}\@\big)$};
\end{tikzpicture}
\end{center}
\vspace{-10pt}%
\caption{Discrete distribution of the connection related to the square loop $\circlearrowleft^{\square}_{(\@q\@)\@(\@r\@)}\!\@\!\big(x\big)$}
\label{cubic_Wilson}
\end{figure}
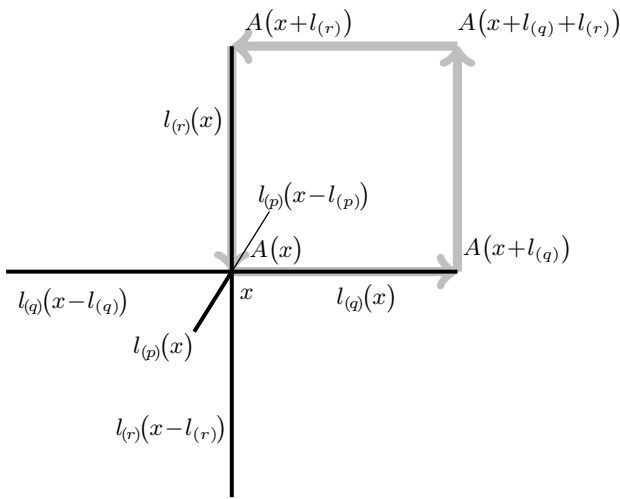
%
%%%	FIGURE	%%%
%
The basic object introducing the discrete quantum formalism is the Wilson loop, i.e. the functional
\begin{align}
\label{Wilson}
W_{\!\uline{A}}(\ell):=\uline{\text{tr}}\bigg[\mathcal{P}\exp\!\bigg(\i\!\oint_{\@\ell}dx^a\uline{A}_a\bigg)\bigg]
\end{align}
of a gauge connection $\uline{A}_a:=\uline{A}_a^I\uline{\sigma}^I$. Here, $a$ labels an Euclidean space direction and $\uline{\sigma}^I$ denotes a traceless Hermitian generator of a considered group representation. The symbol $\mathcal{P}$ denotes path ordering and the trace is taken with respect to the unitary $\mathfrak{su}(N)$ representation that satisfies the commutation relation
\begin{align}
\label{Lie_complex}
[\uline{\sigma}_J,\uline{\sigma}_K]=\i\.C^I_{\ JK}\uline{\sigma}_I\,,
\qquad\text{where }
I,J,K,...=1,2,...,N^2-1
\end{align}
and $C^I_{\ JK}$ are real structure constants. The path $\ell$ is a loop, precisely a closed composition of $\Gamma$ graph's links. The square path example is sketched in FIG~\ref{cubic_Wilson}, where the symbol $l_{(p)}$ expresses a link, and the spatial direction that represents a local link's orientation is labeled by $p$. An important feature of the Wilson loop representation is its invariance under the following gauge transformations \cite{Mandelstam:1968hz,Mandelstam:1978ed},
\begin{align}
\label{Mandelstam_identities}
\begin{split}
W_{\!\uline{A}}\big(\ell_1\@\circ\@\ell_2\big)
&=W_{\!\uline{A}}\big(\ell_2\@\circ\@\ell_1\big)\,,
\\
W_{\!\uline{A}}\big(\ell_1\big)W_{\!\uline{A}}\big(\ell_2\big)
&=W_{\!\uline{A}}\big(\ell_1\@\circ\@\ell_2^{-1}\big)+W_{\!\uline{A}}\big(\ell_1\@\circ\@\ell_2\big)\,,
\\
W_{\!\uline{A}}\big(\ell\big)
&=W^{*\@}_{\!\uline{A}}\big(\ell^{-1}\big)
\quad\Rightarrow\quad W_{\!\uline{A}}\big(\ell\big)\in\mathds{R}\,,
\end{split}
\end{align}
where $^*$ indicates the complex conjugate.

It is worth emphasizing that the dependence of $W_{\!\uline{A}}(\ell)$ on the gauge connection is not local --- see FIG~\ref{cubic_Wilson}. The Wilson loop is related to an oriented path $\ell$, and the corresponding connection is located at certain points on this path. For instance, in the case of a square loop $\ell(s):[0,1]\to M$, embedded in the manifold $M$, these points are $\ell(0)=\ell(1)$, $\ell(1/4)$, $\ell(1/2)$, $\ell(3/4)$. Consequently, to define the relation between the Wilson loop's representation and the vector potential, a procedure of averaging field's locations is needed (for a discussion regarding discretization procedures see Sec.~\ref{Sec_Regularization_/_graph}). The discretization method has to preserve the symmetry transformations of the lattice variables, which form the set of relations in \eqref{Mandelstam_identities}. This issue was theoretically investigated in \cite{Mandelstam:1962mi} and later adjusted to the discretization of the vector potential in \cite{Makeenko:1979pb,Makeenko:1980vm,Gambini:1982bg}. In the original formulation of the gauge field's lattice smearing, the discretization procedure is defined in the momentum representation by postulating a particular discrete operator, known as the `loop derivative'. In this way, the non-locality problem is shifted into the quantum level, where the introduction of customized, effective definitions is more acceptable. Then, the Wilson loop representation is set by the following relation with the gauge connection-related curvature operator,
\begin{align}
\label{curvature_to_Wilson}
\uline{\hat{F}}_{ab}(x):=-\i\uline{\Delta}_{ab}(x)W_{\!\uline{A}}(\ell)\,,
\qquad
x\in\ell\,,
\end{align}
where $\uline{\hat{F}}_{ab}:=\uline{\hat{F}}_{ab}^I\uline{\sigma}^I$.

The first physical application of the Wilson loop representation was proposed regarding electromagnetism in \cite{Gambini:1980wm,diBartolo:1983pt,Gambini:1986ew}. In the case of this Abelian field model, the Wilson loop in \eqref{Wilson} simplifies to
\begin{align}
\label{el_Wilson}
W_{\!\uline{A}}(\ell)\Big|_{\text{U}(1)}=\exp\!\bigg(\i\!\oint_{\@\ell}\mathrm{d}x^a\uline{A}_a\bigg).
\end{align}
The total Hamiltonian for the U$(1)$ vector field $\uline{A}_a$ consists of two constraints. The Abelian Gauss constraint $\int\@\@d^3x\Lambda\partial_a\uline{E}^a$ describes the phase reparametrization. The Hamiltonian constraint
\begin{align}
\label{el_scalar}
H^{(\uline{A})}:=\int\!\!d^3x\,N\mathcal{H}^{(\uline{A})}
=\frac{1}{2}\!\int\!\!d^3x\,N\delta_{ab}
\big(\uline{E}^a\uline{E}^b+\uline{B}^a\uline{B}^b\big)
\end{align}
contains information about the time symmetry and the dynamics of the theory. The quantities $\Lambda$ and $N$ are Lagrange multipliers, the electric field $\uline{E}^a$ is the momentum canonically conjugate to $\uline{A}_a$, and the magnetic field is given by the expression
\begin{align}
\label{el_magnetic}
\uline{B}^a:=\frac{1}{2}\epsilon^{abc}_{\scriptscriptstyle+}\uline{F}_{bc}\,.
\end{align}
The labeling $+$ in $\epsilon^{abc}_{\scriptscriptstyle+}$ indicates that this object is not a Levi-Civita tensor but its density, hence a quantity of a weight $1$.

The loop states are defined by the following transform of the connection wave functional $\Psi[\uline{A}]$,
\begin{align}\label{el_state}
\Psi(\ell):=\int\!\!D\uline{A}\,\text{exp}\bigg(\!-\i\!\int\!\!d^3x\.\uline{X}^{a\@,x}\@(\ell)\uline{A}_a(x)\bigg)\Psi[\uline{A}]\,,
\end{align}
where the loop splitting function
\begin{align}\label{el_delta}
\uline{X}^{a\@,x}\@(\ell):=\!\oint_{\@\ell}dy^a\delta(x-y)
\end{align}
has been introduced. In this way, the continuous-to-discrete transition is implemented at the quantum level by the Fourier-like transform. By postulating then the momentum representation, an analogous loop splitting has to be introduced into the definition of canonical operators that form the states in \eqref{el_state}. The magnetic field operator is going to be defined by using the loop derivative structure in \eqref{curvature_to_Wilson}. The electric operator is the momentum representative, hence, it should act by a multiplication restricted by the symmetry of the lattice structure. Then, the natural form of the actions of electric and magnetic field loop representations is defined by the operator equations:
\begin{align}
\label{el_physical}
\begin{split}
\uline{\hat{E}}^a\@(x)\Psi(\ell):=&\;\uline{X}^{a\@,x}\@(\ell)\Psi(\ell)\,,
\\
\uline{\hat{B}}^a\@(x)\Psi(\ell):=&-\i\.\epsilon^{abc}_{\scriptscriptstyle+}\uline{\Delta}_{bc}(x)\Psi(\ell)\,.
\end{split}
\end{align}

In the Abelian case, the gauge invariance of states is easily implementable. The related symmetry is the phase reparametrization, hence any exponentiated linear functional of $\uline{A}$ appears to be a good candidate for a basis wave state. The natural choice is $\Psi[\uline{A}]=W_{\!\uline{A}}(\ell)$. Consequently, the Gauss constraint becomes automatically satisfied, and the action of the scalar constraint operator takes the following form,
\begin{align}
\label{el_Hamiltonian}
\hat{H}^{(\uline{A})}\Psi(\ell)=\!\int\!\!d^3x\,N
\bigg(
\!-\frac{1}{4}\delta^{ac}\delta^{bd}\uline{\Delta}_{ab}(x)\uline{\Delta}_{cd}(x)
+\frac{1}{2}\delta_{ab}\uline{X}^{a\@,x}\@(\ell)\uline{X}^{b\@,x}\@(\ell)
\bigg)\Psi(\ell)\,.
\end{align}
One can also construct a consistent definition of the ladder operators,
\begin{align}
\label{el_ladder}
\begin{split}
\uline{\hat{a}}^{\dagger}\@\big(\vec{k}\big)&:=
(2\pi)^{-\frac{3}{2}}\!\int\!\!d^3x\Bigg(
\text{exp}\big(-\i\vec{k}\@\cdot\@\vec{x}\big)\epsilon^a_{\uline{A}}\@\big(\vec{k}\big)\frac{k^b}{|\vec{k}|^{\frac{3}{2}}\!}\.
\uline{\Delta}_{ba}(x)
-\i\frac{1}{|\vec{k}|^{\frac{1}{2}}\!}\:\text{exp}\big(-\i\vec{k}\@\cdot\@\vec{x}\big)\epsilon_{a\uline{A}}\@\big(\vec{k}\big)
\uline{X}^{a\@,x}\@(\ell)
\Bigg)\,,
\\
\uline{\hat{a}}\big(\vec{k}\big)&:=
(2\pi)^{-\frac{3}{2}}\!\int\!\!d^3x\Bigg(
\text{exp}\big(-\i\vec{k}\@\cdot\@\vec{x}\big)\epsilon^a_{\uline{A}}\@\big(\vec{k}\big)\frac{k^b}{|\vec{k}|^{\frac{3}{2}}\!}\.
\uline{\Delta}_{ba}(x)
+\i\frac{1}{|\vec{k}|^{\frac{1}{2}}\!}\:\text{exp}\big(-\i\vec{k}\@\cdot\@\vec{x}\big)\epsilon_{a\uline{A}}\@\big(\vec{k}\big)
\uline{X}^{a\@,x}\@(\ell)
\Bigg)\,.
\end{split}
\end{align}
Then, these objects allow one to perform the standard canonical analysis, analogous to the Fock quantization. The detailed review of this method can be found in
\cite{Gambini:1996ik}.

In the following chapters, an analogous representation is going to be constructed concerning the gravitational field. However, this article is restricted only to the classical analysis of the lattice representation. Conversely to the aforementioned Wilson loops representation, the continuous-to-discrete transition problem is not going to be blurred by defining a special effective formalism at the quantum level. A rigorous resolution of the transition problem is the main issue of this article. Another difference regarding the lattice theory of gravity consists of the mentioned quantum level. In the case of the gravitational field, one cannot define the wave function transform analogously to \eqref{el_state}. Here, the difficulty would be related to the presence of the metric tensor in the analogous definition of a state, and it would occur in the scalar product definition. The construction of the quantum theory that is not involved in the mentioned difficulties can be rigorously determined by using the methods elaborated in this article, which are based directly on the analogous formulation of CLQG \cite{Ashtekar:1994mh,Ashtekar:1994wa,Ashtekar:1995zh,Thiemann:1996aw,Thiemann:1996av,Thiemann:1997rv,Fleischhack:2004jc,Lewandowski:2005jk}. Then, the complete model will be presented in the next planned manuscript concerning the whole program of relativistic lattice field theory \cite{Bilski:2021_RCT_III}.

	%%%%%%%%%	%%%%%%%%%	%%%%%%%%%	%%%%%%%%%

	%%%%%%%%%	%%%%%%%%%	%%%%%%%%%	%%%%%%%%%
\section{Discrete regularization}\label{II}

	%%%%%%%%%	%%%%%%%%%	%%%%%%%%%	%%%%%%%%%
\subsection{Gauge bosons}\label{Sec_Regularization_/_Gauge_bosons}

\noindent
The methodology of this project distinguishes the fields that follow Bose-Einstein and Fermi statistics \cite{Bilski:2021_RCT_I}. Gravitational selfinteractions are represented by gauge bosons, hence the lattice smearing of fermionic fields is not going to be discussed in this article. Moreover, in the case of the analysis of the bosonic fields, only the regularization methods are going to be presented. The step-by-step procedure of the lattice smearing of the classical continuous gravitational action will be described in the forthcoming manuscript \cite{Bilski:2021_RCT_III}. Finally, although the discussed methods will be analyzed only in the context of gravity, all of them and all the related results concerning any other vector field's model are analogous.

To follow the notation introduced in Sec.~\ref{Sec_Introduction_/_Wilson}, a general $\mathfrak{su}(N)$-valued vector field's decomposition is expressed by
\begin{align}
\label{vector_field}
\uline{A}:=\uline{A}_a^I\uline{\tau}_Idx^a\,.
\end{align}
The spatial basis element $dx^a$ is related to the differential volume form with the Euclidean signature by the definition ${dx^x\@\wedge\@dx^y\@\wedge\@dx^z}=:d^3x$. The object $\uline{\tau}_I$ denotes an $\mathfrak{su}(N)$ generator. However, conversely to the standard convention concerning the Wilson loops model in which $\uline{\sigma}_I$ represents a Hermitian unitary generator, the CLQG-standardized normalization, $\uline{\tau}_I:={-\i\uline{\sigma}_I/2}$, is applied \cite{Bilski:2020xfq}. Consequently, the $\mathfrak{su}(N)$ traceless antihermitian (skew-Hermitian) representation that satisfies the commutation relation
\begin{align}
\label{Lie_real}
[\uline{\tau}_J,\uline{\tau}_K]=\frac{1}{2}C^I_{\ JK}\uline{\tau}_I\,,
\qquad
I,J,K,...=1,2,...,N^2-1\,,
\end{align}
is postulated, where $C^I_{\ JK}=-C^I_{\ KJ}\in\mathds{R}$. The precise normalization convention for the generators depends on the type of representation, for instance defining, adjoint, etc. In the case of the traceless antihermitian unitary matrices, the defining representation labeled by ${\uline{\tau}_{\@\!^\texttt{d}}\@}$ is uniquely determined. The generators are the ${N\@\times\@N}$ matrices that satisfy $\uline{\text{tr}}\big({\uline{\tau}_{\@\!^\texttt{d}}\@}_J,{\uline{\tau}_{\@\!^\texttt{d}}\@}_K\big)=-\delta_{JK}/2$. Consequently, the vector field coefficients in the defining representation can be derived from the equation
\begin{align}
\label{trace_suN}
\uline{A}_a^J=-2\,\text{tr}\big({\uline{A\:}_{\@\!^\texttt{d}}\@}_a{\uline{\tau}_{\@\!^\texttt{d}}\@}^J\big)\,.
\end{align}
These coefficients are postulated to represent the propagating degrees of freedom of the fundamental fields that follow Bose-Einstein statistics \cite{Bilski:2021_RCT_I}.

The internal space dimension of the fundamental vector potential $\uline{A}_a^J$ varies for different fields. The $N=1$ Abelian field describes electromagnetism, and its isotropic variant can be a model for the inflaton. The $N=2$ group is related to the weak isospin gauge; different representations are models for different particle multiplets. This symmetry group also describes gravity, and the related representation appears to be the defining one\footnote{This \textit{a priori} indeterminacy of the matrix representation selection should be factually dependent on the quantum behavior of the related particles. As long as the eigenstates corresponding to a graviton, antigraviton, and possibly a third particle (if the adjoint representation is selected) are not experimentally indicated, the proper choice of the representation cannot be determined. There are no quantum-gravitational experiments, hence any guessing related to this problem is far from being practicable. In the case of formalism, different representations of the same group require different normalizations. It is not easy to generalize all the constructions into a representation-independent formalism, however, it is easy to repeat all the derivations related to an already described representation. In this article, the defining (physicists often call it fundamental) representation is selected, because it is the simplest one.}. In quantum chromodynamics, the SU$(3)$ model of gluon interactions is related both to the defining flavor representation symmetry and to the adjoint representation of the color octet. The methods elaborated in this article are general. More precisely, although the $\mathfrak{su}(2)$ representation of the internal symmetry is going to be implicitly assumed in several expressions, these methods hold, up to constants, for any Lie group.

For a given action $S^{\scriptscriptstyle\uline{A}}:={S[\uline{A},\partial\@\uline{A},\partial^2\!\uline{A},...]}$ (derivatives can be specified according to any directions), the canonically conjugate momentum coefficient is defined as follows:
\begin{align}
\label{vector_momentum}
\uline{E}^a_J:=\frac{\delta S^{\scriptscriptstyle\uline{A}}}{\delta\partial_t\uline{A}_a^J}\,.
\end{align}
By choosing this definition, the Hamiltonian-Dirac formalism \cite{Hamilton:1834,Hamilton:1835,Dirac:1950pj}, instead of the De Donder-Weyl one \cite{De_Donder:1930,Weyl:1935}, is selected. In the case of gravity, this choice postulates TGA instead of GCA. It is worth emphasizing that the quantity $\uline{E}^t_J$ does not exist concerning any standard action, for instance, the Einstein-Hilbert \cite{Einstein:1916vd}, Palatini \cite{Palatini:1919}, Proca \cite{Proca:1900nv}, Yang-Mills \cite{Yang:1954ek} action, or the Ashtekar-Barbero-Mercuri-Bojowald-Das \cite{Ashtekar:1986yd,Ashtekar:1987gu,Barbero:1994ap,Mercuri:2006um,Bojowald:2007nu} formalism. Moreover, the $\uline{A}_t^J$ internal space vector is the Lagrange multiplier for the constraint, known as the Gauss constraint, which defines the $\mathfrak{su}(N)$ symmetry. Furthermore, the formalism based on expression \eqref{vector_momentum} is not covariant, and the resulting system is independent of so-called ghost degrees of freedom.

Other important quantities are the vector potential curvature $\uline{F}_{ab}^J$ and the related axis-angle vector coefficient (density) $\uline{B}^a_J:=\frac{1}{2}\epsilon^{abc}_{\scriptscriptstyle+}\uline{F}_{bc}^J$. It should be emphasized that both $\uline{E}^a:=\uline{E}^a_J\tau^J$ and $\uline{B}^a:=\uline{B}^a_J\tau^J$ are $\mathfrak{su}(N)$-valued vector densities. The fact they have a weight $1$ results from different reasons. In the case of the momentum, the weight is transferred into this object from the invariant integration measure in the definition in \eqref{vector_momentum}. In the case of the $\uline{B}^a$ field, this density feature is related to the weight $1$ of the Levi-Civita symbol $\epsilon^{abc}_{\scriptscriptstyle+}$. Furthermore, the corresponding Hamiltonian constraint density is also a weight $1$ quantity (a scalar density).

The Poisson brackets for the canonical fields are defined in the standard manner,
\begin{align}
\label{Poisson_brackets}
\{X,Y\}_{\!_{\uline{A},\uline{E}}\!}:=\!\int\!\!d^3x\bigg(
\frac{\delta X}{\delta\uline{A}_a^I(x)\!}\.\frac{\delta Y}{\delta\uline{E}^a_I(x)\!}
-\frac{\delta X}{\delta\uline{E}^a_I(x)\!}\.\frac{\delta Y}{\delta\uline{A}_a^I(x)\!}\.
\bigg),
\end{align}
where $X(y):=X\big[\uline{A}_b^J(y),\uline{E}^c_K(y)\big]$ and $Y(z):=Y\big[\uline{A}_b^J(z),\uline{E}^c_K(z)\big]$ are arbitrary functionals of the canonical variables (and their derivatives of any order). Then, any transformation of these variables or interaction between fundamental fields requires symplectic analysis.

In this project, the gravitational field is described by the pair of the $\mathfrak{su}(2)$-valued fields, similar to the Ashtekar variables\footnote{Actually, the Ashtekar variables are written in the expressions in \eqref{densitized_dreibein} and \eqref{Ashtekar_connection}.} \cite{Ashtekar:1986yd,Ashtekar:1987gu}. The gravitational connection and its canonically conjugated momentum are defined by the formulas:
\begin{align}
\label{ABMBD_connection}
A_a&=A_a^i\tau^i:=\bigg({\!\@}\stackrel{\!\scriptscriptstyle\textsc{\!a\!\.-\!\.b}\!}{A}{\@\@}_a^i+\frac{\kappa}{8}e^i_aJ^0\bigg)\tau^i
=\bigg({\!\@}\stackrel{\!\scriptscriptstyle\textsc{\!t\!\.-\!\.f}\!}{\Gamma}{\!\@}^i_a+\gamma{\!\@}\stackrel{\!\scriptscriptstyle\textsc{\!t\!\.-\!\.f}\!}{K}{\!\@}^i_a+\frac{\kappa}{8}\epsilon^{ijk}e^j_aJ^k\bigg)\tau^i\,,
\\
\label{densitized_dreibein}
E^a&=E^a_i\tau^i:=\sqrt{q}e_i^a\tau^i\,.
\end{align}
Here, $\gamma$ denotes the real Barbero-Immirzi parameter \cite{Barbero:1994ap,Immirzi:1996dr,Rovelli:1997na} and the determinant of the spatial metric tensor related to the Cauchy hypersurface reads
\begin{align}
\label{determinant}
q:=\.\text{det}(q_{ab})=\bigg|\frac{1}{3!}\epsilon_{abc}^{\scriptscriptstyle-}\epsilon^{ijk}E^a_iE^b_jE^c_k\bigg|\,.
\end{align}
The object $\epsilon_{abc}^{\scriptscriptstyle-}$ is the inverse tensor density, hence it is a quantity of a weight $-1$.

The object in \eqref{ABMBD_connection} is the Ashtekar-Barbero-Mercuri-Bojowald-Das (names in the chronological order) real $\mathfrak{su}(2)$-valued torsion-full vector field representative of the gravitational propagating degrees of freedom \cite{Ashtekar:1986yd,Ashtekar:1987gu,Barbero:1994ap,Mercuri:2006um,Bojowald:2007nu}. The Ashtekar-Barbero connection coefficient \cite{Ashtekar:1986yd,Ashtekar:1987gu,Barbero:1994ap} is defined by the formula:
\begin{align}
\label{Ashtekar_connection}
{\!\@}\stackrel{\scriptscriptstyle\textsc{\!a\!\.-\!\.b}\!}{A}{\!\@}_a^i:=\frac{1}{2}\epsilon^{ijk}\Gamma_{jka}+\gamma\Gamma^i_{\ 0a}\,,
\end{align}
where $\Gamma^{\alpha}_{\ \beta a}$ is the (torsion-full in general) Lorentz connection coefficient decomposed in the spatial basis $dx^a$ (and in the internal Minkowski space basis) --- see \eqref{vector_field}. Considering only the internal space components, one can identify two Lorentz connection coefficients that correspond to different symmetry subgroups. The first one is the axis-angle vector $\Gamma^i=\Gamma^i_adx^a:=\frac{1}{2}\epsilon^{ijk}\Gamma_{jka}dx^a$, known as the spin connection, which is the coefficient for the SO$(3)$ group. The second one is the rapidity $K^i=K^i_adx^a:=\Gamma^i_{\ 0a}dx^a$ that parametrizes boosts. The labeling `${{}\stackrel{\!\scriptscriptstyle\textsc{\!t\!\.-\!\.f}\!}{}{\!}}$' in \eqref{ABMBD_connection} denotes torsion-free variables, where ${{\!\@}\stackrel{\!\scriptscriptstyle\textsc{\!t\!\.-\!\.f}\!}{K}{\!\@}^i_a}={e^b_i{\!\@}\stackrel{\!\scriptscriptstyle\textsc{\!t\!\.-\!\.f}\!}{K}{\!\@}_{ab}}$, and ${{\!\@}\stackrel{\!\scriptscriptstyle\textsc{\!t\!\.-\!\.f}\!}{K}{\!\@}_{ab}}$ is the spatial extrinsic curvature.

Another just introduced object is the fermionic axial current, specified as follows,
\begin{align}
\label{current}
J^{\alpha}:=\i\bar{\Psi}\gamma^5\gamma^{\alpha}\Psi\,,
\qquad
\gamma^5:=\i\gamma^0\gamma^1\gamma^2\gamma^3\,.
\end{align}
The complete introduction of the fermionic coupling to the Ashtekar-Barbero connection is scattered across a few publications, \textit{cf.} \cite{Mercuri:2006um,Bojowald:2007nu,Thiemann:1997rt,Thiemann:2007zz,Thiemann:1997rq,Bojowald:2010qpa}.

Finally, the $\mathfrak{su}(2)$ symmetry generator in the defining representation takes the form $\tau^i:=-\frac{\i}{2}\sigma^i$, where $\sigma^i$ is a Pauli matrix. This representation leads to the following normalization regarding the gravitational field representatives derivation (based on equation \eqref{trace_suN}),
\begin{align}
\label{trace_su2}
A_a^i=-2\,\text{tr}(A_a\tau^i)\,,
\qquad
E^a_i=-2\,\text{tr}(E^a\tau^i)\,.
\end{align}
By satisfying the relation in \eqref{Lie_real}, the reality of the canonical variables is determined --- see \cite{Barbero:1994ap,Mercuri:2006um,Bojowald:2007nu,Thiemann:2007zz,Bojowald:2010qpa,Holst:1995pc} for the related Hamiltonian formalism introduction.

The geometrical foundation of the gravitational representation, given in terms of the fields' coefficients in \eqref{trace_su2}, is based on the postulate that the metric tensor is a dynamical field --- see Sec.~\ref{Sec_Regularization_/_ADM}. This assumption coincides with the methodology of this program in which the scales of a distance are considered as the gravitational propagating degrees of freedom, but relativity is a non-dynamical symmetry, \textit{cf.} \cite{Bilski:2021_RCT_I} and Sec.~\ref{Sec_Introduction_/_Equivalence_principle}. By assuming that the spatial metric tensor $q_{ab}$, derived from the spacetime metric $g_{\mu\nu}$ in the Hamilton formalism, is the fundamental canonical field \cite{Arnowitt:1959ah,Arnowitt:1960es,Arnowitt:1962hi}, the following relation holds:
\begin{align}
\label{Poisson_qp}
\big\{A^i_a(x),E_j^b(y)\big\}_{\!\!_{q\@,p}\!\!}=-\frac{\gamma\kappa}{2}\delta_a^b\delta^i_j\.\delta^3\@(x-y)\,.
\end{align}
\textit{cf.} \cite{Barbero:1994ap,Bojowald:2007nu,Bojowald:2010qpa,Holst:1995pc}. Here, the Poisson brackets were derived regarding the metric field $q_{ab}$ and its canonically conjugated momentum $p^{ab}$ --- compare the relation given in the next section, in \eqref{Poisson_ADM}. In this article and in this program in general, all the bosonic fields are represented by the same abstract quantity, the canonical pair of the $\mathfrak{su}(N)$-valued vector fields, given in \eqref{trace_suN} and \eqref{vector_momentum}. Then, since the gravitational variables in \eqref{ABMBD_connection} and \eqref{densitized_dreibein} are the result of the canonical transformation of the ADM variables $q_{ab}$ and $p^{ab}$, the following normalization of the gravitational Poisson brackets is determined from the relation in \eqref{Poisson_qp}, namely
\begin{align}
\label{Poisson_AE}
\{X,Y\}:=-\frac{\gamma\kappa}{2}\{X,Y\}_{\!_{A,E}\!}\,,
\end{align}
where the general definition of the brackets is given in \eqref{Poisson_brackets}.

	%%%%%%%%%	%%%%%%%%%	%%%%%%%%%	%%%%%%%%%
\subsection{Thiemann's regularization of the ADM model in the Ashtekar representation}\label{Sec_Regularization_/_ADM}

\noindent
The application of the Hamiltonian-Dirac formalism \cite{Hamilton:1834,Hamilton:1835,Dirac:1950pj} to the symmetric metric tensor field description of gravity was formulated by Arnowitt, Deser, and Misner \cite{Arnowitt:1959ah,Arnowitt:1960es,Arnowitt:1962hi}. This method defines the direction of this research program. In the resulting TGA, the metric field is decomposed as follows,
\begin{align}
\label{ADM_metric}
g_{\mu\nu}dx^{\mu}dx^{\nu}
=(N^aN_a-N^2)dt^2+2N_adtdx^a+q_{ab}dx^adx^b\,.
\end{align}
The implementation of this formula into the Einstein-Hilbert action leads to the identification of Lagrange multipliers. They are the quantity labeled by $N$ that is known as the lapse function, and the object defined by $N^a$ being called the shift vector. The dynamical quantity is the spatial metric $q_{ab}$. However, the dynamics related to this ${3\@\times\@3}$ symmetric tensor is restricted additionally by gauge invariance. From the twelve-dimensional phase space spanned by the variables $q_{ab}(x)$ and $p^{ab}(x)$ at point $x$, eight dimensions do not correspond to physical propagating degrees of freedom. Four of these dimensions are removed by the constraints solutions (associated with the Lagrange multipliers $N$, $N^x$, $N^y$, $N^z$). Another four correspond to the orbits of the physical configurations, which are distinguishable only by (non-dynamical) gauge transformations. As a result, the gravitational field has four-dimensional physical phase space, hence two propagating degrees of freedom.

The decomposition in \eqref{ADM_metric} has been adjusted to explicitly demonstrate how SE is implemented to the system that is not written explicitly in the covariant formalism. The algebra of constraints reveals that the coordinate systems equivalence splits into the three-diffeomorphism invariance and the time reparametrization symmetry. This splitting should not be surprising. It is a natural consequence of choosing the non-covariant Hamiltonian-Dirac formalism in which equations of motion are the identities derived regarding the fixed coordinate time. Thus, by assuming that the geometry of space or spacetime is a dynamical field, all of its nontrivial temporal transformations are determined by equations of motion. The pure spatial transformations do not modify the form of equations of motion. Thus, the corresponding symmetry is expected to be time-independent, hence be related to the Noether theorem. Only a system with a static geometry results in spacetime invariant under four-diffeomorphisms. It is worth mentioning that the construction of an explicitly covariant formalism, which describes four-diffeomorphisms-invariant transitions between fields' configurations, is \textit{a priori} possible. In this case, however, the transitions would describe different changes than dynamical evolution, which, by definition, is a change relative to some specified time coordinate.

%
%%%	FIGURE	%%%
%
\begin{figure}[h]
\vspace{5pt}%
\begin{center}
\begin{tikzpicture}[scale=1]
\draw[cl15]plot[smooth, tension=0.5] coordinates 
{(-2,2) (-1.5,2.2) (-1,2.3) (-0.5,2.3) (0,2.2) (0.5,2.12) (1,2.08) (1.5,2.1) (2,2.2)};
\draw[cl15]plot[smooth, tension=0.5] coordinates 
{(-3,1) (-2.5,1.2) (-2,1.3) (-1.5,1.3) (-1,1.2) (-0.5,1.12) (0,1.08) (0.5,1.1) (1,1.2)};
\draw[cl15] (-3,1) -- (-2,2);
\draw[cl15] (1,1.2) -- (2,2.2);
\node at (-1.7,1.55) {$\Sigma_{t+\delta t}^{\textsc{a\@d\@m}}$};
\draw[cl15]plot[smooth, tension=0.5] coordinates 
{(-2,0) (-1.5,0.2) (-1,0.3) (-0.5,0.3) (0,0.2) (0.5,0.12) (1,0.08) (1.5,0.1) (2,0.2)};
\draw[cl15]plot[smooth, tension=0.5] coordinates 
{(-3,-1) (-2.5,-0.8) (-2,-0.7) (-1.5,-0.7) (-1,-0.8) (-0.5,-0.88) (0,-0.92) (0.5,-0.9) (1,-0.8)};
\draw[cl15] (-3,-1) -- (-2,0);
\draw[cl15] (1,-0.8) -- (2,0.2);
\node at (-1.75,-0.45) {$\Sigma_t^{\textsc{a\@d\@m}}$};
\draw (-0.9,-0.2) -- node[auto, outer sep=-2.5, pos=0.6] {$Nu^{\mu}\delta t$} (-0.9,1.12);
\draw[dashed, ->] (-0.9,1.25) -- (-0.9,1.75);
\draw (-0.9,-0.2) -- (-0.12,1.05);
\draw[dashed, ->] (-0.06,1.15) -- node[auto, outer sep=-3, pos=0.3] {$v^{\mu}\delta t$} (0.3,1.75);
\draw[->] (-0.9,-0.2) -- node[auto, outer sep=-2, pos=0.6] {$N^{\mu}\delta t$} (0.3,-0.2);
\draw[dotted] (-0.9,1.75) -- (0.3,1.75);
\draw[dotted] (0.3,-0.2) -- (0.3,1.75);
\end{tikzpicture}
\end{center}
\vspace{-10pt}%
\caption{ADM decomposition into Cauchy hypersufaces $\Sigma_t^{\textsc{a\@d\@m}},\Sigma_{t+\delta t}^0,\.$...}
\label{ADM_Sigma_t}
\end{figure}
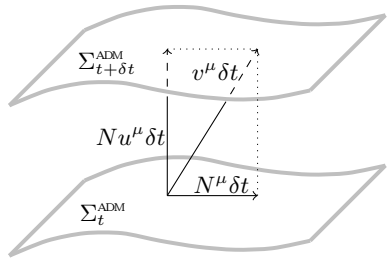
%
%%%	FIGURE	%%%
%
Another worth emphasizing aspect of using the relation in \eqref{ADM_metric}, is the resulting universality of the time coordinate. This coordinate is not related yet to any dynamics. All possible transformations between different time normalizations are trivial gauge transformations. Therefore, this coordinate is the only one, the role of which is the same both in the case of nonperturbative canonically quantized GR and QFT on curved spacetime. The time coordinate does not evolve. In other words, the time flow vector $v^{\mu}$ (the rate of the proper time flow with respect to the coordinate time) decomposes into diffeomorphism transformations and the normalized direction of the hypersurfaces foliation rate $Nu^{\mu}$ --- see FIG~\ref{ADM_Sigma_t}. As a result, if one defines a gauge-invariant scalar product at the quantum level, this product will remain unchanged during the evolution of the system. From the theoretical point of view, this is the main advantage of using TGA.

Finally, in this project, the tensor field formalism is going to be replaced by the gauge-valued vector field formalism. The canonical variables $q_{ab}(x)$ and $p^{ab}(x)$ define the Poisson structure for gravitational kinematics,
\begin{align}
\label{Poisson_ADM}
\big\{q_{ab}(x),p^{cd}(y)\big\}_{\!\!_{q\@,p}\!\!}=\delta_{(a}^c\delta_{b)}^d\.\delta^3\@(x-y)\,,
\end{align}
where the brackets in the factor $\delta_{(a}^c\delta_{b)}^d$ denote a symmetric pair of indices. This definition leads to the relation in \eqref{Poisson_qp}, and consequently sets the normalization of the gravitational Poisson structure in \eqref{Poisson_AE}.

The formulation of the canonically quantizable gravitational field formalism is possible by introduction of the lattice representation. The latter structure is needed to define the aforementioned gauge-invariant scalar product and thus the metric-independent quantum evolution of fundamental fields. The analysis in this article is focused on the maximal precision of the lattice embedding in the spatial Cauchy hypersurface. The SU$(2)$ invariance of this lattice and its diffeomorphism invariance were demonstrated in \cite{Ashtekar:1993wf,Marolf:1994cj,Ashtekar:1994mh,Ashtekar:1994wa,Ashtekar:1995zh} and \cite{Zapata:1997db,Zapata:1997da}, respectively. These aspects will be reviewed in the next installment of the series related to this analysis \cite{Bilski:2021_RCT_III}. Here, we are going to investigate only the continuous-to-discrete transition of the canonical pair of the gravitational variables in \eqref{trace_su2}. This transition defines the lattice structure, specifies the UV regularization\footnote{As usually, the UV limit means the maximal value that an observable can take.} of the Hamiltonian, and provides the geometrical representation of gravitational degrees of freedom.

The basic geometrical object considered in this analysis is the three-dimensional topological manifold $\Sigma_t^{\textsc{a\@d\@m}}$ that has the Euclidean signature and satisfies the axioms of a perfectly normal Hausdorff space. This latter property guarantees that the manifold is metrizable. In this space, the graph structure $\Gamma$, on which the discrete holonomy-flux formulation of the phase space is founded, is embedded. The result of this operation is going to be denoted by $\Sigma_t^{\hash}$ and called a `mosaic manifold' or simply a `mosaic'. This quantity is assumed to be constructed by a tessellation of $\Sigma_t^{\textsc{a\@d\@m}}$, and thus it is composed of cells, the edges of which are identified with links of a piecewise linear lattice. By this assumption, the $\Sigma_t^{\textsc{a\@d\@m}}$ metric space, hence a Hausdorff space, transfers its property of separability into the discrete formalism, which entails divisibility of the piecewise analytic lattice.

The worth mentioning method of tessellation is triangulation. The related graph is denoted by $\Gamma^{\textsc{t\@h}}$ (indicating the tetrahedral form of elementary cells). The triangulation procedure is a technique that was selected to define the lattice for CLQG \cite{Thiemann:2007zz,Regge:1961px,Ambjorn:2012jv}. This technique is applicable to more general manifolds\footnote{Each differentiable manifold is piecewise-linear triangulable \cite{Whitehead:1940}. In the case of two- and three-dimensional topological manifolds, the triangulation is always construable and unique (up to piecewise-linear equivalence) \cite{Moise:1977,Thurston97}. Each of these manifolds admits a unique (up to diffeomorphisms) smooth structure \cite{Munkres:1960,Whitehead:1961,Thurston97}.} than the metrizable spaces (studied in this article) and allows one to form a piecewise analytical graph on which the regularization of the just mentioned model is implemented. The reason for choosing the piecewise analytical graph was the lack of a precise classical relation between the continuous structure of $\Sigma_t^{\textsc{a\@d\@m}}$ and the piecewise linear structure of the lattice in $\Sigma_t^{\hash}$ --- see the discussion in \cite{Thiemann:2007zz}. In the case of CLQG, the graph structure is originally introduced during the quantization step and the continuous property i.e. a smoothness category emerges as a superselection of continuous pieces, each being diffeomorphism invariant in the analytic sense \cite{Thiemann:1997rv}. The disadvantage of this approach is the lack of a precise relation between the quantum and classical theory. For instance, during the derivation of the classical limit of CLQG, one has to define the procedure of a metricity introduction on a triangulable manifold. Up to the knowledge of the author, this problem is not clearly resolved so far. Thus, this article is dedicated to complete the relation between the continuous and discrete formalisms at the classical level. The expected result is the lattice on which the quantum theory is defined and which will be embeddable in a tessellable manifold, not only in a triangulable one. The metric geometry of the mosaic manifold (the tesselated metrizable space) can be consequently interpreted as the affine geometry of the piecewise linear lattice that has been created at the classical level by the manifold's tessellation.

In this article, a particular honeycomb with a higher than the tetrahedral symmetry is going to be investigated in detail. The links of this honeycomb form a quadrilaterally hexahedral graph $\Gamma^{\textsc{q\@h}}$. This quantity provides the most convenient form of cells, which regular structure allows one to define particularly efficient derivation methods and constructions. The invariance of the gravitational variables, which is encoded in the gauge constraints that do not include propagating degrees of freedom, results in specific relations located at cells' boundaries. These relations are determined by solving the corresponding first-class constraint on the lattice. The sum of all quadrilaterally hexahedral cells is then a fiber sum, and the diffeomorphism symmetry implemented at the boundaries specifies the preservation of orientations between bounded regions. From now on we assume that this specification of boundaries is an immanent property of the mosaic manifold. However, the implementation of the diffeomorphism constraints on a preferred lattice $\Gamma$ in $\Sigma_t^{\hash}$ (by the procedure analogous to the one in CLQG \cite{Ashtekar:2004eh,Thiemann:2007zz}) will be described in \cite{Bilski:2021_RCT_III}. Thus, in this manuscript, some specifications of the lattice are only postulated, and their consistency is verified, but their derivation has to be yet demonstrated in the forthcoming article.

In what follows, particular regularization methods will require an additional assumption. This assumption is the local homogeneity, understood as the approximate constancy of the metric tensor determinant within each elementary region. The mosaic manifold satisfying this assumption will be denoted by $\Sigma_t^{\bar{\hash\.}\@}$. This condition, however, is not indispensable. It will be demonstrated that the regularization can be founded both (in general) on $\Sigma_t^{\hash}$ and (under particular assumptions) on $\Sigma_t^{\bar{\hash\.}\@}$\footnote{The approximate constancy of the metric tensor determinant within each elementary cell is redundant concerning the exact regularization, which is proposed in \eqref{e-regularization}. However, in the standard CLQG approach, in which the volume functional is needed to postulate the indispensable relation in \eqref{v-regularization}, one has to \textit{ad hoc} impose this restriction on all the cells in $\Sigma_t^{\hash}$. Thus, one has to select the manifold $\Sigma_t^{\bar{\hash\.}\@}$ instead.}.

Concluding the introduction of the regularization procedure presentation, the mosaic manifold $\Sigma_t^{\hash}$ is a metrizable Cauchy hypersurface with the embedded piecewise linear lattice, which is specified by the structure of the selected tessellation. This lattice\footnote{The lattice determined regarding the gravitaional field is indroduced in Sec.~\ref{Sec_Regularization_/_graph}.} is the quantity on which the loop representation for $A_a$ is founded by following Wilson's idea sketched in Sec.~\ref{Sec_Introduction_/_Wilson}. However, in the case of $E^a$-dependent objects, the lattice representation is defined differently.

One of the motivations to look for a representation different than in the ADM approach was the presence of the $E^a$-dependent functional in a denominator in the Hamiltonian constraint both in the case of the gravitational field's and matter fields' contribution --- see \cite{Barbero:1994ap,Mercuri:2006um,Bojowald:2007nu,Thiemann:2007zz,Bojowald:2010qpa,Holst:1995pc}. This problem concerns in particular the appearance of the negative (and fractional) power of the metric tensor determinant $q$, expressed by the gravitational field in \eqref{determinant}.

The most complete solution to this problem was proposed by Thiemann. The regularization of the UV-divergent factor of the gravitational scalar constraint is realized by using the formula introduced in \cite{Thiemann:1996aw},
\begin{align}
\label{trick_gravity_TT}
\Bigg(\text{sgn}\big(\text{det}(E^a_i)\big)\epsilon_{ijk}\frac{E^b_jE^c_k}{\sqrt{q}}\Bigg)(x)
=\bigg(\epsilon^{abc}_{\scriptscriptstyle+}\frac{\sqrt{q}}{E_i^a}\bigg)(x)\stackrel{?}{=}
2\epsilon^{abc}_{\scriptscriptstyle+}\frac{\delta\mathbf{V}(R)}{\delta E_i^a(x)}
=-\frac{4}{\gamma\kappa}\epsilon^{abc}_{\scriptscriptstyle+}\big\{A^i_a(x),\mathbf{V}(R)\big\}\,,
\end{align}
where
\begin{align}
\label{volume}
\mathbf{V}(R):=\!\int_{\!R}\!\!d^3y\sqrt{q(y)}
\end{align}
is the volume of a region $R$ in a three dimensional space. One of the equalities is labeled by the question mark to indicate that this identity is correct only by assuming $\forall_{\@x}\,x\in R\,\wedge\,\forall_{\@r\in R}\,r=x$. Thus, the expression in \eqref{trick_gravity_TT} introduces some limitation on the continuous quantity on the left-hand side, which for a general formula $F(x)$ takes the form $F(x)\big|_{x\in R}$. This assumption justifies the following identity,
\begin{align}
\label{trick_gravity_delta}
2\frac{\delta}{\delta E_i^a(x)}\int_{\!R}\!\!d^3y\sqrt{q(y)}\.\bigg|_{y\in R}\!=\frac{\sqrt{q(x)}}{E_i^a(x)}\.\bigg|_{x\in R}\,,
\end{align}
which was implicitly applied to \eqref{trick_gravity_TT}; the need for adding this implicit remark was labeled by `$?$'.

Concerning the whole Cauchy hypersurface $\Sigma_t^{\textsc{a\@d\@m}}$, one can either identify this hypersurface with the region $R$, getting
\begin{subequations}
\begin{align}
\label{trick_gravity_whole}
\frac{\sqrt{q(x)}}{E_i^a(x)}=-\frac{4}{\gamma\kappa}\big\{A^i_a(x),\mathbf{V}(\Sigma_t^{\textsc{a\@d\@m}})\big\}\,,
\intertext{or divide the Cauchy hypersurface into a connected sum of smaller regions, $\Sigma_t^{\textsc{a\@d\@m}}\@\to\Sigma_t^{\hash^{\@\textsc{r}}}:=\bighash_{\alpha}R_{\alpha}$. In this latter case, the mosaic domain of the scalar constraint becomes separable,}
\label{trick_gravity_discrete}
H^{\textsc{\.\!g\.\!r\.\!}}\big(\Sigma_t^{\hash^{\@\textsc{r}}}\big):=\sum_{\alpha}H^{\textsc{\.\!g\.\!r\.\!}}(R_{\alpha})\,.
\end{align}
\end{subequations}
It is worth noting that this latter possibility is accessible due to the scalar nature of the energy represented by a scale-independent Hamiltonian constraint $H:=\@\int\!d^3xN\mathcal{H}$ (its density divided by the square root of the metric determinant, $\mathcal{H}/\sqrt{q}$, scales with a volume). Moreover, the absence of singularities is assumed, hence any topological problems in the construction of $\Sigma_t^{\hash}$ are not expected. From the quantum point of view, it is anticipated that all the regions, which are the neighborhoods of the singular solutions in classical continuous GR, should not be described by these solutions. These regions should be rather investigated in a quantum formalism, or at least their semiclassical descriptions should be preferred over the classical ones.

	%%%%%%%%%	%%%%%%%%%	%%%%%%%%%	%%%%%%%%%
\subsection{Gravitational degrees of freedom in the matter sector}\label{Sec_Regularization_/_volume}

\noindent
In this article, the mosaic manifold $\Sigma_t^{\hash}$ is assumed. The elementary cells are the polyhedra, the edges of which are identified with the $\Gamma$ graph that is determined by the tessellation of the metrizable manifold $\Sigma_t^{\textsc{a\@d\@m}}$. In this and the next sections, lattice regularization procedures are introduced. It is important to create procedures, which are well-defined both before and after solving the dynamics-independent gauge constraints that are the components of the total Hamiltonian expressed by the variables $A^i_a$ and $E^a_i$. These constraints are the $\mathfrak{su}(2)$ Gauss constraint and the spatial diffeomorphisms constraint. They are imposed on the lattice structure, in which the directions of graph's edges are determined according to the local orientation of cells-restricted Cartesian systems. This local determination of constant geometry reflects the idea of the existence of local inertial frames. The related directions are going to be introduced by the relation between edges-located SU$(2)$ holonomies (see \eqref{holonomy_general}) and continuously-distributed $\mathfrak{su}(2)$ connections $A^i_a$ --- analogously to the construction of the Wilson-like representation.

The solution of the gauge constraints is known from the formulation of the CLQG model \cite{Ashtekar:2004eh,Thiemann:2007zz} (see \cite{Bilski:2021_RCT_III} regarding the lattice gravity construction) and results in the following restrictions. At nodes (the intersections of links) appear the group averaging integrals of links-located SU$(2)$ irreducible representations (the matrices defining local basis directions in the internal space). Also at the nodes should be placed the group averaging of the spatial diffeomorphisms regarding links-located quantities, which does not change any subgraph's framing integer. This group averaging should be viewed as the set of projectors into equivalence classes of graphs under diffeomorphisms that preserve the discrete oriented structure of $\Gamma$ and do not modify linking numbers of any subgraph. In other words, it is required that the regularization determined in this article is implicitly gauge-invariant. A worth noting advantage of this formulation is the possibility of a simple application of the reduced phase space quantization method. In this context, it is worth mentioning that the recently founded early results concerning the quantizable cosmologically gauge-fixed system are very promising regarding the nonperturbative studies of the CMB anisotropies \cite{Bilski:2021fki}.

The gauge-invariant regularization procedure is included in the graph structure embedding that determines the mosaic manifold $\Sigma_t^{\hash}$. This manifold forms a collection of face-connected convex polyhedra, which faces are convex polygons with a fixed number of edges. The possibly non-Euclidean geometry\footnote{The $\Sigma_t^{\textsc{a\@d\@m}}$ manifold can be tessellated in a huge variety of honeycombs. In this article, however, it is going to be discussed mainly the quadrilaterally hexahedral one. For clarification, it is worth reminding that a quadrilateral hexahedron is defined as a convex polyhedron bounded by six quadrilateral faces, whose polyhedral graph is the same as that of a cube. To specify the nomenclature, it is worth recalling that this mathematical definition determines a more general type of solids than cuboids; these latter objects are rectangular hexahedra. The tessellation into irregular tetrahedra (the polyhedra with four triangular faces), which in the case of any three-dimensional metrizable space is equivalent to a triangulation, is also going to be discussed for comparison. This latter construction plays a significant role in the original regularization method in CLQG \cite{Thiemann:1996aw,Thiemann:2007zz}.} of $\Sigma_t^{\hash}$ is then revealed in the number of cells that share the same edge (a link of $\Gamma$). Concerning all this construction of the gauge-invariant field's representation on a lattice, one should return to the verification of the restrictions postulated by assuming the divergences-removing identity in \eqref{trick_gravity_TT}. Two interpretations of the region $R$ where proposed regarding that identity, namely $\Sigma_t^{\textsc{a\@d\@m}}=R$ in \eqref{trick_gravity_whole} and $\Sigma_t^{\textsc{a\@d\@m}}\@\to\Sigma_t^{\hash^{\@\textsc{r}}}:=\bighash_{\alpha}R_{\alpha}$ in \eqref{trick_gravity_discrete}. In the latter proposal, $\Sigma_t^{\textsc{a\@d\@m}}$ is not identified with the region $R$. Instead, the manifold $\Sigma_t^{\hash^{\@\textsc{r}}}$ is introduced, which is assumed to be decomposable into a connected sum, but this decomposition in \eqref{trick_gravity_discrete} is not \textit{a priori} assumed to be identical with the decomposition of the mosaic manifold $\Sigma_t^{\hash}=\bighash_{\alpha}C_{\alpha}$. This latter structure is consequently denoted as follows:
\begin{align}
\label{gravity_discrete}
H^{\textsc{\.\!g\.\!r\.\!}}\big(\Sigma_t^{\hash}\big):=\sum_{\alpha}H^{\textsc{\.\!g\.\!r\.\!}}(C_{\alpha})\,.
\end{align}
Thus, the size of regions $R_{\alpha}$ is not restricted (so far) to be smaller, bigger, or equal to the size of elementary cells $C_{\alpha}$. However, once specified, this size should be the same regarding any divergences-removing procedure analogous to \eqref{trick_gravity_TT}, which leads to further constraints imposed on the regularization of the $E^a$-dependent UV-divergent factors in the Hamiltonian.

Concerning the scalar constraint of matter fields, $H^{\textsc{\.\!m\@a\@t\.\!t\.\!}}$, this quantity requires the divergences-removing regularization, similarly to the gravitational Hamiltonian. This regularization was introduced in \cite{Thiemann:1997rt}, and it can be recast by using the relation analogous to \eqref{trick_gravity_delta},
\begin{align}
\label{trick_matter_delta}
\frac{2}{n}\frac{\delta}{\delta E_i^a(x)}\bigg(\int_{\!R}\!\!d^3y\sqrt{q(y)}\bigg)^{\!\!n}\.\bigg|_{y\in R}
=\mathbf{V}^{n-1\@}(R)\.\frac{\sqrt{q(x)}}{E_i^a(x)}\.\bigg|_{x\in R}
\stackrel{?}{=}\mathbf{V}_{\!\textsc{\!e\.\!u\@c\.\!l\!}}^{n-1\@}(\@R)\frac{\big[q(x)\big]^{\@\frac{n}{2}}\!}{E_i^a(x)}
\Big[1+\mathcal{O}\Big(\mathbf{V}_{\!\textsc{\!e\.\!u\@c\.\!l\!}}^{\frac{1}{3}\!}(\@R)\Big)\Big]^{\!n-1}\.\bigg|_{x\in R}\,,
\end{align}
where ${\mathbf{V}_{\!\textsc{\!e\.\!u\@c\.\!l\!}}(\!\.R)}$ is the constant spatial coordinate volume of the Euclidean geometry region bounded by $\partial R$. This latter flat region boundary is specified with respect to the boundary of each curved-geometry region $R_{\alpha}$,
\begin{align}
\label{boundary_region}
\forall_{\@\alpha}\,\partial\big(\@R^{\textsc{\.\!e\.\!u\@c\.\!l\.\!}}_{\alpha}\@\big)\approx\partial(R_{\alpha})\,.
\end{align}
To simplify notation, it is useful to define the length corresponding to this constant volume,
\begin{align}
\label{length_region}
\mathfrak{L}_{\alpha}:=\sqrt[3]{\mathbf{V}_{\!\textsc{\!e\.\!u\@c\.\!l\!}}(\@R_{\alpha}\@)}:=\bigg(\int_{\!R_{\alpha}}\!\!\!\!d^3y\bigg)^{\!\!\frac{1}{3}}.
\end{align}
Now, the presence of the question mark in expression \eqref{trick_matter_delta} can be formally specified as being related to the implicit assumption supporting the expansion of volume $\mathbf{V}(R_{\alpha})$ around $\mathfrak{L}_{\alpha}^3$. The possibility to neglect the next-to-the-leading-order terms in this formula has been based on the following approximation:
\begin{align}
\label{approximation_region}
\mathbf{V}(R_{\alpha})\approx\mathbf{V}(\bar{R}_{\alpha}):=\!\int_{\!R_{\alpha}}\!\!\!\!d^3y\sqrt{\bar{q}(y)}=\mathfrak{L}_{\alpha}^3\sqrt{\bar{q}_{\alpha}}\,.
\end{align}
Here, $\bar{R}_{\alpha}$ denotes the constant geometry region, i.e. the region with a constant metric determinant
\begin{align}
\label{constant_determinant}
\bar{q}_{\alpha}:=\frac{1}{\mathbf{V}(R_{\alpha})}\!\int_{\!R_{\alpha}}\!\!\!\!d^3y\,q^{\frac{3}{2}}(y)
\end{align}
that satisfies $\forall_{\@x\in R}\forall_{\@y\in R}\,\bar{q}(x)=\bar{q}(y)$.

It is worth mentioning that in the original construction in \cite{Thiemann:1997rt}, as well as in the improved review of this construction in \cite{Thiemann:2007zz}, an additional restriction on the form of the region $R$ was imposed. This restriction required that $\bar{R}$ is a cubic region\footnote{The author of this article did not find the precise explanation of how the cubic tessellation can approximate the non-Euclidean geometry of $\Sigma_t^{\textsc{a\@d\@m}}$. Furthermore, a much less understandable is the relation between the cubic tessellation in the regularization of the Hamiltonian and the triangulation of the Cauchy hypersurface. In the author's opinion, this point in the original formulation of the regularization of CLQG in \cite{Thiemann:1997rt,Thiemann:2007zz} should be improved, which can be easily done. One can simply assume the specification of the region $R$ analogous to the one given in this article. In the original case, this would be the approximately constant metric determinant within the boundary $\partial(R^{\textsc{t\@h}}_{\alpha})\approx\partial(\bar{R}^{\textsc{t\@h}}_{\alpha})$, where $R^{\textsc{t\@h}}_{\alpha}$ represents a tetrahedral region.}, centered at $x$. A different condition is going to be imposed in this article, by assuming only the approximate local homogeneity in \eqref{approximation_region}, which is less constraining than the original formulation, and the equivalence of the boundaries of the related regions in \eqref{boundary_region}. In other words, the approximately constant metric determinant within the boundary $\partial(R_{\alpha})\approx\partial(\bar{R}_{\alpha})$ is postulated.

After the clarification regarding the formula in \eqref{trick_matter_delta}, one can verify its applicability to the case $\Sigma_t^{\hash^{\@\textsc{r}}}\@\stackrel{?}{\equiv}R$. This issue would additionally require one to assume the approximate global constancy of $q$ on the whole Cauchy hypersurface, i.e. $\forall_{\@x,y\in\Sigma_t^{\hash^{\@\textsc{r}}}}\,q(x)\stackrel{?}{\approx}q(y)$. It is an unacceptable approximation for anyone looking for a universal regularization method applicable to any element of the scalar constraint\footnote{It should be reminded that concerning only the free gravitational sector, one does not need to verify the relation in \eqref{trick_matter_delta}. In this case, by neglecting the matter contribution, only the identity in \eqref{trick_gravity_delta} would need to be satisfied.}. Thus, the general relation between the expression in \eqref{trick_gravity_whole} and the formula in \eqref{trick_matter_delta} regarding the condition $R=\Sigma_t^{\hash^{\@\textsc{r}}}$ is excluded.

By selecting the only specification of the region $R$, which supports both the generally applicable expansion in \eqref{trick_gravity_delta} that justifies the identity in \eqref{trick_gravity_TT} and the extended expansion in \eqref{trick_matter_delta} that is needed to regularize the matter sector of the Hamiltonian constraint in the canonical formulation of GR, one imposes the aforementioned pair of conditions on the form of $\Sigma_t^{\hash^{\@\textsc{r}}}$. It is convenient to rephrase these conditions in a more rigorous way. To decompose the pre-regularized manifold $\Sigma_t^{\textsc{a\@d\@m}}$ into the connected sum, $\Sigma_t^{\hash^{\@\textsc{r}}}\@:=\bighash_{\alpha}R_{\alpha}\approx\bighash_{\alpha}\bar{R}_{\alpha}$, it is sufficient to postulate the following conditions.
\begin{enumerate}[a)]
\item
The geometry of the $R_{\alpha}$ region's boundary must be specified. Moreover, this geometry has to be equal from the perspective of both regions that are connected by the considered fragment of the boundary.
\item
To keep the method generally applicable, one needs to postulate that $\forall_{\@\alpha}\,R_{\alpha}\ll\Sigma_t^{\hash^{\@\textsc{r}}}$. This assumption allows neglecting the higher-order terms in $\mathfrak{L}_{\alpha}$ in \eqref{trick_matter_delta} and \eqref{boundary_region}.
\item
The determinant of the metric tensor has to be approximately constant locally, i.e. $\forall_{\@\alpha}\forall_{\@x,y\in R_{\alpha}}\,q(x)\approx q(y)$. This condition is related to the expressions in \eqref{approximation_region} and \eqref{constant_determinant}. Here, the locality is specified to the inclusion within the same element of the connected sum.
\end{enumerate}

The well-defined procedure of the cellular structure implementation on the metrizable manifold is the tessellation. This has to be used to support point a). Therefore, each region $R_{\alpha}$ should be either identified with a cell $C_{\alpha}$ or with a faces-connected composition of these cells. This identification specifies the existence of the oriented diffeomorphisms at the boundary $\partial(R_{\alpha})$. The metricity of the $\Sigma_t^{\hash^{\@\textsc{r}}}$ manifold entails its separability. This implication allows indicating the lower-dimensional submanifolds within the neighborhood in which manifolds $R_{\alpha}$ (the submanifolds of $\Sigma_t^{\hash^{\@\textsc{r}}}$) join. These submanifolds are the two-dimensional objects embedded as submanifolds into both elements of each pair of three-dimensional manifolds $R_{\alpha}$ and $R_{\beta}$ that share a sector $s$ of their boundary. Let this sector be denoted by $\partial(R_{\alpha})_{\@s}\equiv\partial(R_{\beta})_{\@s}$. Each of these two-dimensional submanifolds satisfies the following relation, ${\forall_{\@\gamma,\alpha\neq\beta}\,\big(\partial(R_{\alpha})_{\@s}\equiv\partial(R_{\beta})_{\@s}\equiv\partial(R_{\gamma})_{\@s}\big)}\Rightarrow{\big(R_{\gamma}\equiv R_{\alpha}\sqcup R_{\gamma}\equiv R_{\beta}\big)}$, i.e. the indicated boundary relates no more than two manifolds $R_{\alpha}$, except the boundaries of the boundaries $\partial(R_{\alpha}$). Finally, the identification of the sections of the boundaries of the pair $R_{\alpha}$ and $R_{\beta}$, along which these manifolds join together, needs to allow also on the separation along the section. This condition requires additional information about the orientation-preserving diffeomorphism transformations between the manifolds through this section (i.e. their common boundary). This clarification completes the list of conditions that specify the decomposition into a connected sum. It is worth emphasizing that the so-determined connected sum becomes also a fiber sum. Moreover, the last restriction regarding the diffeomorphism transformations between regions is satisfied by the mosaic manifold $\Sigma_t^{\hash}$, hence also by any other manifold in which the connected sum decomposition is defined along a submanifold of the honeycomb of $\Sigma_t^{\hash}$. Simply speaking, the regions $R_{\alpha}$ can be determined as the cells $C_{\alpha}$ or, in the less detailed manner, as the face-connected compositions of these cells.

The smaller the regions $R_{\alpha}$ are, the more strictly the condition in point b) is satisfied. The best result is obtainable by assuming that these regions are equal to cells. The smallest size of the cells is not determined. However, it has to exist; otherwise, they would be ill-defined. This restriction can be formalized as follows,
\begin{align}
\label{area_existence}
\forall_{\@\alpha}\,\partial(C_{\alpha})>0
\quad
\text{or}
\quad
\forall_{\@\alpha}\,\mathfrak{L}_{\alpha}>0\,.
\end{align}

The condition in point c) specifies the geometry of each cell to be well enough approximated by the constant geometry. If one considered regions composed of a few cells, the geometry of each one would need to be approximately the same (and constant). Hence the decomposition of any of these regions into cells would be senseless, and the whole region should be considered as a cell. However, due to the reasons that will be clarified in Sec.~\ref{Sec_Regularization_/_revisited}, the regions identified in this section with cells, namely $R_{\alpha}\equiv C_{\alpha}$, will be still considered as a result of a different decomposition and continuously denoted by $R_{\alpha}$.

To formally control the size of $R_{\alpha}\approx\bar{R}_{\alpha}$, the parameter that bounds this size form above has been introduced. To introduce the physical applicability of the theory after setting the approximation, it is practical to restrict the volume of this region to be much smaller than a typical scale of gravitational interactions. It is going to be done by postulating the relation
\begin{align}
\label{limit_region}
\forall_{\@\alpha}\,\mathfrak{L}_{\alpha}\ll \mathbb{L}_0\,,
\end{align}
where the length $\mathfrak{L}_{\alpha}$ has been defined in \eqref{length_region}. The new global parameter $\mathbb{L}_0$ denotes the fixed length, which is comparable with or lower than the distance, concerning which the expression in \eqref{trick_matter_delta} is applicable. Therefore, it is convenient to specify the value of $\mathbb{L}_0$ by setting a physical scale for our analysis. Consequently, $\mathbb{L}_0$ becomes defined as the fiducial unit length of a considered physical system of units. In the case of the International System of Units, this quantity takes value $\mathbb{L}_0^{\@\text{SI}}=1$ m. Then, to perform the expansion and later neglect the higher-order terms in the last brackets in \eqref{trick_matter_delta}, the condition in \eqref{limit_region} must be satisfied. This restriction guarantees the physical applicability of the theory. The formal condition that keeps all the cells well-defined is given in \eqref{area_existence}. It is convenient then reexpressing these restrictions in terms of a dimensionless, locally defined parameters $\lambda_{\alpha}$ that satisfy
\begin{align}
\label{regulator_region}
\forall_{\@\alpha}\ 0<\lambda_{\alpha}:=\frac{\mathfrak{L}_{\alpha}}{\mathbb{L}_0}\ll1\,.
\end{align}
These parameters allows to control the theory independently of the considered system of units.

One should reformulate all the size-related restrictions into dimensionless versions. Each $\lambda_{\alpha}$ parameter specifies the applicability of the approximation in \eqref{approximation_region} as follows,
\begin{align}
\label{trick_matter_corrections}
\mathbf{V}(R_{\alpha})=\lambda_{\alpha}^3\.\mathbb{L}_0^3\.\sqrt{\bar{q}(x)}\.\Big|_{x\in\bar{R}_{\alpha}}\!+\mathcal{O}\big(\lambda_{\alpha}^4\big)
\approx\lambda_{\alpha}^3\.\mathbb{L}_0^3\.\sqrt{\bar{q}_{\alpha}}\approx\lambda_{\alpha}^3\.\mathbb{L}_0^3\.\sqrt{q(x)}\.\Big|_{x\in R_{\alpha}},
\qquad\lambda_{\alpha}\ll1\,.
\end{align}
The introduction of the regularization method in \eqref{trick_matter_delta} requires that the determinant $q$ is almost constant locally, at scales $\lesssim \mathfrak{L}_{\alpha}^3$. This assumption establishes the necessary approximation $\mathbf{V}(R_{\alpha})\approx\mathbf{V}(\bar{R}_{\alpha})$. This condition and the mosaic structure of the manifold entails the following decomposition of the domain of the Hamiltonian constraint $H:=H^{\textsc{\.\!g\.\!r\.\!}}+H^{\textsc{\.\!m\@a\@t\.\!t\.\!}}$,
\begin{align}
\label{trick_decomposition}
H\big(\Sigma_t^{\hash^{\@\textsc{r}}}\big)\approx H\big(\Sigma_t^{\bar{\hash\.}\@}\big)=\int_{\Sigma_t^{\bar{\hash\.}\@}}\!\!\!\!d^3x\,N\mathcal{H}
=\sum_{\alpha}\!\int_{\!\bar{R}_{\alpha}}\!\!\!\!d^3x\,N\mathcal{H}
=\sum_{\alpha}H(\bar{R}_{\alpha})\,,
\qquad
\forall_{\@\alpha}\,\lambda_{\alpha}\ll1\,.
\end{align}
Consequently, for each region $R_{\alpha}$, the identity
\begin{align}
\label{v-regularization}
\frac{q^n\@(x)\!}{E_i^a(x)}\bigg|_{x\in R_{\alpha}}\!\!
=\frac{\lambda_{\alpha}^{3(1-2n)}\mathbb{L}_0^{3(1-2n)}\!\!}{n}\.
\big(1+\mathcal{O}(\lambda_{\alpha})\big)^{\@1-2n}\.
\frac{\delta\.\mathbf{V}^{2n\@}(\bar{R}_{\alpha})}{\delta E_i^a(x)}\bigg|_{x\in\bar{R}_{\alpha}}\!\!
\approx-\frac{2\.\lambda_{\alpha}^{3(1-2n)}\mathbb{L}_0^{3(1-2n)}\!\!}{\gamma\kappa\.n}\.
\big\{A^i_a(x),\mathbf{V}^{2n}(\bar{R}_{\alpha})\big\}\Big|_{x\in\bar{R}_{\alpha}}\!\!
\end{align}
holds. According to the convention of this article, the application of this expression, often so-called the `Thiemann's trick', is going to be named the volume regularization.

	%%%%%%%%%	%%%%%%%%%	%%%%%%%%%	%%%%%%%%%
\subsection{Graph structure from the holonomy representation}\label{Sec_Regularization_/_graph}

\noindent
In the previous section, the lattice regularization of the gravitational momentum $E^a$ has been introduced. The resulting formula in \eqref{v-regularization} is dependent on the Poisson brackets of the connection field located inside of a small region and on some power of the volume of this region. Consequently, the method of application of this formula to the Hamiltonian constraint, originally introduced in \cite{Thiemann:1997rt}, has been named the volume regularization. The indicated small regions are the elementary cells of the mosaic manifold $\Sigma_t^{\hash}$ or collections of these cells. This indication specifies the notion of a locality to regions $R_{\alpha}$ (bigger or equal to the size of elementary cells $C_{\alpha}$) within which all dynamical and gauge degrees of freedom are approximately uniform. The piecewise linear structure is constructed in a manner that preserves both gauge symmetries of the gravitational sector. Moreover, the graph $\Gamma$, which determines the discreteness of $\Sigma_t^{\hash}$, is also invariant under all the gauge symmetries of the matter sector, which are defined as different representations of the SU$(N)$ group.

This discrete structure at the classical level, which approximates the smooth manifold arbitrarily well, allows one to define canonical quantization on a lattice. It is a significant difference in comparison to CLQG. The latter theory is not constructed by the canonical quantization method. In the formulation of CLQG, the lattice quantum theory \cite{Ashtekar:2004eh,Thiemann:2007zz}, which is nonperturbative and preserves SE at the quantum level, is postulated. However, this formulation is not defined by assuming the Dirac-Heisenberg-DeWitt \cite{Dirac:1925jy,Heisenberg:1929xj,DeWitt:1967yk} form of the operators,
\begin{align}
\label{DeWitt_representation}
\hat{A}^i_a|\ldots\rangle=A^i_a|\ldots\rangle\,,
\qquad
\hat{E}_i^a|\ldots\rangle=-\i\,\kbar\frac{\delta}{\delta A^i_a}|\ldots\rangle\,.
\end{align}
Moreover, the (semi)classical limit of the postulated theory, which should be continuous, has not been so far derived. Furthermore, in the studies related to this issue, it has been revealed that the introduction of additional \textit{ad hoc} modifications is needed to obtain the continuous limit of the model \cite{Dittrich:2014ala}.

In the program presented in this article, the gauge-invariant lattice structure is introduced at the classical level, therefore the quantization procedure is the canonical quantization. Then, the (semi)classical limit should be founded on the lattice. Finally, the continuous limit will be obtained by reversing the classical discretization method. It is expected that as a result of this program, many problems, presented in the analogous research regarding CLQG, can be avoided. A worth recalling comment concerning both the CLQG formalism and the one presented in this article is the lack of the implicit covariance of indices. This issue, mentioned in the introduction (in Sec.~\ref{Sec_Introduction_/_Equivalence_principle}), is discussed in detail in \cite{Bilski:2021_RCT_I}.

To represent both canonical fields of gravity in terms of geometrical objects, one still needs to elaborate the lattice representation of $A_a$. The natural strategy is to construct an object analogous to \eqref{el_Wilson}, which will be expandable not only in terms of the curvature of $A_a$ but also in terms of the connection. In this way, the $A_a^i$ field in \eqref{v-regularization} will be regularizable. Notice, however, that the Wilson representation has not been introduced according to the mentioned natural strategy. Instead, the densitized functional $\uline{B}^a$ of the one-form $\uline{A}_a$ was initially constructed, see \eqref{el_magnetic}. This construction allowed one to express the scalar constraint in \eqref{el_scalar} only in terms of the densitized vector objects. Thus, in the case of the gravitational field, the densitized curvature $B^a$ of the connection $A_a$ is also going to be defined.

To follow Wilson's methodology, this section is initiated by the introduction of the reciprocal of the SU$(2)$ holonomy of the gravitational connection $A_a$ defined in \eqref{ABMBD_connection}, known as the Ashtekar-Barbero-Mercuri-Bojowald-Das connection, \textit{cf.} \cite{Ashtekar:1986yd,Barbero:1994ap,Mercuri:2006um,Bojowald:2007nu}. This reciprocal holonomy is defined as
\begin{align}
\label{holonomy_general}
h_{\ell}\@:=h_{\ell}\@\Big[A\Big[{\!\@}\stackrel{\!\scriptscriptstyle\textsc{\!t\!\.-\!\.f}\!}{A}{\!\@},E\Big]\Big]\@:=\mathcal{P}\exp\!\bigg(\!\int_{\@\ell}\!ds\,\dot{\ell}^a\@(s)\.A^i_a\@\big(\ell(s)\big)\tau^i\!\bigg).
\end{align}
The indicated dependence of the densitized dreibein $E^a$\footnote{Both pairs of the square brackets $[\,]$ in \eqref{holonomy_general} denote a functional dependence, not a commutator.} is recognizable from the definition of the connection in \eqref{ABMBD_connection}. However, this dependence does not modify the symplectic structure of the real Ashtekar variables in \cite{Ashtekar:1986yd,Ashtekar:1987gu,Barbero:1994ap}, \textit{cf.} \cite{Bojowald:2007nu,Bojowald:2010qpa}. Moreover, as suggested in \cite{Bilski:2021_RCT_I}, the presence of the momentum degrees of freedom in the explicit expression of the connection coefficient,
\begin{align}
\label{ABMBD_connection_coefficient}
A_a^i=\;\stackrel{\!\scriptscriptstyle\textsc{\!a\!\.-\!\.b}\!}{A}{\@\@}_a^i+\frac{\kappa}{8E^a_i}\sqrt{q}J^0
={\!\@}\stackrel{\!\scriptscriptstyle\textsc{\!t\!\.-\!\.f}\!}{\Gamma}{\!\@}^i_a
+\gamma{\!\@}\stackrel{\!\scriptscriptstyle\textsc{\!t\!\.-\!\.f}\!}{K}{\!\@}^i_a
+\frac{\kappa}{8E^a_j}\epsilon^{ijk}\sqrt{q}J^k\,,
\end{align}
is inevitable --- the multiplication by $\sqrt{q}$ redefines the weight of a multiplied quantity by $+1$. This operation determines the weight of the scalar representatives of the fermionic field in GR\footnote{The fact that the classical observables should be represented by objects of a weight $1$ was mentioned just before formula \eqref{Poisson_brackets}. This issue, discussed in \cite{Bilski:2021_RCT_I}, was also noticed in \cite{Thiemann:2007zz}. The factual physical problems that arise by neglecting the densitization of the classical observables were recognized in \cite{Bilski:2020fmn}.}. The dimensional analysis required that the objects $\kappa\sqrt{q}J^0/8$ and $\kappa\sqrt{q}\epsilon^{ijk}J^k/8$ had to be multiplied by the inverse of either $E^a_i$ or $B^a_i$ (the latter vector density coefficient is defined later in this section, namely, in \eqref{densitized_curvature}). The multiplication by $1/B^a_i$ would lead to a modification of the Poisson brackets in \eqref{Poisson_qp} and \eqref{Poisson_AE}. Remarkably, the analysis in \cite{Bojowald:2007nu} concerning the coupling between gravity and fermions demonstrated that the densitized currents $\sqrt{q}J^0$ and $\sqrt{q}J^k$ are multiplied by the inverse of the gravitational momentum. Therefore the symplectic structure remained preserved, and the apparently non-minimal coupling (see \eqref{ABMBD_connection} and \eqref{ABMBD_connection_coefficient}) has been verified to be actually the minimal coupling.

It is worth mentioning that the original definition of the holonomy of the torsion-free Ashtekar-Barbero connection reads
\begin{align}
\label{holonomy_Ashtekar-Barbero}
{\!\@}\stackrel{\!\scriptscriptstyle\textsc{\!t\!\.-\!\.f}\!}{h}{\!\@}_{\ell}^{-1}\@:={\!\@}\stackrel{\!\scriptscriptstyle\textsc{\!t\!\.-\!\.f}\!}{h}{\!\@}_{\ell}^{-1}\@\Big[{\!}\stackrel{\!\scriptscriptstyle\textsc{\!t\!\.-\!\.f}\!}{A}{\@\@}\Big]\@:=\mathcal{P}\exp\!\bigg(\!-\!\int_{\@\ell}\!ds\,\dot{\ell}^a\@(s)\.{\!\@}\stackrel{\!\scriptscriptstyle\textsc{\!t\!\.-\!\.f}\!}{A}{\!\@}^i_a\@\big(\ell(s)\big)\tau^i\!\bigg).
\end{align}
This object is the solution of the parallel transport equation for a vector along the path $\ell$ in the absence of fermions and torsion. As in \eqref{Wilson} and \eqref{holonomy_general}, $\mathcal{P}$ represents the path ordering. Moreover, in the absence of fermionic degrees of freedom and torsion, the holonomy $h_{\ell}$ coincides with ${\!\@}\stackrel{\!\scriptscriptstyle\textsc{\!t\!\.-\!\.f}\!}{h}{\!\@}_{\ell}$.

As a next step, the identification of the path $\ell$ with the structure of $\Gamma$ should be implemented. This starts by selecting the path $l_{(p)}\pos{v}\in\Gamma$ between the points $v$ and ${v\@+\@l_{(p)}}$, oriented along the $p$-direction, where the angle brackets define the position of a given quantity distributed along the whole path initiated at $v$. This path is the object, which determines the lattice embedding into the $\Sigma_t^{\textsc{a\@d\@m}}$ manifold. Consequently, $l_{(p)}$ is going to be called the link, which connects two nearest neighbor nodes along the $p$-direction of some unspecified yet lattice. The inverse link ${l_{(p)}^{-1}\pos{v}}:={l_{(p^{-1})}\pos{v\@+\@l_{(p)}}}$ is defined along the same (oriented) path, in the opposite direction.

To plan the discussion concerning the lattice dynamics, it is useful to introduce the ratio between the distance ${l_{(p)}\pos{v}}$ between the nearest neighbor nodes along the path and some fixed fiducial length. For simplicity, the distance is denoted by the same symbol as the path itself. This ratio is a local positive and real dimensionless parameter defined by
\begin{align}
\label{regulator}
\forall_p\forall_v\ 0\leq\varepsilon_{(p)}\pos{v}:=\frac{l_{(p)}\pos{v}\@}{\mathbb{L}_0}\ll1\,.
\end{align}
One should pay attention that so far the ratio between the range of the local constancy of $q$ and lengths of links, namely, $l/\mathfrak{L}=\varepsilon/\lambda$, remains unspecified (but it is bounded from above).

By expanding then the holonomy
\begin{align}
\label{holonomy}
h_{(p)}\pos{v}:=h_{l_{(p)}\pos{v}}=\mathcal{P}\exp\!\bigg(\!\int_{\@v}^{v+l_{(p)}}\!\!\!\!\!\!\!\!\!\!ds\,\dot{l}^a\@(s)\.A_a\@\big(l(s)\big)\!\bigg)
:=\mathcal{P}\exp\!\bigg(\!\int_{\@0}^{\mathbb{L}_0\.\varepsilon_{(p)}\@\pos{v}}\!\!\!\!\!\!\!\!ds\,\dot{l}^a\pos{v}\@(s)\.A_a^i\@\big(l(s)\big)\tau^i\!\bigg)
\end{align}
around the infinitesimal limit of the length of $l_{(p)}\pos{v}$ (formally, around the zeroth value of the regulator $\varepsilon_{(p)}\pos{v}$), one finds the relation that links the holonomy directly with the gravitational connection,
\begin{align}
\begin{split}
\label{holonomy_expansion}
h_{(p)}^{\pm1}\pos{v}=&\;h_{(p^{-1})}^{\mp1}\pos{v\@+\@l_{(p)}}
=\mathds{1}\pm\mathbb{L}_0\.\varepsilon_{(p)}\pos{v}A_{(p)}\@(v)
+\frac{1}{2}\mathbb{L}_0^2\.\varepsilon_{(p)}^2\pos{v}
\big(
A^2_{(p)}\@(v)\pm\partial_{(p)}A_{(p)}\@(v)
\big)+\mathcal{O}(\varepsilon^3)
\\
=&\;
\mathds{1}\mp\mathbb{L}_0\.\varepsilon_{(p)}\pos{v}A_{(p)}\@\big(v\@+\@l_{(p)}\big)
+\frac{1}{2}\mathbb{L}_0^2\.\varepsilon_{(p)}^2\pos{v}
\Big(
A^2_{(p)}\@\big(v\@+\@l_{(p)}\big)\mp\partial_{(p)}A_{(p)}\@\big(v\@+\@l_{(p)}\big)
\Big)+\mathcal{O}(\varepsilon^3)
\\
=&\;\mathds{1}\pm\mathbb{L}_0\.\varepsilon_{(p)}\pos{v}A_{(p)}\@(v)+\mathcal{O}(\varepsilon^2)
=\mathds{1}\mp\mathbb{L}_0\.\varepsilon_{(p)}\pos{v}A_{(p)}\@\big(v\@+\@l_{(p)}\big)+\mathcal{O}(\varepsilon^2)\,.
\end{split}
\end{align}
The same expression written in the abstract indexless convention is given in Appendix~\ref{Appendix_holonomy_path}. To explain the notation, it should be indicated that the position dependence in ${h_p\pos{v}}$ labels the initial point of the link ${l_{(p)}\pos{v}}$, hence the inverse holonomy is denoted as ${h_{(p^{-1})}\pos{v\@+\@l_{(p)}}}$\footnote{The reader should not confuse the quantity ${h_{(p^{-1})}\pos{v\@+\@l_{(p)}}}$ with ${h_{(p^{-1})}\pos{v}}$. The latter one is the inverse of ${h_{(p)}\pos{v\@-\@l_{(p)}}}$.}. It is also worth mentioning that, in the second line in \eqref{holonomy_expansion} and in the last part of this equality, the alternative expansion of the holonomy is presented. This expansion is not located at $v$ but at ${v\@+\@l_{(p)}}$, however, it is derived around the zeroth value of the same regulator ${\varepsilon_{(p)}\pos{v}}$ (that is associated with the same link). This equality indicates that any holonomy along a link is related to its connection representations located at both endpoints of this link. It should be emphasized that an analogous problem does not appear when comparing the volume of a region centered at some point with its densitized dreibein representation at this point --- see Sec.~\ref{Sec_Regularization_/_ADM}.

%
%%%	FIGURE	%%%
%
\begin{figure}[h]
\vspace{5pt}%
\begin{center}
\begin{tikzpicture}[scale=1]
%holonomy
\draw[cl35,->]{(0.5,0.8) -- (3.5,0.8)};
\draw[cl35,->]{(3.5,0.8) -- (3.5,3.8)};
\draw[cl35,->]{(3.5,3.8) -- (0.5,3.8)};
\draw[cl35,->]{(0.5,3.8) -- (0.5,0.8)};
%secondary horizontal
\draw[cb15]{(-2.5,0.8) -- (3.5,0.8)};
%secondary diagonal
\draw[cb15]{(0.5,0.8) -- (0,0)};
%secondary vertical
\draw[cb15]{(0.5,3.8) -- (0.5,-2.2)};
%tertiary diagonal
\draw[cb05]{(1,1.6) -- (0.5,0.8)};
%%%	labeling		%%%
\node at (-0.1,1.05) {$A_{(q)}\@(v)$};
\node at (4.45,1.05) {$A_{(r)}\@\big(v\@+\@l_{(q)}\@\big)$};
\node at (-1.0,4.05) {$A_{(q^{-1})}\@\big(v\@+\@l_{(q)}\!+\@l_{(r)}\@\big)$};
\node at (4.6,4.05) {$A_{(r^{-1})}\@\big(v\@+\@l_{(r)}\@\big)$};
\node[text=gray] at (2.1,1.05) {$h_{(q)}\pos{v}$};
\node[text=gray] at (2.1,4.05) {$h_{(q^{-1})}\pos{v\@+\@l_{(q)}\!+\@l_{(r)}\@}$};
\node[text=gray] at (1.6,2.8) {$h_{(r^{-1})}\pos{v\@+\@l_{(r)}\@}$};
\node[text=gray] at (4.45,2.8) {$h_{(r)}\pos{v\@+\@l_{(q)}\@}$};
%%%	dressing		%%%
\node at (0.7,0.55) {$v$};
\node at (-0.4,-0.2) {$l_{(p)}\pos{v}$};
\node at (1.6,1.8) {$l_{(p)}\pos{v\@-\@l_{(q)}\@}$};
\node at (2.7,0.45) {$l_{(q)}\pos{v}$};
\node at (-1.2,0.4) {$l_{(q)}\pos{v\@-\@l_{(q)}\@}$};
\node at (0,2.8) {$l_{(r)}\pos{v}$};
\node at (-0.35,-1.3) {$l_{(r)}\pos{v\@-\@l_{(r)}\@}$};
\end{tikzpicture}
\end{center}
\vspace{-10pt}%
\caption{Decomposition of holonomy $h_{(q)\@(r)}\pos{v}$ (gray) around the oriented loop,
$\circlearrowleft^{\square}_{(\@q\@)\@(\@r\@)}\@\!\pos{v}=\big(\!\!\circlearrowleft^{\square}_{(\@r\@)\@(\@q\@)}\!\!\big)^{\@-1}\pos{v}$}
\label{square_holonomy}
\end{figure}
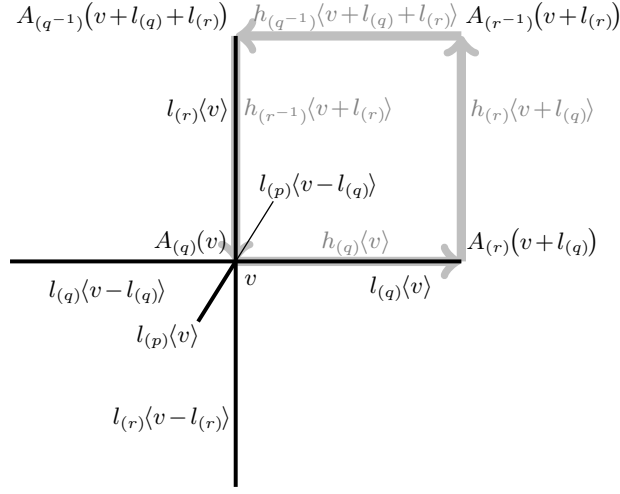
%
%%%	FIGURE	%%%
%
Considering next the oriented closed quadrilateral path
\begin{align}
\label{squared_loop}
\circlearrowleft^{\square}_{(\@q\@)\@(\@r\@)}\@\!\pos{v}
:=l_{(q)}\pos{v}\circ
l_{(r)}\pos{v\@+\@l_{(q)}}\circ
\big(l_{(q)}\big)^{\@-1}\pos{v\@+\@l_{(q)}\@+\@l_{(r)}}\circ
\big(l_{(r)}\big)^{\@-1}\pos{v\@+\@l_{(r)}}\,,
\end{align}
one can construct an object analogous to the expression under the trace in the definition of the Wilson loop in \eqref{Wilson}. To do that one should expand the holonomy $h_{(q)\@(r)}:=h_{\circlearrowleft_{(\@q\@)\@(\@r\@)}}$ of the quadrilateral\footnote{It is worth noting that the expression in \eqref{loop_holonomy_expansion} is not universal, but it is derived for a quadrilateral loop. Considering a different, for instance, a triangular loop, the factor in front of the term of order $\varepsilon^2$ would be different, namely, $\mathbb{L}_0^2/2$ instead of $\mathbb{L}_0^2$ --- see \cite{Vines:2014uoa}.} loop $\circlearrowleft_{(\@q\@)\@(\@r\@)}$ (see FIG.~\ref{square_holonomy}) around the limit of the length of the loop-composing links. This expansion leads to
\begin{align}
\label{loop_holonomy_expansion}
h_{(q)\@(r)}^{\pm1}\pos{v}:=\mathcal{P}\exp\!\Bigg(\!\pm\!
\oint_{_{\scriptstyle\!\!\ell\equiv\circlearrowleft^{\square}_{\@(\@q\@)\@(\@r\@)}\!\pos{v}}}\!\!\!\!\!\!\!\!\!\!\!\!\!\!\!\!\!\!\!ds
\.\dot{\ell}^a\@(s)\.A_a\@\big(\ell(s)\big)
\@\Bigg)
=h_{(r)\@(q)}^{\mp1}\pos{v}
=\mathds{1}\pm\mathbb{L}_0^2\.\varepsilon_{(q)}\pos{v}\.\varepsilon_{(r)}\pos{v}F_{(q)\@(r)}\@(v)+\mathcal{O}(\varepsilon^3)
=\mathds{1}+\mathcal{O}(\varepsilon^2)\,.
\end{align}
Analogously as the form of the expansion in \eqref{holonomy_expansion}, the form of the one written above is not unique. There is more than one possibility of choosing the point around which the loop's length is expanded. Precisely, in the case of the quadrilateral loop, there are four such points, namely, the vertices of the polygon at which the expansion can be taken. They are marked in FIG.~\ref{square_holonomy}. Another thing that should be emphasized is the identification of the loop's orientation with the right-hand rule. This identification, however, does not specify the internal orientation of links. It only indicates the order in the set of connected internally-oriented links that compose the loop and are invariant under cyclic position permutations (from the perspective of the loop, not the whole graph).

The curvature of the gravitational connection is defined in the standard way,
\begin{align}
\label{Ashtekar_curvature}
F_{ab}=F_{ab}^i\tau^i:=\big(\partial_aA^i_b-\partial_bA^i_a+\epsilon^{ijk}A^j_aA^k_b\big)\tau^i\,.
\end{align}
Then, the magnetic field analogue concerning the gravitational connection, called in this article the densitized curvature, is constructed in the same manner as the $\uline{B}^a$ field in \eqref{el_magnetic}, namely,
\begin{align}
\label{densitized_curvature}
B^a:=\frac{1}{2}\epsilon^{abc}_{\scriptscriptstyle+}F_{bc}\,.
\end{align}
It is worth noting that this definition clarifies why the right-hand rule naturally indicates the `positive' orientation of loops: the same rule sets the orientation of $B^a$. This quantity is a vector density --- the same category of objects as the densitized dreibein. Therefore, both $E^a$ and $B^a$ are related to the links of the same lattice, which is dual to the direct lattice that supports the representation of the connection (introduced by the expansion in \eqref{holonomy_expansion}). This direct/reciprocal lattice relation is equivalent to (more precisely, taken from) a crystallographic framework. This corresponding frameworks argument indicates the necessity of formulating a theory in terms of densities instead of zero-weighted functionals. As a result, all degrees of freedom can be smeared over the same lattice, hence the relation between regulators $\varepsilon$ and $\lambda$ can be easily postulated. This relation leads to the cancellation of regularization parameters in the lattice Hamiltonian expressed in terms of the holonomies and fluxes. Thus, the unified regulator does not have to be \textit{a priori} specified, and it can determine the value of quantum corrections (analogously to the determination of the fine-structure constant in quantum electrodynamics).

	%%%%%%%%%	%%%%%%%%%	%%%%%%%%%	%%%%%%%%%
\subsection{Piecewise linear graph and a graph with short edges}\label{Sec_Regularization_/_Wigner}

\noindent
The geometrical introduction of the holonomy representation via the short path expansions along a link and around a loop has been defined in \eqref{holonomy_expansion} and \eqref{loop_holonomy_expansion}, respectively. By applying these expansions to the piecewise linear lattice that specifies the mosaic manifold $\Sigma_t^{\hash}$, the first of these expressions simplifies.

The gravitational connection $A_a$ is constant along a line segment in the Euclidean geometry. Thus, it is constant along each linear link, but it varies along a piecewise linear path composed of several links. Considering a holonomy along a linear path, the expansion in \eqref{holonomy_expansion} becomes
\begin{align}
\label{holonomy_expansion_linear}
\bar{h}_{(p)}^{\pm1}\pos{v}
=\exp\!\big(\!\pm\mathbb{L}_0\.\varepsilon_{(p)}\pos{v}\bar{A}^i_{(p)}\pos{v}\.\tau^i\big)
=\mathds{1}\pm\mathbb{L}_0\.\varepsilon_{(p)}\pos{v}\bar{A}_{(p)}\pos{v}
+\frac{1}{2}\mathbb{L}_0^2\.\varepsilon_{(p)}^2\pos{v}\bar{A}^2_{(p)}\pos{v}
+\mathcal{O}(\varepsilon^3)\,.
\end{align}
The position $\pos{v}$ in ${\bar{A}_{(p)}\pos{v}}$ denotes the initial point of the linear link\footnote{The classical connection $A_{(p)}\@(v)$ is the object defined at a given point $v$. The spatially constant connection along a linear link ${\bar{A}_{(p)}\pos{v}}$ is the spatially extended object defined at the link ${l_{(p)}\pos{v}}$, between its endpoints $v$ and ${v\@+\@l_{(p)}}$.}, analogously as in the case of $h_{(p)}^{\pm1}\pos{v}$ and $\varepsilon_{(p)}\pos{v}$. As a result, all the locally defined variables and parameters in \eqref{holonomy_expansion_linear} are located along the same spatially extended object, namely the link ${l_{(p)}\pos{v}}$.

It is worth temporarily neglecting the postulate of piecewise linearity. So far, this assumption was introduced as a possible simplification of the system. It was justified by the existence of the geometry of $\Sigma_t^{\textsc{a\@d\@m}}$ that is non-Euclidean in general. The aim of this and the next section is to prove that the lattice, on which one can define a Hilbert space \textit{\`a la} the one in CLQG \cite{Ashtekar:1994mh,Ashtekar:1994wa,Ashtekar:1995zh}, can be simplified to the piecewise linear case without loss of its applicability. One will find that this simplification is actually more precise than CLQG. The regularization in the latter model can be then considered as a more complicated, less accurate, and less universal approximation of the better symmetry-preserving theory of gravity regularized over a piecewise linear lattice.

The demonstration of this idea requires going ahead with the analysis in this article and concerning the Hilbert space construction. The structure of this object regarding a gauge field was specified by Wigner \cite{Wigner:1931,Wigner:1939cj}. Two aspects are worth direct investigation. However, let first a few facts be recalled. The gauge symmetry of a classical field is determined by the choice of the matrix representation of generators. At the quantum level, symmetry transformations are the transformations of the normalized non-zero vectors that belong to a Hilbert space. The transformations of these vectors, which are called rays, are represented by unitary or antiunitary transformations of the Hilbert space. For a given gauge field, the normalization is determined for the pair: a group element and the parameter of its representation. This normalization depends also on the type of a Lie group, the type of the group representation, and the choice of the Hermitian or antihermitian generators. A general method of application of gauge symmetries to QFT has been established in \cite{Weinberg:1995mt}. In \cite{Bilski:2020xfq}, this formalism concerning the unitary and linear representation generators has been adjusted to the representation of SU$(2)$ in CLQG, where the $\mathfrak{su}(2)$ generators are antihermitian and are normalized as follows,
\begin{align}
\label{antihermitian}
[\tau^j,\tau^k]=\epsilon^{ijk}\tau^i\,.
\end{align}
At this stage, the defining type of the representation, namely $\tau^i=-\frac{\i}{2}\sigma^i$, where $\sigma^i$ is a Pauli matrix, does not have to be postulated yet.

The quantum states space assumed in this program (see \cite{Bilski:2021_RCT_III}) is going to be the one on which CLQG is defined \cite{Ashtekar:1995zh}. Moreover, it is required that each link-related sector of this space, is a Hilbert space that coincides with Wigner's construction. Furthermore, gauge operators are assumed to be Wigner operators \cite{Marolf:1994wh,Wigner:1957ep,Salecker:1957be}. Finally, the mentioned aspects, needed to be investigated regarding the SU$(2)$ symmetry after the quantization of the connection $A_a$, are the following. One has to verify whether the holonomy, which is going to be quantized after the regularization of the Hamiltonian, is the element of the group. Then, one must check if the quantum symmetry representation of the holonomy on the Hilbert space is the symmetry representation of the connection operator $\hat{A}_a$.

The verification of the first condition is easy. By comparing both the exponential definition in \eqref{holonomy} and its short link expansion in \eqref{holonomy_expansion}, one finds that the group multiplication law is satisfied,
\begin{align}
\begin{split}
\label{holonomy_group_element}
h_{(p)}\pos{v}\@[A]\.h_{(p)}\pos{v}\@[A'\,\@]=&\;\mathds{1}
+\mathbb{L}_0\.\varepsilon_{(p)}\pos{v}\big(A_{(p)}\@(v)\@+\@A'_{(p)}\@(v)\big)
+\frac{1}{2}\mathbb{L}_0^2\.\varepsilon_{(p)}^2\pos{v}\.\partial_{(\@p\@)}\@\big(A_{(p)}\@(v)\@+\@A'_{(p)}\@(v)\big)
\\
&+\frac{1}{2}\mathbb{L}_0^2\.\varepsilon_{(p)}^2\pos{v}\big(A_{(p)}\@(v)\@+\@A'_{(p)}\@(v)\big)^{\@2}
\@+\mathcal{O}(\varepsilon^3)
=h_{(p)}\pos{v}\@[A\@+\@A'\,\@]\,.
\end{split}
\end{align}
For simplicity, the calculations in this section are formulated regarding the reciprocal holonomy. For analogous derivations concerning the original definition of the holonomy of the Ashtekar-Barbero connection in \eqref{holonomy_Ashtekar-Barbero}, see \cite{Bilski:2020xfq}.

To verify the second condition, one needs to expand the group element around the identity. By choosing $\tau^i$ as the representation generator, and postulating that the holonomy is an exponential map from a representation of some quantum field, the expansion around the unit element reads
\begin{align}
\label{holonomy_exponential_map}
h_{(p)}\pos{v}\@[\,\@\theta\,\@]=\mathds{1}+\theta_{(p)}\pos{v}
+\frac{1}{2}\theta^2_{(p)}\pos{v}
-\frac{1}{2}[\.\theta_{(p)}\pos{v},\theta_{(p)}\pos{v}]
+\mathcal{O}\big(\theta_{(p)}^3\pos{v}\big)\,.
\end{align}
In the chosen normalization\footnote{Analogous analysis regarding Hermitian and unitary generators is given in \cite{Bilski:2020xfq}. This standard normalization is then adjusted to antihermitian generators in \eqref{Lie_real}, thus, in the case of $\mathfrak{su}(2)$, into the formula in \eqref{antihermitian}. This adjustment determines the structure of the expansion in \eqref{holonomy_exponential_map}.}, the representation of the gauge transformation equals
\begin{align}
\label{abstract_representation}
\theta_{(p)}\pos{v}:=\theta_{(p)}^i\pos{v}\.\tau^i
:=\!\int_0^{\mathbb{L}_0\.\varepsilon_{(p)}\@\pos{v}}\!\!\!\!\!\!\!\!ds\.\dot{\ell}^a\@(s)\.A_a\@\big(\ell(s)\big)\,.
\end{align}
It is convenient to introduce the probability distribution ${\tilde{A}_{(p)}\pos{v}}$ of the gravitational connection, which is the so-normalized representation of the gauge transformation:
\begin{align}
\label{connection_distribution}
\tilde{A}_{(p)}\pos{v}:=\frac{\theta_{(p)}\pos{v}}{\mathbb{L}_0\.\varepsilon_{(p)}\pos{v}\!}
=\frac{1}{\mathbb{L}_0\.\varepsilon_{(p)}\pos{v}\!}
\sum_{n=0}^{\infty}\frac{1}{n!}\frac{\mathrm{d}^{\!\.n}\theta_{(p)}\pos{v}\!}{\mathrm{d}\varepsilon_{(\@p\@)}^n\@\pos{v}\!}
\bigg|_{\@\varepsilon_{(\@p\@)}\@\pos{v}\!\.=\,\@0}\varepsilon_{(\@p\@)}^{n\mathstrut}\pos{v}\,.
\end{align}
The right-hand side of this formula, namely the power series representation, is determined by assuming the constraints in \eqref{regulator} on the value of the regulator. Analogously to the notation regarding ${h_{(p)}^{\pm1}\pos{v}}$, ${\varepsilon_{(p)}\pos{v}}$, and the spatially constant connection in \eqref{holonomy_expansion_linear}, the position $\pos{v}$ in ${\tilde{A}_{(p)}\pos{v}}$ denotes the initial point of a link. In other words, the expression in \eqref{abstract_representation} is the definition of the link-located representation of the gauge transformation, independent of the local specification of a link. Concerning then the linear link, the quantities $\bar{A}_{(p)}\pos{v}$ and $\tilde{A}_{(p)}\pos{v}$ become equal.

Analogously to \eqref{connection_distribution}, one can define the probability distribution of the densitized curvature of the gravitational connection around a loop,
\begin{align}
\label{curvature_distribution}
\tilde{B}^{(p)}\@\big(\@S_{\@(\@v\@+\@l_{\@(\@q\@)}\@\@/\@2\@+\@l_{\@(\@r\@)}\@\@/\@2\@)}\!\big)
:=\frac{1}{2}\epsilon^{(p)qr}_{\scriptscriptstyle+}\tilde{F}_{qr}\pos{v}:=
\frac{\epsilon^{(p)\@(q)\@(r)}_{\scriptscriptstyle+}}{\mathbb{L}_0\big(
\varepsilon_{(q)}\pos{v}\!+\@\varepsilon_{(r)}\pos{v\@+\@l_{(q)}\@}\!+\@\varepsilon_{(q)}\pos{v\@+\@l_{(r)}\@}\!+\@\varepsilon_{(r)}\pos{v}
\big)\!}
\@\oint_{_{\scriptstyle\!\!\ell\equiv\circlearrowleft^{\square}_{\@(\@q\@)\@(\@r\@)}\!\pos{v}}}\!\!\!\!\!\!\!\!\!\!\!\!\!\!\!\!\!\!\!ds\.F(s)\,.
\end{align}
Here, ${S_{\@(\@v\@+\@l_{\@(\@q\@)}\@\@/\@2\@+\@l_{\@(\@r\@)}\@\@/\@2\@)}}$ is the surface centered at ${v\@+\@l_{(p)}\@/2\@+\@l_{(r)}\@/2}$. The quantity in the denominator on the right-hand side is the sum of four regulators associated to the links composing the quadrilateral loop $\circlearrowleft^{}_{(\@q\@)\@(\@r\@)}\@\!\pos{v}$. As in the standard definition in \eqref{densitized_curvature}, the Levi-Civita symbols $\epsilon^{(p)qr}_{\scriptscriptstyle+}$ and $\epsilon^{(p)\@(q)\@(r)}_{\scriptscriptstyle+}$ are tensor densities of a weight $1$. The adaptation of this definition to the loops composed of different number of links is straightforward by using an appropriate algebraic mean.

It is easy to see that the last element in \eqref{holonomy_exponential_map} vanishes. As noticed in \cite{Bilski:2020xfq}, this element expresses the Lie brackets of symmetric quantities. Hence, the expansion around the unit element simplifies to
\begin{align}
\label{holonomy_exponential_expansion}
h_{(p)}\pos{v}\@[\,\@\theta\,\@]=\mathds{1}
+\mathbb{L}_0\.\varepsilon_{(p)}\pos{v}\tilde{A}_{(p)}\pos{v}
+\frac{1}{2}\mathbb{L}_0^2\.\varepsilon_{(p)}^2\pos{v}\tilde{A}^2_{(p)}\pos{v}
+\mathcal{O}\big(\theta_{(p)}^3\pos{v}\big)\,,
\end{align}
where the normalization from \eqref{connection_distribution} has been implemented. By comparing this result with \eqref{holonomy_expansion_linear}, one finds that the states, constructed from the representation of the holonomy along a linear link, automatically preserve symmetry transformations with the required quadratic precision. Then, by expanding the probability distribution of the gravitational connection ${\tilde{A}_{(p)}\pos{v}}$ around the zero value of the regulator, one obtains the following approximation,
\begin{align}
\label{connection_distribution_expansion}
\tilde{A}_{(p)}\pos{v}=\frac{1}{\mathbb{L}_0\.\varepsilon_{(p)}\pos{v}\!}
\!\int_0^{\mathbb{L}_0\.\varepsilon_{(p)}\@\pos{v}}\!\!\!\!\!\!\!\!ds\.A(s)=A_{(p)}\@(v)+\frac{1}{2}\mathbb{L}_0\.\varepsilon_{(p)}\pos{v}\.\partial_{(p)}A_{(p)}\@(v)
+\mathcal{O}(\varepsilon^2)\,.
\end{align}
By applying this result into \eqref{holonomy_exponential_expansion}, one finds 
\begin{align}
\label{cubic_order}
\mathcal{O}(\theta^3)=\mathcal{O}(\varepsilon^3)
\end{align}
and can easily identify the pairs of all the other explicitly corresponding terms when comparing \eqref{holonomy_expansion} with \eqref{holonomy_exponential_expansion}. Hence, the representations of the holonomy located at initial points of short links are also good candidates for symmetry-preserving states. However, the only representation of the gauge symmetry group of holonomy, which is independent of properties of links, is ${\tilde{A}_{(p)}\pos{v}}$. The connections ${\bar{A}_{(p)}\pos{v}}$ and $A_{(p)}\@(v)$ require additional assumptions either regarding links' geometry or their lengths.

In this point, the problem indicated in Sec.~\ref{Sec_Regularization_/_graph}, which is regarding the relation between the continuous gravitational field $A(x)$ and one of the three discrete candidates, reappears. Holonomies are link-related objects, hence the formal statement of the gravitational connection smearing should be specified as follows,
\begin{align}
\label{connection_smearing}
A_{a}(x)
\ \longrightarrow\ \frac{1}{\mathbf{V}_{\!\textsc{\!e\.\!u\@c\.\!l\!}}(\@C_{\@x}\pos{v}\@)\!}
\int_{\!S^{l\pos{\@v\@}}_{\@\pos{\@v\@}}}\!\!\!d^2x
\tilde{A}_{p}\pos{v}\,.
\end{align}
Here, the quantity in the denominator is the coordinate volume ${\mathbf{V}_{\!\textsc{\!e\.\!u\@c\.\!l\!}}(\@C_{\@x}\pos{v}\@)}:=\!{\int_{\!S^{l\pos{\@v\@}}_{\@\pos{\@v\@}}}\!\!\int_0^{\mathbb{L}_0\.\varepsilon_{(\@p\@)}\@\pos{\@v\@}}\!\!d^3\@s}$ of an elementary cell $C_{\@x}$, centered at $x$. The surface and path integrals are assumed to be never tangential and intersect inside both integration intervals. Then, it is assumed that the direction $p$ at point $x$ equals $a$. Several specific methods of the determination of the right-hand side of the relation in \eqref{connection_smearing} are discussed in the next sections. This determination entails the structure of the Hamiltonian constraint on a lattice, thus the structure of the related quantum theory. In particular, a different choice of elementary cells than quadrilateral hexahedra requires appropriate adjustment of the volume element ${d^3\@s}$. Moreover, the selection of the short links or linear links approximations differently simplifies the integrals in \eqref{connection_smearing}.

The analysis in the next sections is not considered as being complete, i.e. there might be many more possibilities to determine the right-hand side of \eqref{connection_smearing}. Still, the several structures presented in this article are as precise as the adaptation of the classical symmetry transformations into the quantum level in quantum mechanics and in QFT. Hence, conversely to the Hamiltonian constraint operator in CLQG \cite{Thiemann:1996ay,Thiemann:1996aw,Thiemann:1997rt,Ashtekar:2004eh,Thiemann:2007zz}, the quantized Hamiltonian of the lattice gravity, constructed by using the methods presented in this article (see \cite{Bilski:2021_RCT_III}), can be considered as a candidate for the Hamiltonian of a quantum gravitational field. However, alternative propositions of the lattice smearing are theoretically possible. One only needs to remember that the quadratic-order precision in expansions of all gauge-valued quantities is not a supplementary but a necessary condition.

	%%%%%%%%%	%%%%%%%%%	%%%%%%%%%	%%%%%%%%%
\subsection{Implementation of the holonomy representation}\label{Sec_Regularization_/_holonomy}

\noindent
The holonomy representations of $A_a$ and $F_{bc}$ have been defined. Reversing the expression in \eqref{loop_holonomy_expansion} is straightforward. By taking the trace of this expression one can remove the unit element and reformulate $F_{bc}^i$ as $\text{tr}(h_{bc}\tau^i)$ (multiplied by non-dynamical factors). The only problem is regarding the location of the curvature and the related loop holonomy. An analogous operation concerning the holonomy representation of $A_a$ appears to lead to a similar problem in choosing the position of a short links expansion --- see \eqref{holonomy_expansion}.

The formula that is needed is an $\varepsilon^2$-order power series of holonomies. The first step toward this relation between $A_a$ and $h_a$ is the following expression,
\begin{align}
\label{double_holonomy_expansion}
h_{(p)}^{\mathstrut}\pos{v}-h_{(p)}^{-1}\pos{v}=2\.\mathbb{L}_0\.\varepsilon_{(p)}\pos{v}\tilde{A}_{(p)}\pos{v}
+\mathcal{O}(\varepsilon^3)\,.
\end{align}
For clarity, from now on, all the cubic-order corrections are going to be expressed by the quantity $\mathcal{O}(\varepsilon^3)$ --- this unification is guaranteed by the relation in \eqref{cubic_order}. It is easy to see that in the linear links approximation, the quadratic-order terms vanish. As a result, one obtains the direct relation between a gravitational connection and holonomies; precisely, ${\tilde{A}_{(p)}\pos{v}}$ in \eqref {double_holonomy_expansion} is replaced by ${\bar{A}_{(p)}\pos{v}}$. In the short links approximation, the result is not directly applicable. In this case, the occurring problem is the presence of the derivative of the connection that contributes to a quadratic-order term in the regulator, namely
\begin{align}
\label{double_holonomy_short}
\begin{split}
h_{(p)}^{\mathstrut}\pos{v}-h_{(p)}^{-1}\pos{v}
=&\;2\.\mathbb{L}_0\.\varepsilon_{(p)}\pos{v}A_{(p)}\@(v)
+\mathbb{L}_0^2\.\varepsilon_{(p)}^{2}\pos{v}\.\partial_{(p)}A_{(p)}\@(v)
+\mathcal{O}(\varepsilon^3)
\\
=&-\@2\.\mathbb{L}_0\.\varepsilon_{(p)}\pos{v}A_{(p)}\@\big(v\@+\@l_{(p)}\big)
-\mathbb{L}_0^2\.\varepsilon_{(p)}^{2}\pos{v}\.\partial_{(p)}A_{(p)}\@\big(v\@+\@l_{(p)}\big)
+\mathcal{O}(\varepsilon^3)\,.
\end{split}
\end{align}

It is worth recalling that the short links approach is the method assumed in the original construction of CLQG \cite{Thiemann:1996aw,Thiemann:1997rt,Ashtekar:2004eh,Thiemann:2007zz}. This method requires an additional procedure of the approximation of the ${\partial_{(p)}A_{(p)}\@(v)}$ quantity in \eqref{double_holonomy_short}. This procedure has been proposed in \cite{Bilski:2020poi} to increase the precision of the holonomy-connection relation up to the quadratic-order terms in the regularization parameter. However, in the original construction of CLQG, only the approximation up to the order $\varepsilon$ is assumed. In this case, the connection in the volume regularization procedure in \eqref{v-regularization} is lattice-smeared by using the formula
\begin{align}
\label{link_connection_2}
h_{(p)}^{-1\mathstrut}\pos{v}\big\{h^{\mathstrut}_{(p)}\pos{v},\mathrm{f}\@\,[E^a\pos{v}]\big\}
=\mathbb{L}_0\.\varepsilon_{(p)}\pos{v}\big\{A_{(p)}\@(v),\mathrm{f}\@\,[E^a\pos{v}]\big\}+\mathcal{O}\big(\varepsilon^2\big)\,.
\end{align}
The object $\mathrm{f}\@\,[E^a\pos{v}]$ is some functional of the gravitational momentum. The consistently accurate expansion (i.e. also up to the order $\varepsilon$) of the loop holonomy equals
\begin{align}
\label{loop_connection_2}
\Big(h_{(q)\@(r)}\pos{v}-h_{(q)\@(r)}^{-1}\pos{v}\Big)
=0+\mathcal{O}(\varepsilon^2)\,.
\end{align}
This relation is not applicable to formulate any regularization of the curvature $F_{(q)\@(r)}\@(v)$, \textit{cf.} \eqref{loop_holonomy_expansion}. Another problem, indicated in \cite{Bilski:2020poi}, is regarding the expression in \eqref{link_connection_2} --- this expression is not unique. Moreover, the lattice-regularized Hamiltonian in the approximation up to the order $\varepsilon$ does not describe the physics defined by the classical continues Hamiltonian with appropriate precision. In the case of gauge field theories, this precision is determined by the relation between group elements and their representations, thus it allows neglecting the terms of order $\varepsilon^3$ and higher. Therefore, the omission of several (but not all) terms of order $\varepsilon^2$ in the original formulation of CLQG \cite{Thiemann:1996aw,Thiemann:1997rt,Ashtekar:2004eh,Thiemann:2007zz} is not precise enough neither consistent. However, in the limit $\varepsilon\to0$, the classical analog of this lattice model coincides with the continuous expression.

%
%%%	FIGURE	%%%
%
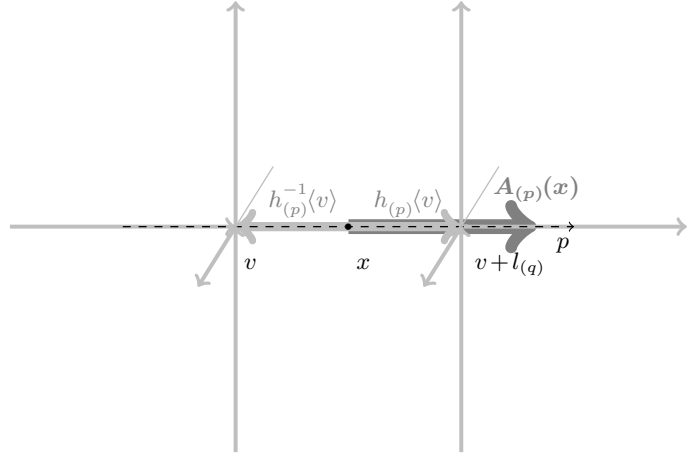
\begin{figure}[h]
\vspace{5pt}%
\begin{center}
\begin{tikzpicture}[scale=1]
%holonomy
\draw[cg25,double,->]{(0.5,0.8) -- (3,0.8)};
\draw[cl35,<->]{(-1,0.8) -- (2,0.8)};
\node[text=gray] at (-0.1,1.15) {$h_{(p)}^{-1\mathstrut}\pos{v}$};
\node[text=gray] at (1.3,1.15) {$h_{(p)}^{\mathstrut}\pos{v}$};
%secondary horizontal
\draw[cl15,->]{(-4,0.8) -- (-1,0.8)};
\draw[cl15,->]{(2,0.8) -- (5,0.8)};
\draw[db05,->]{(-2.5,0.8) -- (3.5,0.8)};
\node at (3.35,0.55) {$p$};
%secondary diagonal
\draw[cl15,->]{(-1,0.8) -- (-1.5,0)};
\draw[cl15,->]{(2,0.8) -- (1.5,0)};
%secondary vertical
\draw[cl15,<-]{(-1,3.8) -- (-1,0.8)};
\draw[cl15,<-]{(2,3.8) -- (2,0.8)};
\draw[cl15,<-]{(-1,0.8) -- (-1,-2.2)};
\draw[cl15,<-]{(2,0.8) -- (2,-2.2)};
%tertiary diagonal
\draw[cl05,->]{(-0.5,1.6) -- (-1,0.8)};
\draw[cl05,->]{(2.5,1.6) -- (2,0.8)};
%%%	dressing		%%%
\node at (0.5,0.8) {\tiny\textbullet};
\node at (0.7,0.3) {$x$};
\node at (-0.8,0.3) {$v$};
\node at (2.65,0.3) {$v\@+\@l_{(q)}$};
\node[text=gray] at (3,1.3) {$\bm{A_{(p)}\@(x)}$};
\end{tikzpicture}
\end{center}
\vspace{-10pt}%
\caption{Holonomy representation (light gray) of the gravitational connection probability distribution (dark gray)\\
in the short length approximation $A_{(p)}\@(x)\to\tilde{A}_{(p)}\pos{v}\approx\frac{1}{2}\Big(A_{(p)}\@(v)+A_{(p)}\@\big(v\@+\@l_{(p)}\big)\Big)$}
\label{holonomy_A}
\end{figure}
%
%%%	FIGURE	%%%
%
The consistent inclusion of the loop holonomy concerning the piecewise analytical lattice with short links should entail the following procedure regarding the link holonomy. One begins from the formula in \eqref{double_holonomy_short}. The derivative ${\partial_{(\@p\@)}A_{(\@p\@)}\@(v)}$ is the limit ${\mathbb{L}_0\.\varepsilon_{(p)}\pos{v}}\to0$ of the connections located at the link's endpoints, namely,
\begin{align}
\label{derivative-difference}
\frac{1}{\mathbb{L}_0\.\varepsilon_{(p)}\pos{v}\!}
\Big(A_{(p)}\@\big(v\@+\@l_{(p)}\big)-A_{(\@p\@)}\@(v)\Big)
=\partial_{(\@p\@)}A_{(\@p\@)}\@(v)+\mathcal{O}(\varepsilon)\,.
\end{align}
By applying this result into \eqref{double_holonomy_short}, one solves two piecewise analytical lattice-related problems at once. The outcome has the form linear in $A_{(\@p\@)}$, thus it is an easily applicable relation, and it is link-symmetric in the distribution of locations of the connection,
\begin{align}
\label{double_holonomy_difference}
h_{(p)}^{\mathstrut}\pos{v}-h_{(p)}^{-1}\pos{v}=\mathbb{L}_0\.\varepsilon_{(p)}\pos{v}\Big(A_{(p)}\@\big(v\@+\@l_{(p)}\big)+A_{(\@p\@)}\@(v)\Big)
+\mathcal{O}(\varepsilon^3)\,.
\end{align}
It is worth mentioning that the second problem was regarding the connection distribution in elementary cells --- see \eqref{connection_smearing}. The result in \eqref{double_holonomy_difference} is the short length approximation of the probability distribution of the gravitational connection $\tilde{A}_{(p)}\pos{v}$ in \eqref{connection_distribution_expansion} (see FIG.~\ref{holonomy_A}). More precisely, it is the arithmetic mean of the expansions of $\tilde{A}_{(p)}\pos{v}$ in \eqref{double_holonomy_expansion} located at both endpoints of the link ${l_{(p)}\pos{v}}$.

Then, to elaborate the problem with the symmetrization completely, one should introduce the equivalents of ${A_{(p)}\@(v)}$ and ${B^{(p)}\@(v)}$, defined as the short links approximations of $\tilde{A}_{(p)}\pos{v}$ and $\tilde{B}^{(p)}\pos{v}$, symmetrized concerning nodes distributions inside a link and a loop, respectively. These symmetrizations (of the quantities defined in \eqref{connection_distribution} and \eqref{curvature_distribution}, respectively) are the following:
\begin{align}
\label{symmetric_A}
\stackrel{\scriptscriptstyle\textsc{m\!\.e\!\.a\!\.n}\!}{A}{\!\@}_{(p)}\pos{v}:=&\;
\frac{1}{2}\Big(A_{(p)}\@(v)+A_{(p)}\@\big(v\@+\@l_{(p)}\big)\Big)
=\tilde{A}_{(p)}\pos{v}+\mathcal{O}(\varepsilon^2)\,,
\\
\label{symmetric_B}
\!\!\stackrel{\scriptscriptstyle\textsc{m\!\.e\!\.a\!\.n}\!}{B}{\!\@}^{(p)}
\@\big(\@S_{\@(\@v\@+\@l_{\@(\@q\@)}\@\@/\@2\@+\@l_{\@(\@r\@)}\@\@/\@2\@)}\!\big)\@:=&\;
\frac{1}{4}\Big(\@
B^{(p)}\@(v)\@+\@B^{(p)}\@\big(v\!+\!l_{(p)}\@\big)\@+\@
B^{(p)}\@\big(v\!+\!l_{(p)}\!+\!l_{(r)}\@\big)\@+\@
B^{(p)}\@\big(v\!+\!l_{(r)}\@\big)
\@\Big)
=\tilde{B}^{(p)}\@\big(\@S_{\@(\@v\@+\@l_{\@(\@q\@)}\@\@/\@2\@+\@l_{\@(\@r\@)}\@\@/\@2\@)}\!\big)
\@+\@\mathcal{O}(\varepsilon^3)\,.\!\!
\end{align}
The point that corresponds to the arithmetic mean of the nodes is the centroid of these nodes' positions. The centroid of the link ${l_{(p)}\pos{v}}$ is located at ${v\@+\@l_{(p)}\@/2}$. The surface ${S_{\@(\@v\@+\@l_{\@(\@q\@)}\@\@/\@2\@+\@l_{\@(\@r\@)}\@\@/\@2\@)}}$ has been labeled by its vertex centroid\footnote{The vertex centroid of a polygon is defined at the point, which is the arithmetic mean of the coordinates of this polygon's vertices.}, hence this centroid is located at ${v\@+\@l_{(p)}\@/2\@+\@l_{(r)}\@/2}$. Thus, it is natural to consider the lattice-symmetrization of the connection $A_{a}(x)$ by the distribution ${\tilde{A}_{a}\pos{x\@-\@l_{(a)}\@/\@2}}\approx{\stackrel{\scriptscriptstyle\textsc{m\!\.e\!\.a\!\.n}\!}{A}{\!\@}_{(p)}\pos{x\@-\@l_{(a)}\@/2}}$ and the symmetrization of the densitized curvature $B^{a}(x)$ by the distribution ${\tilde{B}^{a}\@\big(\@S_{\@(\@x\@)}\!\big)}\approx{\stackrel{\scriptscriptstyle\textsc{m\!\.e\!\.a\!\.n}\!}{B}{\!\@}^{(p)}\@\big(\@S_{\@(\@x\@)}\!\big)}$.

%
%%%	FIGURE	%%%
%
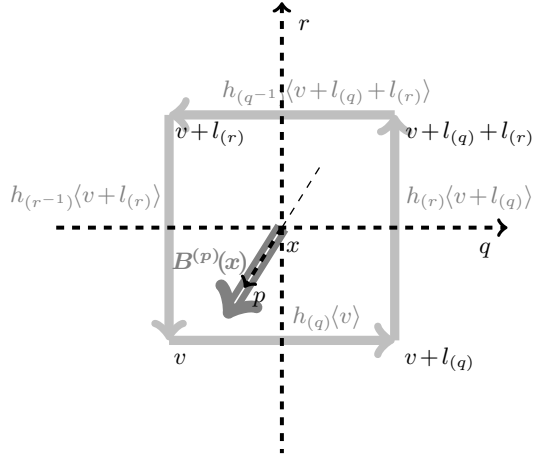
\begin{figure}[h]
\vspace{5pt}%
\begin{center}
\begin{tikzpicture}[scale=1]
%holonomy
\draw[cl35,->]{(-1,-0.7) -- (2,-0.7)};
\draw[cl35,->]{(2,-0.7) -- (2,2.3)};
\draw[cl35,->]{(2,2.3) -- (-1,2.3)};
\draw[cl35,->]{(-1,2.3) -- (-1,-0.7)};
\draw[cg25,double,->]{(0.5,0.8) -- (-0.25,-0.4)};
\node[text=gray] at (1.1,-0.4) {$h_{(q)}\pos{v}$};
\node[text=gray] at (3.0,1.2) {$h_{(r)}\pos{v\@+\@l_{(q)}}$};
\node[text=gray] at (1.1,2.6) {$h_{(q^{-1})}\pos{v\@+\@l_{(q)}\@+\@l_{(r)}}$};
\node[text=gray] at (-1.4,1.2) {\hspace{-40pt}$h_{(r^{-1})}\pos{v\@+\@l_{(r)}}$};
%secondary horizontal
\draw[db15,->]{(-2.5,0.8) -- (3.5,0.8)};
\node at (3.2,0.5) {$q$};
%secondary diagonal
\draw[db15,->]{(0.5,0.8) -- (0,0)};
\node at (0.2,-0.15) {$p$};
%secondary vertical
\draw[db15,<-]{(0.5,3.8) -- (0.5,-2.2)};
\node at (0.8,3.5) {$r$};
%tertiary diagonal
\draw[db05]{(1,1.6) -- (0.5,0.8)};
%%%	dressing		%%%
\node at (0.65,0.55) {$x$};
\node at (-0.85,-0.95) {$v$};
\node at (2.6,-0.95) {$v\@+\@l_{(q)}$};
\node at (3.0,2.05) {$v\@+\@l_{(q)}\@+\@l_{(r)}$};
\node at (-0.45,2.05) {$v\@+\@l_{(r)}$};
\node[text=gray] at (-0.45,0.35) {$\bm{B^{\@(p)\@}\@(\@x\@)}$};
\end{tikzpicture}
\end{center}
\vspace{-10pt}%
\caption{Quadrilateral loop holonomy representation of\\$B^{(p)}\@(x)
\to\tilde{B}^{(p)}\@\big(\@S_{\@(\@v\@+\@l_{\@(\@q\@)}\@\@/\@2\@+\@l_{\@(\@r\@)}\@\@/\@2\@)}\!\big)
\approx\frac{1}{4}\Big(\@
B^{(p)}\@(v)\@+\@B^{(p)}\@\big(v\!+\!l_{(p)}\@\big)\@+\@
B^{(p)}\@\big(v\!+\!l_{(p)}\!+\!l_{(r)}\@\big)\@+\@
B^{(p)}\@\big(v\!+\!l_{(r)}\@\big)
\@\Big)$ in \eqref{symmetric_B}}
\label{holonomy_B}
\end{figure}
%
%%%	FIGURE	%%%
%
Consequently, the application of the $\varepsilon^2$-order expansion in \eqref{loop_holonomy_expansion} of the following loop holonomy,
\begin{align}
\label{loop_holonomy_linear}
h_{(q)\@(r)}\pos{v}=
h^{\mathstrut}_{(q)}\pos{v}\.
h^{\mathstrut}_{(r)}\pos{v\@+\@l_{(q)}}\.
h^{\mathstrut}_{(q^{-1})}\pos{v\@+\@l_{(q)}\@+\@l_{(r)}}\.
h^{\mathstrut}_{(r^{-1})}\pos{v\@+\@l_{(r)}}
=
h^{\mathstrut}_{(q)}\pos{v}\.
h^{\mathstrut}_{(r)}\pos{v\@+\@l_{(q)}}\.
h^{-1\mathstrut}_{(q)}\pos{v\@+\@l_{(r)}}\.
h^{-1\mathstrut}_{(r)}\pos{v}\,,
\end{align}
leads to the result
\begin{align}
\label{loop_connection_3}
\epsilon^{pqr}_{\scriptscriptstyle+}\Big(h_{qr}\pos{v}-h_{qr}^{-1}\pos{v}\Big)
=4\.\mathbb{L}_0^2\.\epsilon^{(p)\@(q)\@(r)}\.\varepsilon_{(q)}\pos{v}\.\varepsilon_{(r)}\pos{v}\.\stackrel{\scriptscriptstyle\textsc{m\!\.e\!\.a\!\.n}\!}{B}{\@\@}^{p}\@\big(S_{\@(\@v\@+\@l_{(q)}\@\@/2\@+\@l_{(r)}\@\@/2\@)}\@\big)+\mathcal{O}(\varepsilon^3)\,.
\end{align}
Here, $\epsilon^{(p)\@(q)\@(r)}$ is a Levi-Civita tensor and $\stackrel{\scriptscriptstyle\textsc{m\!\.e\!\.a\!\.n}\!}{B}{\@\@}^{p}\@\big(S_{\@(\@v\@+\@l_{(q)}\@\@/2\@+\@l_{(r)}\@\@/2\@)}\@\big)$ is a quantity of a weight $1$ --- see definition \eqref{curvature_distribution}. The formula in \eqref{loop_connection_3} provides a resolution in the shift between the position of the $B$-field and the path-located loop holonomy representation --- see FIG.~\ref{holonomy_B}. To similarly smear the connection in \eqref{v-regularization} on a lattice with the same $\varepsilon^2$ accuracy, one should apply the relation in \eqref{double_holonomy_difference}. It leads to the following expression,
\begin{align}
\label{link_connection_3}
\Big\{\Big(h^{\mathstrut}_{p}\pos{v}-h_{p}^{-1\mathstrut}\pos{v}\Big),\mathrm{f}\@\,[E^a\pos{v}]\Big\}
=2\.\mathbb{L}_0\.\varepsilon_{(p)}\pos{v}\Big\{\!\stackrel{\scriptscriptstyle\textsc{m\!\.e\!\.a\!\.n}\!}{A}{\!\@\!}_{p}\pos{v},\mathrm{f}\@\,[E^a\pos{v}]\Big\}
+\mathcal{O}\big(\varepsilon^3\big)\,.
\end{align}
The result is not only more accurate than the CLQG's relation in \eqref{link_connection_2}, but it resolves the half link's length shift between the locations of connection and holonomy representations (this issue is sketched in FIG.~\ref{holonomy_A}, see also alternative expansions in \eqref{double_holonomy_short}).

The methods in \eqref{loop_connection_3} and \eqref{link_connection_3} improve the regularization proposed in CLQG \cite{Thiemann:1996aw,Thiemann:1997rt,Ashtekar:2004eh,Thiemann:2007zz}. They require the short-links expansion of holonomies at the nodes that contribute to the paths (links or loops) related to these holonomies. Another needed assumption is the symmetrization of the positions of these nodes in \eqref{symmetric_A} and \eqref{symmetric_B}. As a result, one obtains a Hamiltonian constraint expressed in terms of gauge-invariant quantities located on lattice's links. The other two constraints can be solved by implementing gauge-invariant projectors at nodes, namely the link's intersections. As a result, one is able to define Hilbert spaces at links, however, there is a problem: the Hamiltonian does not act on single links but on loops and regions. The result of this action is then dependent on the projectors on gauge-invariant states. Therefore, the classical invariants of a pair of first-class constraints modify the quantum solutions of another first-class constraint, namely the quantized Hamiltonian. Thus, the states space in CLQG is not a Hilbert space neither a tensor product of these spaces. It is the configuration of Hilbert spaces that are correlated by active transformations, which is a rather vague construction. Moreover, there are no analogous constructions in any well-tested physical theories.

Fortunately, a bit different approximations lead to much more solid results. Instead of expanding holonomies at all paths-related nodes and averaging the outcome, one can expand them regarding the mean values of connections along links. As it is demonstrated in the following expressions, namely in \eqref{holonomy_regularized} and \eqref{loop_regularized}, the results of this method equal to the linear links assumption. This assumption simplifies and clarifies several steps toward quantization, and, as it is explained in Sec.~\ref{Sec_Regularization_/_volume}, is rigorously construable. As a result, the simplification does not only lead to the redundancy of an additional symmetrization. An appropriate construction of the lattice allows indicating the projector-independent elementary cells on which the Hamiltonian acts (see Sec.~\ref{Sec_Regularization_/_Elementary_cell}). Moreover, the actions on these cells of all other operators or combinations of operators are well-defined and gauge-invariant. Furthermore, the cells-associated states spaces can be defined as a sum of tensor products of Hilbert spaces. Finally, from the perspective of the short-links expansion, the linearity of links can be also supported by the equation, which holds in the limit $\varepsilon\to0$,
\begin{align}
\label{constant_connection}
\bar{A}_{(p)}\pos{v}
=\!\lim_{\varepsilon_{\@(\@p\@)\@}\pos{\@v\@}\to0}\!\!\stackrel{\scriptscriptstyle\textsc{m\!\.e\!\.a\!\.n}\!}{A}{\!\@\!}_{(p)}\pos{v}
=\!\lim_{\varepsilon_{\@(\@p\@)\@}\pos{\@v\@}\to0}\!A_{(p)}\@(v)
=\!\lim_{\varepsilon_{\@(\@p\@)\@}\pos{\@v\@}\to0}\!A_{(p)}\@\big(v\@+\@l_{(p)}\big)\,.
\end{align}

The piecewise linear lattice structure entails the spatial constancy of connections along links. As a result, the holonomy-connection relation becomes simplified to the one in \eqref{holonomy_expansion_linear}. Consequently, the lattice smearing of the gravitational connection is defined by the relation
\begin{align}
\label{holonomy_regularized}
\Big\{\Big(\bar{h}_{p}^{\mathstrut}\pos{v}-\bar{h}_{p}^{-1}\pos{v}\Big),\mathrm{f}\@\,\big[\bar{E}^a\pos{v}\big]\Big\}
=2\.\mathbb{L}_0\.\varepsilon_{(p)}\pos{v}\Big\{\bar{A}_{p}\pos{v},\mathrm{f}\@\,[\bar{E}^a\pos{v}\big]\Big\}
+\mathcal{O}\big(\varepsilon^3\big)\,.
\end{align}
The quantity $\mathrm{f}\@\,[\bar{E}^a\pos{v}\big]$ indicates that the momentum functional is also defined regarding a piecewise linear lattice. The similarity between \eqref{link_connection_3} and \eqref{holonomy_regularized} is evident.

The loop-related formula is analogously similar to \eqref{loop_connection_3}, namely,
\begin{align}
\label{loop_regularized}
\epsilon^{pqr}_{\scriptscriptstyle+}\Big(\bar{h}_{qr}\pos{v}-\bar{h}_{qr}^{-1}\pos{v}\Big)
=4\.\mathbb{L}_0^2\.\epsilon^{(p)\@(q)\@(r)}\.\varepsilon_{(q)}\pos{v}\.\varepsilon_{(r)}\pos{v}
\.\bar{B}^{p}\@\big(S_{\@(\@v\@+\@l_{(q)}\@\@/2\@+\@l_{(r)}\@\@/2\@)}\@\big)
+\mathcal{O}(\varepsilon^3)\,,
\end{align}
where $\epsilon^{(p)\@(q)\@(r)}$ is a Levi-Civita tensor. The linear links-related loop holonomy equals the composition of four linear holonomies and can be expanded at any of the related nodes, for instance,
\begin{align}
\label{loop_linear}
\bar{h}_{(q)\@(r)}\pos{v}=
\bar{h}^{\mathstrut}_{(q)}\pos{v}\.
\bar{h}^{\mathstrut}_{(r)}\pos{v\@+\@l_{(q)}}\.
\bar{h}^{-1\mathstrut}_{(q)}\pos{v\@+\@l_{(r)}}\.
\bar{h}^{-1\mathstrut}_{(r)}\pos{v}
=\mathds{1}+\mathbb{L}_0^2\.\varepsilon_{(q)}\pos{v}\.\varepsilon_{(r)}\pos{v}\bar{F}_{(q)\@(r)}\pos{v}+\mathcal{O}(\varepsilon^3)\,.
\end{align}
The expansion on the right-hand side, which is not unique, has the same structure as the one in \eqref{loop_holonomy_expansion}. Hence, the holonomy $\bar{h}_{(q)\@(r)}\pos{v}$ is defined along a non-trivial quadrilateral loop embedded in a non-Euclidean space in general. This demonstrates that the piecewise linear structure of the lattice does not modify the notion of the spatial curvature. Consequently, the densitized gravitational curvature $\bar{B}^{p}\@\big(S_{(v)}\@\big)$ becomes defined in the same manner as the quantities in \eqref{curvature_distribution} and \eqref{symmetric_B} but regarding linear links,
\begin{align}
\label{curvature_distribution_linear}
\bar{B}^{p}\@\big(\@S_{\@(\@v\@+\@l_{\@(\@q\@)}\@\@/\@2\@+\@l_{\@(\@r\@)}\@\@/\@2\@)}\!\big)
:=\frac{1}{8}\epsilon^{pqr}_{\scriptscriptstyle+}\@\Big(\@
\bar{F}_{qr}\pos{v}\@+\@
\bar{F}_{rq}\pos{v\@+\@l_{(q)}}\@+\@
\bar{F}_{qr}\pos{v\!+\!l_{(p)}\!+\!l_{(r)}}\@+\@
\bar{F}_{rq}\pos{v\!+\!l_{(r)}}
\@\Big)\,.
\end{align}
Here, analogously to \eqref{Ashtekar_curvature} and \eqref{densitized_curvature}, the curvature along linear links is specified as follows,
\begin{align}
\label{densitized_curvature_linear}
\bar{F}_{qr}\pos{v}
:=\bar{F}\!\.\Big(l^{\mathstrut}_{(q)}\pos{v},l^{-1\mathstrut}_{(r)}\pos{v}\@\Big)
:=\big(\partial_q\bar{A}^i_r\pos{v}-\partial_r\bar{A}^i_q\pos{v}+\epsilon_{ijk}\bar{A}^j_q\pos{v}\bar{A}^k_r\pos{v}\big)\tau^i\,.
\end{align}

	%%%%%%%%%	%%%%%%%%%	%%%%%%%%%	%%%%%%%%%
\subsection{Regularization revisited}\label{Sec_Regularization_/_revisited}

\noindent
In the previous sections, two different regularization issues have been discussed: one in Sec.~\ref{Sec_Regularization_/_volume} concerning the geometrical smearing of the degrees of freedom related to $E^a$, another in Sec.~\ref{Sec_Regularization_/_graph}, \ref{Sec_Regularization_/_Wigner}, and \ref{Sec_Regularization_/_holonomy} regarding the holonomy representations of the degrees of freedom related to $A_a$. It is worth comparing the approximations used in the mentioned regularization steps.

The first regularization is assumed by applying the identity in \eqref{v-regularization} (called the volume regularization, or colloquially `Thiemann's trick'), and it introduces the expansion in terms of the dimensionless parameter $\lambda$. The terms of order $\lambda^2$ and higher are neglected. The second group of regularization steps is linked to the introduction of the holonomy representation, with the precision controlled by the dimensionless parameter $\varepsilon$. In this case, it is possible to consider two alternative procedures. The standard CLQG approach assumes the regularization via a pair of differently approximated expressions. In the one in \eqref{link_connection_2}, all the terms of order $\varepsilon^2$ and higher are neglected, while in the expression in \eqref{loop_connection_3}, the neglected terms are of order $\varepsilon^3$ and higher. This inconsistency should be corrected. In the consistent approximation, one considers the application of the pair of formulas \eqref{link_connection_3} and (as previously) \eqref{loop_connection_3}. Consequently, in both expressions, the terms of order $\varepsilon^3$ and higher are consistently neglected, which determines a uniform order of approximation. The second procedure is motivated by the possibility of defining the link-related states spaces as the Hilbert spaces for all gauge-invariant operators. In this procedure, the expansion is elaborated on linear links, and it results in the relations in \eqref{holonomy_regularized} and \eqref{loop_regularized}. These formulas are also the expansions up to the quadratic order in the regularization parameter, namely, up to the terms of order $\varepsilon^2$.

It would be an interesting possibility to correlate two \textit{a priori} disconnected scales represented by the parameters $\lambda$ and $\varepsilon$. These parameters were introduced in expressions \eqref{regulator_region} and \eqref{regulator}, respectively. The analysis in Sec.~\ref{Sec_Regularization_/_volume} suggests the identification $R_{\alpha}=C_{\alpha}$, thus the following relation,
\begin{align}
\label{parameters_ratio}
\forall_{\@\alpha}\,\frac{\varepsilon_{\alpha}}{\lambda_\alpha}=\frac{l_\alpha}{\mathfrak{L}_\alpha}\lesssim1\,,
\end{align}
where 
\begin{align}
\label{averaged link}
l_\alpha:=\sqrt[3]{\mathbf{V}_{\!\textsc{\!e\.\!u\@c\.\!l\!}}(\@C_{\alpha}\@)}
:=\bigg(\!\int_{\!C_{\alpha}}\!\!\!\!d^3s\!\bigg)^{\!\!\frac{1}{3}}\@
=\mathbb{L}_0\.\varepsilon_{\alpha}
\end{align}
and $\varepsilon_{\alpha}:=l_\alpha/\mathbb{L}_0$. However, the estimation in \eqref{parameters_ratio} is natural only regarding the piecewise linear lattice with the elementary cells constructed by the tessellation of $\Sigma_t^{\textsc{a\@d\@m}}$. Concerning the cells composed of analytical links, their geometry needs to approximate Euclidean geometry neither at the scale of a cell nor at much smaller or much bigger scales. If one wanted to correlate the regulators $\varepsilon_{\alpha}$ and $\lambda_\alpha$ on the piecewise analytical lattice, one would need to do it as an \textit{ad hoc} assumption. This conclusion is another argument suggesting to assume the linearity of links.

From now on, the piecewise linear structure of lattice becomes implicitly postulated. The short analytical links approximation would require several subsequent assumptions, which are not needed regarding the lattice introduced in the standard tessellation of the space with a non-Euclidean (in general) geometry. Moreover, the quantization of the short analytical links approximation would lead to a non-separable Fock-like space in which the Hilbert-like spaces would be connected by gauge transformation projectors. The choice of the piecewise linear lattice allows constructing quantum gravity \textit{\`a la} QFT for a gauge vector field.

It is worth emphasizing that the relation in \eqref{parameters_ratio} is independent of the selected structure of tessellation. As a result of the regularization, the gravitational theory becomes located on the lattice that coincides with the edges of cells of the mosaic manifold $\Sigma_t^{\hash}$. This discrete classical system approximates the continuous one only up to the linear order in $\lambda$. Moreover, by increasing the inequality in \eqref{parameters_ratio} (by decreasing the elementary cells' sizes), the lattice approximation does not increase its accuracy. The only possibility to increase this accuracy up to the terms of order $\varepsilon^2$ is a reformulation of the volume regularization in \eqref{v-regularization} by the construction of at least a $\lambda^2$-order approximation. Fortunately, this improvement is not only possible, but the approximate volume regularization can be replaced by an exact relation.

The $\lambda$-corrections are removed from the regularized expressions by introducing the following functional of the metric tensor determinant,
\begin{align}
\label{W_volume}
\mathbf{W}(R,n):=\int_{\!R}\!\!d^3x\.q^n\@(x)\,.
\end{align}
It should be emphasized that this object shares all the spatial symmetries with the volume defined in \eqref{volume}. Moreover, both quantities decompose equally concerning the decomposition of the manifold. Finally, by taking $n=1/2$, they coincide. Consequently, the objects $\mathbf{W}(R,1/2)$, $\mathbf{W}(R,1)$, and $\mathbf{W}(R,n)$ are a scalar, a scalar density, and a $(2n\@-\@1)$-density (a quantity of weight $2n\@-\@1$). It is also easy to verify that the functional $\mathbf{W}(R,n)$ is related to the volume by the following approximation
\begin{align}
\label{W_trick_matter_corrections}
\mathbf{W}(R_{\alpha},n)=\lambda_{\alpha}^3\.\mathbb{L}_0^3\.\bar{q}^{\.n}\@(x)\.\Big|_{x\in\bar{R}_{\alpha}}\!+\mathcal{O}\big(\lambda^4\big)
=\lambda_{\alpha}^{3(1-2n)}\.\mathbb{L}_0^{3(1-2n)}\.\mathbf{V}^{2n}\@(R_{\alpha})\.\Big|_{x\in R_{\alpha}}\!
+\mathcal{O}\big(\lambda_{\alpha}^4\big)\,,
\qquad\lambda_{\alpha}\ll1\,.
\end{align}
The region $R$ represents any part of $\Sigma_t^{\hash}$. In particular, one can consider the natural choice, i.e. set this region as coinciding with an elementary cell, thus getting $\mathbf{W}(C_{\alpha},n)$.

The main reason to define the new functional of the metric tensor determinant in \eqref{W_volume} was the construction of the regularization-introducing identity analogous to the approximate expression in \eqref{v-regularization}. This identity reads
\begin{align}
\label{e-regularization}
\frac{q^n\@(x)\!}{E_i^a(x)}\bigg|_{x\in R_{\alpha}}\!\!
=-\frac{2}{\gamma\kappa\.n}
\big\{A^i_a(x),\mathbf{W}(R_{\alpha},n)\big\}\Big|_{x\in R_{\alpha}}
\end{align}
and is going to be called the exact regularization procedure (in analogy to the volume regularization in \eqref{v-regularization}). It is worth noting that the same equation can be defined regarding any spatial region $R$. Therefore, this expression is also directly applicable concerning the embedding of a piecewise linear lattice in the spatial manifold.

The introduction of the identity in \eqref{e-regularization} completes the list of formulas that define the gravitational field's lattice smearing. By constructing the gauge-invariant lattice over the graph determined by a selected honeycomb, one obtains the classical model of lattice gravity \cite{Bilski:2021_RCT_III}. The most important achievement regarding this construction is the accuracy of the discrete approximation of the classical continuous gravitational Hamiltonian. The approximation up to the quadratic terms in the regularization parameter is as accurate as the quantum approximation of gauge symmetries by their representations. Therefore, the procedure in this article has the precision of the standard gauge field quantization in QFT. The presented formulation is an order more precise than the formalism of CLQG \cite{Thiemann:1996aw,Thiemann:1997rt,Ashtekar:2004eh,Thiemann:2007zz}, which has only a linear accuracy. Moreover, by lattice smearing the system at the classical level one obtains the model directly quantizable by the Dirac-Heisenberg canonical procedure \cite{Dirac:1925jy,Heisenberg:1929xj}. Furthermore, its construction on the separable piecewise linear lattice allows defining the DeWitt representation of operators (namely the representation in \eqref{DeWitt_representation}) in the standard Dirac-Heisenberg-Wigner method \cite{Dirac:1925jy,Heisenberg:1929xj,Wigner:1939cj,DeWitt:1967yk}. Finally, the operators will act on links-located Hilbert spaces or tensor products of these spaces. Each Hilbert space will be rigorously defined, yet metric-independent by using the construction of quantum spaces proposed in CLQG in \cite{Ashtekar:1995zh}.

The main part of the remaining analysis in this article concerns honeycombs. In what follows, three tessellations are discussed, namely, the quadrilaterally hexahedral, quadrilaterally dodecahedral, and tetrahedral honeycombs. They determine the manifolds $\Sigma_t^{\hash^{\@\textsc{qh}}}$, $\Sigma_t^{\hash^{\@\textsc{qd}}}$, and $\Sigma_t^{\hash^{\@\textsc{th}}}$, decomposable into $C^{\textsc{q\@h}}_{\alpha}$, $C^{\textsc{q\@d}}_{\alpha}$, and $C^{\textsc{t\@h}}_{\alpha}$ elementary cells, respectively. The related graphs are going to be consistently denoted by $\Gamma^{\textsc{q\@h}}$, $\Gamma^{\textsc{q\@d}}$, and $\Gamma^{\textsc{t\@h}}$, respectively. The first of the mentioned mosaic manifold candidates is preferable by its direct relation with the volume element and an easily construable slicing between any pair of opposite faces of each quadrilaterally hexahedral cell $C^{\textsc{q\@h}}_{\alpha}$. The quadrilaterally dodecahedral tessellation is a worth considering choice, however, the related elementary cells can be dissected into four quadrilateral hexahedra. The tetrahedral honeycomb is discussed because it is the linear limit of the piecewise analytical lattice formulated by the triangulation of a topological space in CLQG \cite{Thiemann:1996aw,Thiemann:2007zz}.

	%%%%%%%%%	%%%%%%%%%	%%%%%%%%%	%%%%%%%%%
\subsection{Tessellations}\label{Sec_Regularization_/_Tessellations}

\noindent
The aim of the lattice smearing of the gravitational propagating degrees of freedom is the construction of a particular structure embedded in the physical space. On the one hand, it should be independent of any coordinate system. In the Hamilton-Dirac formalism \cite{Dirac:1950pj}, this condition means the independence of all gauge transformations. In the case of the gravitational gauge field, these transformations are spatial diffeomorphisms and rotations in the internal space. On the other hand, the SE of the construction of this structure has to allow defining a states space \textit{\`a la} Fock space in which the scalar product is independent of the metric tensor and the operators are the Wigner representations of symmetry transformations.

The second group of the mentioned requirements is going to be precisely formulated in the next article \cite{Bilski:2021_RCT_III}. Regarding this analysis, it is worth recalling the fact that if one associates the gauge-invariant symmetry elements with a discrete structure, one can define the analogously discrete structure of Hilbert spaces in which states are irreducible representations of the symmetry transformations \cite{Ashtekar:1995zh,Wigner:1931}. In this article, the gravitational connection $A$ is the representation of the symmetry group of holonomies at linear links. These links will be later the locations of Hilbert spaces. The preservation of the gauge invariance in this structure requires defining appropriate projectors into equivalence classes of diffeomorphisms \cite{Ashtekar:1995zh,Ashtekar:2004eh,Zapata:1997db,Zapata:1997da} and projectors into irreducible representations of internal rotations \cite{Ashtekar:1995zh,Fleischhack:2004jc,Lewandowski:2005jk}. Both types of these projectors are placed at the intersections of the links, which are called nodes. The first-class system of constraints related to the implementation of different symmetries and degrees of freedom requires that the propagating degrees of freedom are independent of the gauge transformations. In other words, the dynamics located at linear links has to be independent of the symmetries invariance implemented at nodes. The only relation between the representatives of different symmetries should be included in the relative positions of these objects. In this way, the symmetry-independent degrees of freedom become locally positioned and oriented without specification of their coordinates. This description has BI by construction. The canonical quantization of such formulated model, its quantum evolution, and derivation of measurable quantities will preserve BI if it will not be additionally modified by any \textit{ad hoc} added background-dependent methods.

The piecewise linear lattice is the graph $\Gamma$ formed by the edges of the tessellated mosaic manifold $\Sigma_t^{\hash}$. The regularization formula in \eqref{loop_regularized} depends on the number of links in a loop. This dependence occurs in the same way also regarding the short analytical links regularization via the expression in \eqref{loop_connection_3}. To preserve the unique normalization in these formulas, which describe the smearing of the densitized gravitational curvature, it is required that the lattice is composed of loops with a constant number of links. The simplest loops are triangles, the next ones are quadrilaterals, etc. As a result, the elementary cells are the polyhedra, which faces are all polygons of the same type.

The next restriction on the lattice construction assumes the possibility of the Gauss constraint implementation. The related gauge invariance of the SU$(2)$ structure distributed in terms of holonomies along links has to be projectable into irreducible representations of this gauge group. In other words, the SU$(2)$ holonomy elements that meet at links' intersections, have to be decomposable in the local orthonormal basis. The rotational invariance of SU$(2)$ ensures that the group elements projected along the links emanated from a single node sum to zero. This condition allows to construct a convex polyhedron, the faces of which are dual to these links \cite{Charles:2010,Bianchi:2010gc}. The lattice is then dual to the convex polyhedra at nodes, hence also consists of convex polyhedral cells. Consequently, the group averaging that implements the gauge invariance at nodes can be viewed as the selection of the equivalence classes of the dual polyhedra. Thus, the polyhedral graph related to each elementary cell is a convex isohedral $1$-skeleton.

Besides that, observational arguments suggest that the whole Universe or its large fragments are approximately flat. Thus, a candidate for the physical theory of geometry should include the construction of a model of Euclidean space. The description of homogeneous, isotropic, and flat geometry both locally and at large scales appears to be naturally represented by the elementary cells in which physical features (at least their volumes and faces that define fluxes of fields) are all equal. The aforementioned restrictions specify this description into the mosaic manifold composed of convex isohedral and isochoric (cell-transitive) solids. The only polyhedra that satisfy these restrictions are a cube and a rhombic dodecahedron. Moreover, the assumption that the low energy limit of the gravitational theory is the Newtonian gravity implies that the Euclidean flat space is the natural candidate for the ground states vacuum limit of quantum geometry; hence the opportunity for choosing this uniform mosaic manifold is also motivated theoretically. Furthermore, the choice of the picture in which only operators evolve and the states are time-independent \cite{Heisenberg:1925zz} entails the constant valency of nodes. What are the state vectors in this so-called Heisenberg picture? These are the vectors on which operators act. The simplest object, which is `large enough' that all observables (holonomies, fluxes, volumes, curvatures, and the Hamiltonian itself) will have well-defined actions on this object, is the elementary cell. Therefore, after quantization, elementary cells will be called elementary state vectors or simply wave functions. The related states spaces will be the superpositions of the Hilbert spaces at all links being the edges of corresponding polyhedra. Consequently, the convex isohedral and isochoric cells should remain the polyhedra with a constant number of faces that are the polygons composed of a constant number of links. However, the distribution of cells does not need to be regular, neither their number needs to remain constant. All these implications restrict the choice of the elementary cell candidate for a vector state for the model of lattice gravity, and thus for the whole model of a discrete relativistic theory. As a result, the only tessellations that satisfy all the just mentioned restrictions consist of quadrilateral hexahedra or quadrilateral dodecahedra. The related mosaic manifolds are denoted by $\Sigma_t^{\hash^{\@\textsc{qh}}}$ and $\Sigma_t^{\hash^{\@\textsc{qd}}}$, respectively.

There are two arguments why only $\Sigma_t^{\hash^{\@\textsc{qh}}}$ should be considered. The first argument was already mentioned earlier; quadrilateral hexahedra have a direct relation with the volume element $dx\wedge dy\wedge dz$. The second one is even stronger; any quadrilateral dodecahedron can be dissected into four quadrilateral hexahedra. Therefore, the mosaic manifold $\Sigma_t^{\hash^{\@\textsc{qd}}}$ can be dissected into $\Sigma_t^{\hash^{\@\textsc{qh}}}$\footnote{It is worth noting that regarding a uniform honeycomb, the rhombic dodecahedron dissects into four trigonal trapezohedra. Thus, each cell $C^{\textsc{q\@h}}$ can be simplicially decomposed into a union of 4 tetrahedra $C^{\textsc{t\@h}}$.}. The latter tessellation is then more elementary.

%
%%%	FIGURE	%%%
%
\begin{figure}[h]
\vspace{5pt}%
\begin{center}
\begin{tikzpicture}[scale=0.5]
%	secondary quadrilateral
\draw[cl05]{(1.0,11.4) -- (8.4,10.4)};
\draw[cl05]{(3.0,4.0) -- (9.0,3.0)};
\draw[cl05]{(1.0,11.4) -- (3.0,4.0)};
\draw[cl05]{(8.4,10.4) -- (9.0,3.0)};
%	lower diagonals
\draw[cl15]{(1.0,0.4) -- (3.0,4.0)};
\draw[cl15]{(6.4,0.0) -- (9.0,3.0)};
%	upper diagonals
\draw[cl15]{(0.0,10.0) -- (1.0,11.4)};
\draw[cl15]{(7.0,7.0) -- (8.4,10.4)};
%	primary quadrilateral
\draw[cl25]{(0.0,10.0) -- (7.0,7.0)};
\draw[cl25]{(1.0,0.4) -- (6.4,0.0)};
\draw[cl25]{(0.0,10.0) -- (1.0,0.4)};
\draw[cl25]{(7.0,7.0) -- (6.4,0.0)};
%	dressing	I
\node at (3.2,2.6) {$l_{\textsc{i}}\pos{v_{\@\alpha}}$};
\node at (9.4,1.6) {$l_{\textsc{i}}\pos{v_{\@\alpha}\@\@+\@l_{\textsc{ii}}}$};
\node at (2.0,10.5) {$l_{\textsc{i}}\pos{v_{\@\alpha}\@\@+\@l_{\textsc{iii}}}$};
\node at (9.8,8.7) {$l_{\textsc{i}}\pos{v_{\@\alpha}\@\@+\@l_{\textsc{ii}}\!+\@l_{\textsc{iii}}}$};
%	dressing	II
\node at (5.4,4.0) {$l_{\textsc{ii}}\pos{v_{\@\alpha}}$};
\node at (5.7,11.2) {$l_{\textsc{ii}}\pos{v_{\@\alpha}\@\@+\@l_{\textsc{iii}}}$};
\node at (4.2,0.6) {$l_{\textsc{ii}}\pos{v_{\@\alpha}\@\@+\@l_{\textsc{i}}}$};
\node at (5.0,9.0) {$l_{\textsc{ii}}\pos{v_{\@\alpha}\@\@+\@l_{\textsc{iii}}\!+\@l_{\textsc{i}}}$};
%	dressing	III
\node at (3.3,6.6) {$l_{\textsc{iii}}\pos{v_{\@\alpha}}$};
\node at (2.1,4.9) {$l_{\textsc{iii}}\pos{v_{\@\alpha}\@\@+\@l_{\textsc{i}}}$};
\node at (10.3,6.0) {$l_{\textsc{iii}}\pos{v_{\@\alpha}\@\@+\@l_{\textsc{ii}}}$};
\node at (8.85,4.2) {$l_{\textsc{iii}}\pos{v_{\@\alpha}\@\@+\@l_{\textsc{i}}\!+\@l_{\textsc{ii}}}$};
\end{tikzpicture}
\end{center}
\vspace{-10pt}%
\caption{Quadrilateral hexahedron $C^{\textsc{q\@h}}_{\alpha}$}
\label{QH_links}
\end{figure}
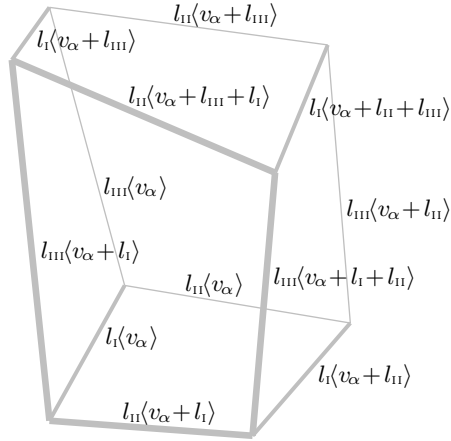
%
%%%	FIGURE	%%%
%
The edges of an elementary cell $C^{\textsc{q\@h}}_{\alpha}$ of the postulated tessellation are sketched in FIG.~\ref{QH_links}. In a local coordinate system, these edges and their lengths are denoted as follows:
${l_{\textsc{i}}\pos{v_{\@\alpha}}}$,
${l_{\textsc{i}}\pos{v_{\@\alpha}\@\@+\@l_{\textsc{ii}}}}$,
${l_{\textsc{i}}\pos{v_{\@\alpha}\@\@+\@l_{\textsc{iii}}}}$,
${l_{\textsc{i}}\pos{v_{\@\alpha}\@\@+\@l_{\textsc{ii}}\!+\@l_{\textsc{iii}}}}$,
${l_{\textsc{ii}}\pos{v_{\@\alpha}}}$,
${l_{\textsc{ii}}\pos{v_{\@\alpha}\@\@+\@l_{\textsc{iii}}}}$,
${l_{\textsc{ii}}\pos{v_{\@\alpha}\@\@+\@l_{\textsc{i}}}}$,
${l_{\textsc{ii}}\pos{v_{\@\alpha}\@\@+\@l_{\textsc{iii}}\!+\@l_{\textsc{i}}}}$,
${l_{\textsc{iii}}\pos{v_{\@\alpha}}}$,
${l_{\textsc{iii}}\pos{v_{\@\alpha}\@\@+\@l_{\textsc{i}}}}$,
${l_{\textsc{iii}}\pos{v_{\@\alpha}\@\@+\@l_{\textsc{ii}}}}$,
${l_{\textsc{iii}}\pos{v_{\@\alpha}\@\@+\@l_{\textsc{i}}\!+\@l_{\textsc{ii}}}}$.
The faces $F_{\@\alpha}$ of this quadrilateral hexahedron are going to be labeled by the same indices as the links that share only their endpoints with these faces. For instance, regarding the notation in FIG.~\ref{QH_links}, $F^{\textsc{i}}_{\@\alpha}$ and $F'^{\textsc{i}}_{\@\alpha}$ are the back and front faces, $F^{\textsc{ii}}_{\@\alpha}$ and $F'^{\textsc{ii}}_{\@\alpha}$ are the left and right ones. The bottom and the top faces are denoted by $F^{\textsc{iii}}_{\@\alpha}$ and $F'^{\textsc{iii}}_{\@\alpha}$, respectively. In what follows, it will also be needed to describe the structures composed of several cells or faces. From now on, a face-connected cluster of cells is going to be called a region and denoted by $R$. An edge-connected layer of faces is going to be called a surface and denoted by $S$.

%
%%%	FIGURE	%%%
%
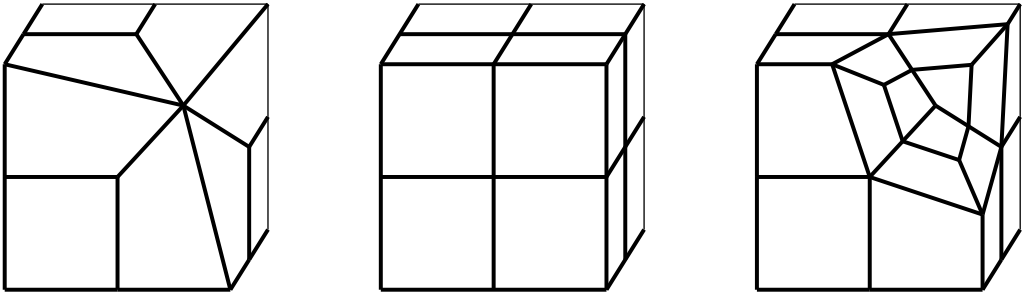
\begin{figure}[h]
\vspace{5pt}%
\begin{center}
\begin{tikzpicture}[scale=1]
		%%%	%%%	%%%
		%		tesseractic	      %
		%%%	%%%	%%%
%	%	%	%	horizontal	%	%	%	%
\draw[cb05]{(-5.75,2.7) -- (-4.25,2.7)};
\draw[cb05]{(-4.25,2.7) -- (-2.75,2.7)};
\draw[cb15]{(-4.5,2.3) -- (-6,2.3)};
\draw[cb15]{(-3.875,1.35) -- (-4.5,2.3)};
\draw[cb15]{(-6.25,1.9) -- (-3.875,1.35)};
\draw[cb15]{(-4.75,0.4) -- (-6.25,0.4)};
\draw[cb15]{(-6.25,-1.1) -- (-4.75,-1.1)};
\draw[cb15]{(-4.75,-1.1) -- (-3.25,-1.1)};
%	%	%	%	vertical	%	%	%	%
\draw[cb15]{(-6.25,1.9) -- (-6.25,0.4)};
\draw[cb15]{(-6.25,0.4) -- (-6.25,-1.1)};
\draw[cb15]{(-4.75,0.4) -- (-4.75,-1.1)};
\draw[cb15]{(-4.75,0.4) -- (-3.875,1.35)};
\draw[cb15]{(-3.875,1.35) -- (-3.25,-1.1)};
\draw[cb15]{(-3,-0.7) -- (-3,0.8)};
\draw[cb15]{(-3,0.8) -- (-3.875,1.35)};
\draw[cb05]{(-2.75,2.7) -- (-2.75,1.2)};
\draw[cb05]{(-2.75,1.2) -- (-2.75,-0.3)};
%	%	%	%	diagonal	%	%	%	%
\draw[cb15]{(-6.25,1.9) -- (-6.0,2.3)};
\draw[cb15]{(-6.0,2.3) -- (-5.75,2.7)};
\draw[cb15]{(-4.25,2.7) -- (-4.5,2.3)};
\draw[cb15]{(-3.875,1.35) -- (-2.75,2.7)};
\draw[cb15]{(-2.75,1.2) -- (-3.0,0.8)};
\draw[cb15]{(-3.25,-1.1) -- (-3.0,-0.7)};
\draw[cb15]{(-3.0,-0.7) -- (-2.75,-0.3)};
		%%%	%%%	%%%
		%		cubic		      %
		%%%	%%%	%%%
\draw[cb15]{(2,-0.7) -- (2,2.3)};
\draw[cb15]{(2,2.3) -- (-1,2.3)};
\draw[cb15]{(0.75,2.7) -- (0.25,1.9)};
\draw[cb15]{(0.25,1.9) -- (0.25,-1.1)};
\draw[cb15]{(2.25,1.2) -- (1.75,0.4)};
\draw[cb15]{(1.75,0.4) -- (-1.25,0.4)};
%secondary horizontal
\draw[cb15]{(-1.25,1.9) -- (1.75,1.9)};
\draw[cb15]{(-1.25,-1.1) -- (1.75,-1.1)};
%secondary vertical
\draw[cb15]{(-1.25,1.9) -- (-1.25,-1.1)};
\draw[cb15]{(1.75,1.9) -- (1.75,-1.1)};
%secondary diagonal
\draw[cb15]{(-1.25,1.9) -- (-0.75,2.7)};
\draw[cb15]{(1.75,1.9) -- (2.25,2.7)};
\draw[cb15]{(1.75,-1.1) -- (2.25,-0.3)};
%tertiary horizontal
\draw[cb05]{(-0.75,2.7) -- (2.25,2.7)};
\draw[cb05]{(2.25,2.7) -- (2.25,-0.3)};
		%%%	%%%	%%%
		%	regular order-5	      %
		%%%	%%%	%%%
%secondary horizontal
	%
\draw[cb15]{(5.5,2.3) -- (4,2.3)};
\draw[cb15]{(6.125,1.35) -- (5.5,2.3)};
\draw[cb15]{(3.75,1.9) -- (4.75,1.9)};
\draw[cb15]{(5.5,2.3) -- (4.75,1.9)};
\draw[cb15]{(5.25,0.4) -- (4.75,1.9)};
\draw[cb15]{(5.438,1.625) -- (5.813,1.825)};
\draw[cb15]{(5.438,1.625) -- (5.688,0.875)};
\draw[cb15]{(5.438,1.625) -- (4.75,1.9)};
\draw[cb15]{(5.25,0.4) -- (3.75,0.4)};
\draw[cb15]{(3.75,-1.1) -- (5.25,-1.1)};
\draw[cb15]{(5.25,-1.1) -- (6.75,-1.1)};
	%
%secondary vertical
	%
\draw[cb15]{(3.75,1.9) -- (3.75,0.4)};
\draw[cb15]{(3.75,0.4) -- (3.75,-1.1)};
\draw[cb15]{(5.25,0.4) -- (5.25,-1.1)};
\draw[cb15]{(5.25,0.4) -- (6.125,1.35)};
\draw[cb15]{(6.75,-0.1) -- (6.75,-1.1)};
\draw[cb15]{(6.75,-0.1) -- (5.25,0.4)};
\draw[cb15]{(6.75,-0.1) -- (7,0.8)};
\draw[cb15]{(5.688,0.875) -- (6.438,0.625)};
\draw[cb15]{(6.563,1.075) -- (6.438,0.625)};
\draw[cb15]{(6.75,-0.1) -- (6.438,0.625)};
\draw[cb15]{(7,-0.7) -- (7,0.8)};
\draw[cb15]{(7,0.8) -- (6.125,1.35)};
	%
%secondary diagonal
	%
\draw[cb15]{(3.75,1.9) -- (4.0,2.3)};
\draw[cb15]{(4.0,2.3) -- (4.25,2.7)};
\draw[cb15]{(5.75,2.7) -- (5.5,2.3)};
\draw[cb15]{(7.083,2.433) -- (7.25,2.7)};
\draw[cb15]{(7.083,2.433) -- (5.5,2.3)};
\draw[cb15]{(7.083,2.433) -- (7,0.8)};
\draw[cb15]{(5.813,1.825) -- (6.604,1.892)};
\draw[cb15]{(6.563,1.075) -- (6.604,1.892)};
\draw[cb15]{(7.083,2.433) -- (6.604,1.892)};
\draw[cb15]{(7.25,1.2) -- (7.0,0.8)};
\draw[cb15]{(6.75,-1.1) -- (7.0,-0.7)};
\draw[cb15]{(7.0,-0.7) -- (7.25,-0.3)};
	%
%tertiary horizontal
	%
\draw[cb05]{(4.25,2.7) -- (5.75,2.7)};
\draw[cb05]{(5.75,2.7) -- (7.25,2.7)};
\draw[cb05]{(7.25,2.7) -- (7.25,1.2)};
\draw[cb05]{(7.25,1.2) -- (7.25,-0.3)};
\end{tikzpicture}
\end{center}
\vspace{-10pt}%
\caption{Honeycombs $\{4,3,r\}$ with fragments labeled by $r$ taking values $3$, $4$, and $5$, respectively}
\label{honeycombs}
\end{figure}
%
%%%	FIGURE	%%%
%
The quadrilaterally hexahedral tessellation into the mosaic manifold $\Sigma_t^{\hash^{\@\textsc{qh}}}$ can be formally described as follows. One should first notice that due to the diffeomorphisms continuation through faces, each elementary cell must have six nearest neighbors (connected by faces). However, the entire structure of $\Gamma^{\textsc{q\@h}}$ does not need to be regular. The number of cells $r$ around each link may vary (and so the number of them around each node). It equals $4$ regarding the cuboidal honeycomb of an Euclidean space, takes value $3$ or $2$ concerning elliptic geometry, and $5$, $6$, or higher value for hyperbolic geometry. These tessellations can be represented by the Schl\"{a}fli symbol $\{4,3,r\}$\footnote{The Schl\"{a}fli symbol $\{p,q,r\}$ is an object characterizing a tessellation of a three-dimensional space or a structure of a regular four-dimensional polytope. The first symbol, $p$, specifies the number of edges of each elementary face (polygon), where $3$ denotes triangles, $4$ --- quadrilaterals, $5$ --- pentagons, etc. The second one, $q$, specifies the number of faces around each vertex of each elementary region (polyhedron). For instance, a tetrahedron has 3 triangles around each vertex and is represented by  $\{3,3\}$, while a quadrilateral hexahedron (or a cube in the regular case) has 3 quadrilaterals around each vertex and is represented by  $\{4,3\}$. A convex four-dimensional polytope having r $\{p,q\}$-convex polyhedral cells around each edge is represented by $\{p,q,r\}$.}. To describe a model corresponding to all cosmological and astrophysical data collected so far, one should be able to simplify the analysis to three types of honeycombs labeled by $\{4,3,3\}$, $\{4,3,4\}$, and $\{4,3,5\}$ --- see FIG.~\ref{honeycombs}. It is worth mentioning that the regular structures with $r=3$, $4$, and $5$ represent the tesseractic-like, cubic-like, and the regular order-5 cubic-like (it has no special name) tessellations, respectively. A fragment of the tesseractic-like honeycomb is sketched on the left-hand side in FIG.~\ref{honeycombs}, and a fragment of the order-5 cubic-like structure is given on the right-hand side. If one considered the regular tessellation of the entire space, the former structure would be finite, and the latter would be compact\footnote{Typical homogeneous models of space, which are implied by observations or postulated in leading gravitational theories, are restricted to a tessellation specified by the parameter $r=4$ in the Schl\"{a}fli symbol $\{4,3,r\}$. In models with inhomogeneous geometry, fragments of a honeycomb can be described by $r=3$ or $r=5$. The whole manifold $\Sigma_t^{\hash^{\@\textsc{qh}}}$ represented by the regular tesseractic-like or the regular order-5 cubic-like structure could be studied only as theoretical models.}.

Finally, it is worth noting that the tessellation into $\Sigma_t^{\hash^{\@\textsc{th}}}$, postulated by CLQG (see \cite{Thiemann:2007zz,Bianchi:2010gc} for details), can be analogously described by the Schl\"{a}fli symbol $\{3,3,r\}$.

	%%%%%%%%%	%%%%%%%%%	%%%%%%%%%	%%%%%%%%%
\subsection{Quadrilaterally hexahedral elementary cell}\label{Sec_Regularization_/_Elementary_cell}

\noindent
The analysis of gauge holonomies' expansions led to conclusions that the smearing of these objects along linear links of a lattice is preferable over the smearing along analytical links. Then, only the former procedure allows one to construct a quantum gauge theory of the gravitational field in the standard Dirac-Heisenberg-Wigner-DeWitt method \cite{Dirac:1925jy,Heisenberg:1929xj,Wigner:1939cj,DeWitt:1967yk}, which additionally has SE and leads to predictions that satisfy BI. In the discussion in Sec.~\ref{Sec_Regularization_/_Tessellations}, it was pointed out that the simplest candidates for state vectors are elementary cells. However, this suggestion has to be verified in a direct calculation. One needs to check how the discrete distribution of holonomies interacts with the functional $\mathbf{W}(R,n)$ in \eqref{e-regularization} (or $\mathbf{V}(R)$ in \eqref{v-regularization}).

The starting point is going to be a precise determination of two correlations regarding the locations of continuous fields and their discrete lattice equivalents. The first relation is between a holonomy along a link and a connection at a point --- see FIG.~\ref{holonomy_A}. The second one is between a holonomy around a loop and a curvature at a point --- see FIG.~\ref{holonomy_B}. Considering links and loops in the quadrilaterally hexahedral lattice, the related locations of connections and curvatures are not the same. Hence, the lattice smearing of the gravitational field cannot be precisely determined from the form of the reversed maps of holonomies expansions along links and loops.

To elaborate the indicated problem, one should focus on the results of the analysis in Sec.~\ref{Sec_Regularization_/_Wigner}, namely, on the fact that Hilbert spaces are going to be located at linear links. This perspective sheds new light on the relation regarding the preferred quadrilateral structure of loops. Each loop constructed by a tessellation is flat and consists of four linear links. Connections are spatially constant along links, hence the center of the linear distribution of a point-located connection is the midpoint of a link. This linear distribution equals the connection's distribution in the definition of a holonomy --- compare FIG.~\ref{holonomy_A}. This equality also explains the proposed averaging of the CLQG's connection in \eqref{symmetric_A} (applicable to \eqref{link_connection_3}), which is needed to increase the linear precision of the approximation in CLQG to the quadratic order in the regularization parameter.

Considering next the quadrilateral face $F^{\textsc{i}}_{\@\alpha}$ of a cell $C^{\textsc{q\@h}}_{\@\alpha}$, illustrated in FIG.~\ref{QH_links}, one can precisely indicate the directly related loop holonomy. Thus, the densitized curvature ${\bar{B}^{\textsc{i}}\@\big(\@F^{\textsc{i}}_{\@\alpha}\big)}:={\bar{B}^{\textsc{i}}\@\big(S^{\textsc{i}}_{\@(v_{\@\alpha\@})}\@\big)}$, which uniquely corresponds to the face $F^{\textsc{i}}_{\@\alpha}$, can be determined. By expanding the loop holonomy, one can locate the resulting quantity at any node in the loop. Then the linear constancy of links and connections allows one to extend this location anywhere on the loop. Moreover, the loop's flatness, being a consequence of a tessellation, involves the constant distribution of the densitized curvature over the entire surface, which has been labeled by ${\bar{B}^{\textsc{i}}\@\big(\@F^{\textsc{i}}_{\@\alpha}\big)}$. Therefore, it is natural to assume that the loop holonomy representation corresponds directly to the lattice smearing of the curvature around its location at the center of the mentioned distribution. How to indicate this center? It should be the geometric center of the points at which the formal expansion can be defined, thus the vertex centroid --- see FIG.~\ref{holonomy_B}. This conclusion is true regarding both the piecewise analytical lattice with short links expansion (the CLQG's regularization) and the postulated piecewise linear lattice with the loop representation defined in \eqref{symmetric_B} (applicable to \eqref{loop_connection_3}).

%
%%%	FIGURE	%%%
%
\begin{figure}[h]
\vspace{5pt}%
\begin{center}
\begin{tikzpicture}[scale=0.5]
%	primary quadrilateral
\draw[cg25,double,->]{(3.5,8.5) -- (7.0,7.0)};
\draw[cl25]{(0.0,10.0) -- (7.0,7.0)};
\draw[cg25,double,->]{(3.7,0.2) -- (6.4,0.0)};
\draw[cl25]{(1.0,0.4) -- (6.4,0.0)};
\draw[cg25,double,->]{(0.5,5.2) -- (0.0,10.0)};
\draw[cl25]{(0.0,10.0) -- (1.0,0.4)};
\draw[cg25,double,->]{(6.7,3.5) -- (7.0,7.0)};
\draw[cl25]{(7.0,7.0) -- (6.4,0.0)};
%	centr./bim.
\draw[cg25]{(3.5,8.5) -- (3.7,0.2)};
\node at (4.75,2.5) {$b_{\textsc{iii}}(\@F^{\textsc{i}}_{\@\alpha}\@)$};
\draw[cg25]{(0.5,5.2) -- (6.7,3.5)};
\node at (5.1,4.45) {$b_{\textsc{ii}}(\@F^{\textsc{i}}_{\@\alpha}\@)$};
\node at (4.8,0.7) {$\bar{A}_{\textsc{ii}}\pos{\@v_{\@\alpha}\@}$};
\node at (5.8,8.6) {$\bar{A}_{\textsc{ii}}\pos{\@v_{\@\alpha}\@\@+\@l_{\textsc{iii}}}$};
\node at (1.45,7.2) {$\bar{A}_{\textsc{iii}}\pos{\@v_{\@\alpha}\@}$};
\node at (8.55,4.8) {$\bar{A}_{\textsc{iii}}\pos{\@v_{\@\alpha}\@\@+\@l_{\textsc{ii}}}$};
\end{tikzpicture}
\end{center}
\vspace{-10pt}%
\caption{Locations of gravitational connections of on $F^{\textsc{i}}_{\@\alpha}$ and the related bimedians}
\label{face_bimedians}
\end{figure}
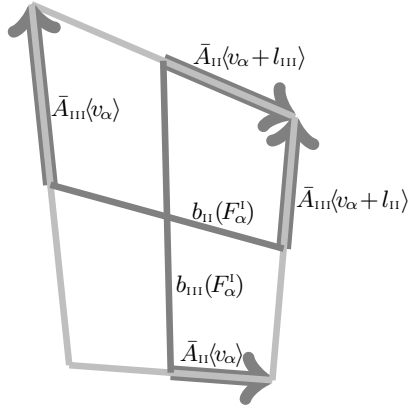
%
%%%	FIGURE	%%%
%
Concerning a face $F^{\textsc{i}}_{\@\alpha}$, one can indicate four corresponding link holonomies, thus also four constant connections. Their locations at the midpoints of four links are indicated in FIG.~\ref{face_bimedians}. Although the connections along opposite links are different, the resultant value and orientation of these objects regarding the whole face can be simplified. Due to the loop's flatness, one can define the following quantity
\begin{align}
\label{bimedian_connection}
\bar{A}_{b_{\@(\@q\@)}\@(\@F^{\@(\@p\@)}_{\@\alpha}\@)}
:=\frac{1}{2}\Big(\bar{A}_{(q)}\pos{v_{\@\alpha}}+\bar{A}_{(q)}\pos{v_{\@\alpha}\@+\@l_{(r)}}\Big)\,,
\end{align}
where the triple $p\neq q\neq r$ indicates the three directions that become orthogonal in the cuboidal limit of the quadrilaterally hexahedral cell. The connection ${\bar{A}_{b_{\@(\@q\@)}\@(\@F^{\@(\@p\@)}_{\@\alpha}\@)}}$ is located along the bimedian ${b_{(q)}\@\big(\@F^{(p)}_{\@\alpha}\big)}$. Therefore, the center of its distribution is the bimedian's midpoint. This point is located at the intersection of bimedians, which is the definition of the vertex centroid. As a result, considering the cell's face as a whole, the center of the lattice smearing of both a curvature and connections is the same, namely, it is the vertex centroid of this face.

%
%%%	FIGURE	%%%
%
\begin{figure}[h]
\vspace{5pt}%
\begin{center}
\begin{tikzpicture}[scale=0.75]
%	secondary quadrilateral
\draw[cl05]{(1.0,11.4) -- (8.4,10.4)};
\draw[cl05]{(3.0,4.0) -- (9.0,3.0)};
\draw[cl05]{(1.0,11.4) -- (3.0,4.0)};
\draw[cl05]{(8.4,10.4) -- (9.0,3.0)};
%	centr./bim.
\draw[cg05]{(4.7,10.9) -- (6.0,3.5)};
\node at (6.25,6.15) {$b_{\textsc{iii}}(F^{\textsc{i}}_{\@\alpha})$};
\draw[cg05]{(2.0,7.7) -- (8.7,6.7)};
\node at (8.4,7.0) {$b_{\textsc{ii}}(F^{\textsc{i}}_{\@\alpha})$};
%	lower diagonals
\draw[cl15]{(1.0,0.4) -- (3.0,4.0)};
\draw[cl15]{(6.4,0.0) -- (9.0,3.0)};
%	centr./bim.
\draw[cg15]{(0.5,10.7) -- (2.0,2.2)};
\node at (1.65,8.45) {$b_{\textsc{iii}}(F^{\textsc{ii}}_{\@\alpha})$};
\draw[cg15]{(2.0,7.7) -- (0.5,5.2)};
\node at (2.1,6.7) {$b_{\textsc{i}}(F^{\textsc{ii}}_{\@\alpha})$};
%%%	%%%	%%%	%%%
\node at (4.475,5.775) {\color{gray}\textbullet};
\node at (4.2,5.6) {$x_{\alpha}$};
%%%	%%%	%%%	%%%
\draw[cg15]{(2.0,2.2) -- (7.7,1.5)};
\node at (2.9,2.35) {$b_{\textsc{ii}}(F^{\textsc{iii}}_{\@\alpha})$};
\draw[cg15]{(6.0,3.5) -- (3.7,0.2)};
\node at (5.1,1.05) {$b_{\textsc{i}}(F^{\textsc{iii}}_{\@\alpha})$};
%	upper diagonals
\draw[cl15]{(0.0,10.0) -- (1.0,11.4)};
\draw[cl15]{(7.0,7.0) -- (8.4,10.4)};
%	centr./bim.
\draw[cg15]{(0.5,10.7) -- (7.7,8.7)};
\node at (6.5,9.4) {$b_{\textsc{ii}}(F'^{\textsc{iii}}_{\@\alpha})$};
\draw[cg15]{(4.7,10.9) -- (3.5,8.5)};
\node at (5.2,10.25) {$b_{\textsc{i}}(F'^{\textsc{iii}}_{\@\alpha})$};
\draw[cg15]{(7.7,8.7) -- (7.7,1.5)};
\node at (8.5,3.9) {$b_{\textsc{iii}}(F'^{\textsc{ii}}_{\@\alpha})$};
\draw[cg15]{(8.7,6.7) -- (6.7,3.5)};
\node at (8.8,5.6) {$b_{\textsc{i}}(F'^{\textsc{ii}}_{\@\alpha})$};
%	primary quadrilateral
\draw[cl25]{(0.0,10.0) -- (7.0,7.0)};
\draw[cl25]{(1.0,0.4) -- (6.4,0.0)};
\draw[cl25]{(0.0,10.0) -- (1.0,0.4)};
\draw[cl25]{(7.0,7.0) -- (6.4,0.0)};
%	centr./bim.
\draw[cg25]{(3.5,8.5) -- (3.7,0.2)};
\node at (4.33,6.5) {$b_{\textsc{iii}}(F'^{\textsc{i}}_{\@\alpha})$};
\draw[cg25]{(0.5,5.2) -- (6.7,3.5)};
\node at (2.35,5.05) {$b_{\textsc{ii}}(\!F'^{\textsc{i}}_{\@\alpha}\!)$};
\end{tikzpicture}
\end{center}
\vspace{-10pt}%
\caption{Bimedians of the following faces of the elementary cell $C^{\textsc{q\@h}}_{\alpha}$:\\
back $F^{\textsc{i}}_{\@\alpha}$, front $F'^{\textsc{i}}_{\@\alpha}$,
left $F^{\textsc{ii}}_{\@\alpha}$, right $F'^{\textsc{ii}}_{\@\alpha}$,
bottom $F^{\textsc{iii}}_{\@\alpha}$, and top $F'^{\textsc{iii}}_{\@\alpha}$}
\label{QH_bimedians}
\end{figure}
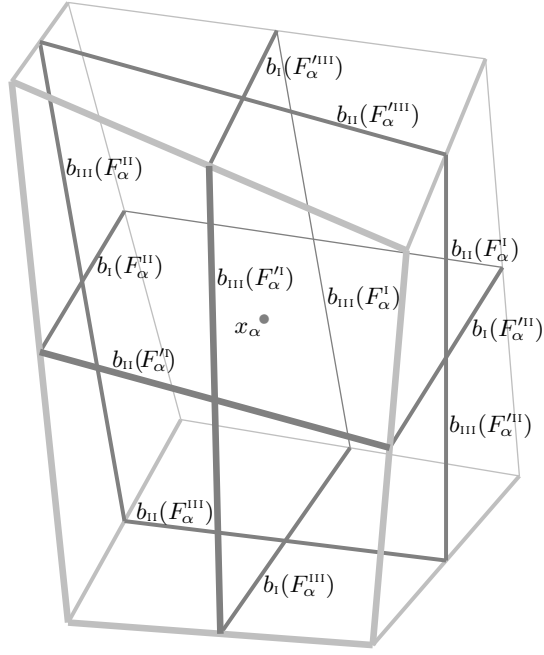
%
%%%	FIGURE	%%%
%
Consequently, in the case of flat faces bounded by linear links that are the outcome of a tessellation, one can simplify the lattice distribution of gravitational connections and holonomies to faces' bimedians, as sketched in FIG.~\ref{face_bimedians}. This simplification concerning the objects related to the opposite links of a face is rigorous due to its flatness. This feature ensures that the length of each bimedian is exactly the arithmetical mean of the lengths of the opposite links, and the bimedian's direction is the arithmetical mean of the links' directions. This determination of bimedians can be extrapolated into six pairs of bimedians indicated for the faces of a cell. It is illustrated in FIG.~\ref{QH_bimedians}. In this figure, the orientation of bimedians became selected in the way, which indicates the right-hand-sided orientation of loops in \eqref{squared_loop} (that determine locations and orientations of the corresponding loop holonomies).

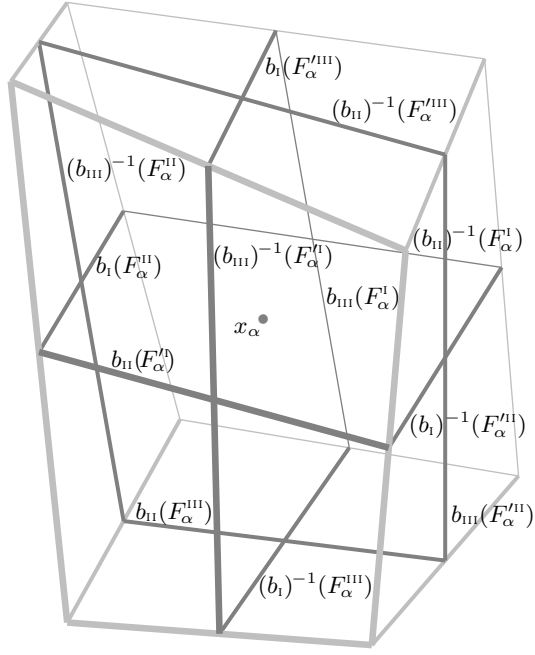
\begin{figure}[h]
\vspace{5pt}%
\begin{center}
\begin{tikzpicture}[scale=0.75]
%	secondary quadrilateral
\draw[cl05]{(1.0,11.4) -- (8.4,10.4)};
\draw[cl05]{(3.0,4.0) -- (9.0,3.0)};
\draw[cl05]{(1.0,11.4) -- (3.0,4.0)};
\draw[cl05]{(8.4,10.4) -- (9.0,3.0)};
%	centr./bim.
\draw[cg05]{(4.7,10.9) -- (6.0,3.5)};
\node at (6.25,6.15) {$b_{\textsc{iii}}(F^{\textsc{i}}_{\@\alpha})$};
\draw[cg05]{(2.0,7.7) -- (8.7,6.7)};
\node at (8.1,7.2) {$(b_{\textsc{ii}})^{\@-1}(F^{\textsc{i}}_{\@\alpha})$};
%	lower diagonals
\draw[cl15]{(1.0,0.4) -- (3.0,4.0)};
\draw[cl15]{(6.4,0.0) -- (9.0,3.0)};
%	centr./bim.
\draw[cg15]{(0.5,10.7) -- (2.0,2.2)};
\node at (2.05,8.45) {$(b_{\textsc{iii}})^{\@-1}(F^{\textsc{ii}}_{\@\alpha})$};
\draw[cg15]{(2.0,7.7) -- (0.5,5.2)};
\node at (2.1,6.7) {$b_{\textsc{i}}(F^{\textsc{ii}}_{\@\alpha})$};
%%%	%%%	%%%	%%%
\node at (4.475,5.775) {\color{gray}\textbullet};
\node at (4.2,5.6) {$x_{\alpha}$};
%%%	%%%	%%%	%%%
\draw[cg15]{(2.0,2.2) -- (7.7,1.5)};
\node at (2.9,2.35) {$b_{\textsc{ii}}(F^{\textsc{iii}}_{\@\alpha})$};
\draw[cg15]{(6.0,3.5) -- (3.7,0.2)};
\node at (5.4,1.05) {$(b_{\textsc{i}})^{\@-1}(F^{\textsc{iii}}_{\@\alpha})$};
%	upper diagonals
\draw[cl15]{(0.0,10.0) -- (1.0,11.4)};
\draw[cl15]{(7.0,7.0) -- (8.4,10.4)};
%	centr./bim.
\draw[cg15]{(0.5,10.7) -- (7.7,8.7)};
\node at (6.8,9.5) {$(b_{\textsc{ii}})^{\@-1}(F'^{\textsc{iii}}_{\@\alpha})$};
\draw[cg15]{(4.7,10.9) -- (3.5,8.5)};
\node at (5.2,10.25) {$b_{\textsc{i}}(F'^{\textsc{iii}}_{\@\alpha})$};
\draw[cg15]{(7.7,8.7) -- (7.7,1.5)};
\node at (8.55,2.3) {$b_{\textsc{iii}}(F'^{\textsc{ii}}_{\@\alpha})$};
\draw[cg15]{(8.7,6.7) -- (6.7,3.5)};
\node at (8.1,3.9) {$(b_{\textsc{i}})^{-1}(F'^{\textsc{ii}}_{\@\alpha})$};
%	primary quadrilateral
\draw[cl25]{(0.0,10.0) -- (7.0,7.0)};
\draw[cl25]{(1.0,0.4) -- (6.4,0.0)};
\draw[cl25]{(0.0,10.0) -- (1.0,0.4)};
\draw[cl25]{(7.0,7.0) -- (6.4,0.0)};
%	centr./bim.
\draw[cg25]{(3.5,8.5) -- (3.7,0.2)};
\node at (4.65,6.9) {$(b_{\textsc{iii}})^{\@-1}(F'^{\textsc{i}}_{\@\alpha})$};
\draw[cg25]{(0.5,5.2) -- (6.7,3.5)};
\node at (2.35,5.05) {$b_{\textsc{ii}}(\!F'^{\textsc{i}}_{\@\alpha}\!)$};
\end{tikzpicture}
\end{center}
\vspace{-10pt}%
\caption{Bimedians forming $C^{\textsc{q\@h}}_{\alpha}$-related triplet of loops, $\squarearrowleft^{\textsc{i}}_{\alpha}$, $\squarearrowleft^{\textsc{ii}}_{\alpha}$, and $\squarearrowleft^{\textsc{iii}}_{\alpha}$ in (\ref{loop_I}--\ref{loop_III}), and their vertex centroid $x_{\alpha}$}
\label{loops_bimedians}
\end{figure}

It is worth recalling that the bimedians of the quadrilateral faces of $C^{\textsc{q\@h}}_{\alpha}$ intersect at the faces' midpoints. These intersections specify the positions of the vertex centroids of all quadrilaterals. Following the notation introduced in FIG.~\ref{QH_links} and FIG.~\ref{face_bimedians}, the relative position of the faces of $C^{\textsc{q\@h}}_{\alpha}$ becomes labeled according to FIG.~\ref{QH_bimedians}. Let a right-hand-oriented quadrilateral loop, set along four bimedians, be defined in analogy to the general expression in \eqref{squared_loop}, namely
\begin{subequations}
\begin{align}
\label{loop_I}
\begin{split}
\squarearrowleft^{\textsc{i}}_{\alpha}
\;:=&\,\circlearrowleft^{\square}_{(\@b_{\textsc{ii}}\@)\@(\@b_{\textsc{iii}}\@)}\!\!\pos{v_{\alpha}\@\@+\@l_{\textsc{i}}\@/2}
=l_{\textsc{ii}}\pos{v_{\alpha}\@\@+\@l_{\textsc{i}}\@/2}\circ
l_{\textsc{iii}}\pos{v_{\alpha}\@\@+\@l_{\textsc{i}}\@/2\@+\@l_{\textsc{ii}}}\circ
(l_{\textsc{ii}})^{\@-1}\pos{v_{\alpha}\@\@+\@l_{\textsc{i}}\@/2\@+\@l_{\textsc{ii}}\!+\@b_{\textsc{iii}}}\circ
(l_{\textsc{iii}})^{\@-1}\pos{v_{\alpha}\@\@+\@l_{\textsc{i}}\@/2\@+\@l_{\textsc{iii}}}
\\
=&\,\;b_{\textsc{ii}}(F^{\textsc{iii}}_{\@\alpha})\circ
b_{\textsc{iii}}(F'^{\textsc{ii}}_{\@\alpha})\circ
(b_{\textsc{ii}})^{\@-1}(F'^{\textsc{iii}}_{\@\alpha})\circ
(b_{\textsc{iii}})^{\@-1}(F^{\textsc{ii}}_{\@\alpha})\,.
\end{split}
\intertext{
Here, $b_{\textsc{ii}}(F^{\textsc{iii}}_{\@\alpha}):=l_{\textsc{ii}}\pos{v_{\alpha}\!+\@\frac{1}{2}l_{\textsc{i}}}$ specifies the position of the bimedian between the midpoints of the opposite edges of $F^{\textsc{iii}}_{\@\alpha}$. This bimedian is oriented as follows. It originates at $v_{\alpha}\!+\@\frac{1}{2}l_{\textsc{i}}$ and directs toward $v_{\alpha}\!+\@\frac{1}{2}l_{\textsc{i}}\@+\@l_{\textsc{ii}}$. Its length equals $|b_{\textsc{ii}}(F^{\textsc{iii}}_{\@\alpha})|=|(b_{\textsc{ii}})^{-1}(F^{\textsc{iii}}_{\@\alpha})|$. For simplicity, both bimedians and their lengths are going to be denoted by the same object (the same simplification was assumed regarding links). It also should be emphasized that the order of the evaluation of bimedians in \eqref{loop_I} dictates the orientation of the bimedian-composing loop in FIG.~\ref{loops_bimedians}. In this figure, the orientation of bimedians became modified to indicate the right-hand-sided orientation of loops in \eqref{squared_loop} that determine locations and orientations of loop holonomies. This orientation, however, is not related to the graph-composing bimedians, which are oriented as links --- see FIG.~\ref{QH_bimedians}. The remaining pair of bimedian loops is constructed in analogy to \eqref{loop_I},}
\label{loop_II}
\squarearrowleft^{\textsc{ii}}_{\alpha}
\;=&\ 
b_{\textsc{iii}}(F^{\textsc{i}}_{\@\alpha})\circ
b_{\textsc{i}}(F'^{\textsc{iii}}_{\@\alpha})\circ
(b_{\textsc{iii}})^{\@-1}(F'^{\textsc{i}}_{\@\alpha})\circ
(b_{\textsc{i}})^{\@-1}(F^{\textsc{iii}}_{\@\alpha})\,,
\\
\label{loop_III}
\squarearrowleft^{\textsc{iii}}_{\alpha}
\;=&\ 
b_{\textsc{i}}(F^{\textsc{ii}}_{\@\alpha})\circ
b_{\textsc{ii}}(F'^{\textsc{i}}_{\@\alpha})\circ
(b_{\textsc{i}})^{\@-1}(F'^{\textsc{ii}}_{\@\alpha})\circ
(b_{\textsc{ii}})^{\@-1}(F^{\textsc{i}}_{\@\alpha})\,.
\end{align}
\end{subequations}
It is worth noting that the symbols $\squarearrowleft^{\textsc{i}}_{\alpha}$, $\squarearrowleft^{\textsc{ii}}_{\alpha}$, and $\squarearrowleft^{\textsc{iii}}_{\alpha}$ uniquely specify all properties of each loop. They indicate its handedness (the mentioned ones are right-handed, hence denoted by $\squarearrowleft$), the position of the related cell (in this case $C^{\textsc{q\@h}}_{\alpha}$), and their symmetry axes $x^{\textsc{i}}$, $x^{\textsc{ii}}$, and $x^{\textsc{iii}}$, respectively (each loop is invariant under cyclic perturbations of bimedians). Analogously, the left-handed loops would be denoted by $\squarearrowright$. It is easy to see that each pair of the triplet of the loops $\squarearrowleft^{\texttt{\#}}_{\alpha}$ intersects twice at the vertex centroids of opposite faces, $c^{\texttt{\#}}_{\alpha}$ and $c'^{\texttt{\#}}_{\alpha}$ (see the notation in Sec.~\ref{Sec_Regularization_/_holonomy}). Finally, it should be emphasized that the center of these centroids coincides with the vertex centroid of the whole hexahedron $C^{\textsc{q\@h}}_{\alpha}$, denoted by $x_{\alpha}:=x_{\alpha}(C^{\textsc{q\@h}}_{\alpha})$ in FIG.~\ref{QH_bimedians} and \ref{loops_bimedians}.
%
%%%	FIGURE	%%%
%
%
%%%	FIGURE	%%%
%

	%%%%%%%%%	%%%%%%%%%	%%%%%%%%%	%%%%%%%%%
\subsection{Holonomy-flux algebra distribution on a graph}\label{Sec_Regularization_/_Holonomy-flux_algebra}

\noindent
Before postulating the form of the distribution of functionals $\mathbf{W}(R,n)$ and $\mathbf{V}(R)$ in the mosaic manifold $\Sigma_t^{\hash^{\@\textsc{qh}}}\!$, thus before defining $\mathbf{W}(C^{\textsc{q\@h}}_{\alpha}\@,n)$ and $\mathbf{V}(C^{\textsc{q\@h}}_{\alpha})$, it is convenient to begin by deriving the following quantity, $\big\{\bar{h}_l\.\pos{v},\bar{E}^a_i(x)\big\}$. This object is the element of expression \eqref{holonomy_regularized}, which can be then applied into formula \eqref{e-regularization} regarding the exact regularization, or into \eqref{v-regularization} regarding the significantly less accurate approximation of the volume regularization. By locating the momentum $\bar{E}^a_i$ at a point $w$ on a lattice, the derivation of the mentioned Poisson brackets simplifies to the calculation of the functional derivative ${\delta\bar{h}_{l\.\pos{v}}[A]/\delta A^i_a(w)}$. The result is significantly different for $w\in l$ positioned at a link's endpoint or between the endpoints \cite{Lewandowski:1993zq,Dona:2010hm} --- see Appendix~\ref{Appendix_derivative} for details. However, in all the cases, i.e. for the momentum located between the endpoints, at the initial point, or the final point of a link, the result (given in \eqref{variation_trick_inside}, \eqref{variation_trick_source}, and \eqref{variation_trick_target}, respectively) is proportional to the quadratic Dirac delta in the $2$-dimensional space, which is oriented orthogonally to the momentum's orientation. Thus, this result is ill-defined. This problem suggests the next step toward the lattice regularization of $E^a$.

Let $\bar{S}^{l\.\pos{v}}_{\@(w)}$ denotes the flat surface intersecting a linear link ${l\.\pos{v}}$ at $w$ and let $\mathbf{A}\big(\bar{S}^{l\.\pos{v}}_{\@(w)}\big)$ denotes the area of this surface. Let also the whole considered system be restricted to the region ${R\@\,\pos{v}}$, where $\mathbf{A}\big(\bar{S}^{l\.\pos{v}}_{\@(w)}\big)$ equals ${l^2_{\!\perp}\pos{v}}:={|l^{\phantom{\prime}}_{\!\perp\@}\pos{v}}\!\.\wedge {l^{\prime}_{\!\perp\@}\pos{v}\@|}$, and the following relation holds, ${\mathbf{V}\big(R\@\,\pos{v}\big)}:={\mathbf{V}\big(l\@\,\pos{v}\!\.\wedge\bar{S}^{l\.\pos{v}}_{\@(w)}\big)}={\mathbf{V}\big(l\@\,\pos{v}\!\.\wedge l^{\phantom{\prime}}_{\!\perp\@}\pos{v}\!\.\wedge l^{\prime}_{\!\perp\@}\pos{v}\big)}$. The link-oriented flux of the gravitational momentum through the area $\mathbf{A}\big(\bar{S}^{l\.\pos{v}}_{\@(w)}\big)$ is given by the expression
\begin{align}
\label{flux_momentum}
\bar{f}_i\@\Big(\@\bar{S}^{l\.\pos{v}}_{\@(w)}\@\Big)
:=\int_{\!\bar{S}^{l\pos{\@v\@}}_{\@(\@w\@)}}\!\!\!\!\!\!\!\enn_a\bar{E}^a_i
:=\!\int_{\!\bar{S}^{l\pos{\@v\@}}_{\@(\@w\@)}}\!\!\!\!\!dy\.dz\.\epsilon_{abc}^{\scriptscriptstyle-}\bar{E}^a_i\partial_yx^b\partial_zx^c
=l^2_{\!\perp}\pos{v}\.\bar{E}^{l}_i\pos{v}\,,
\end{align}
where $\epsilon_{abc}^{\scriptscriptstyle-}$, as previously, denotes the inverse tensor density, and $\enn_a$ projects $E^a_i$ into the directions normal to the surface at each point. Thus, in the case of the flat surface, the momentum is projected into the direction orthogonal to flat surface $\bar{S}^{l\.\pos{v}}_{\@(w)}$. It is worth noting that the analogous definition regarding an analytical link and a possibly curved surface, ${f_i\@\big(\@S^{l\.\pos{v}}_{\@(w)}\@\big)}:={\int_{\!S^{l\pos{\@v\@}}_{\@(\@w\@)}}\!\enn_aE^a_i}$, is not equal to ${l^2_{\!\perp}\pos{v}\.E^{l}_i\pos{v}}$. A more general analysis analysis for CLQG's applications concerning the holonomy-flux algebra on a lattice with the piecewise linear links, which approximate analytical links arbitrarily well, is given in \cite{Bilski:2021ysc}.

The Poisson brackets between the link holonomy $\bar{h}_l$ and the flux in \eqref{flux_momentum} are well-defined,
\begin{align}
\label{holonomy_flux}
\Big\{\bar{h}_l\.\pos{v},\bar{f}_i\@\Big(\@\bar{S}^{l\.\pos{v}}_{\@(w)}\@\Big)\@\Big\}
=\left\{\begin{array}{ll}
-\frac{\gamma\kappa}{2}h_{l\.\pos{v,w}}\tau^ih_{l\.\pos{w,v+l}}
\quad&\text{ for }w\in(v,v+l)
\\
-\frac{\gamma\kappa}{4}\tau^ih_{l\.\pos{v}}
\quad&\text{ for }w=v
\\
-\frac{\gamma\kappa}{4}h_{l\.\pos{v}}\tau^i
\quad&\text{ for }w=v+l\,.
\end{array}\right.
\end{align}
These results include all non-trivially distinct intersections between ${l\.\pos{v}}$ and $\bar{S}^{l\.\pos{v}}_{\@(w)}$, corresponding to the outcomes of the functional derivatives in \eqref{variation_trick_inside}, \eqref{variation_trick_source}, and \eqref{variation_trick_target}, respectively. It is worth noting that the factor in front of each result is a consequence of the symplectic structure normalization in \eqref{Poisson_AE}. It also should be emphasized that these results are regarding the cooriented intersecting pair of a link and a surface. In the case of the reversed orientation, the outcomes in \eqref{holonomy_flux} change their signs according to the following calculation,
\begin{align}
\label{delta_surface_integral}
\int_{\!\!_{\!\scriptstyle\bar{S}^{l_{\@(\@a\@)}\@\pos{v}}_{\@(\@w\@)}}}\!\!\!\!\!\!\!\!\!\!d^2\sigma
\.\delta^2\@\big(l_{_{\@\perp a}}\@(w)-w_{_{\@\perp a}}\big)\Big|_{w\in[v,v+l]}\!\!
=\prod_{b\neq a}^2\!
\Bigg(\!\int_{\@v\@-\@w_{\@(\.\!b\.\!)}}^0\!\!+\!\int_0^{v\@+\@l\@-\@w_{\@(\.\!b\.\!)}}\!\Bigg)
d\sigma^{\perp(\@a\@)\@,(\.\!b\.\!)}d\sigma^{(\.\!b\.\!)}\.
\delta\big(l_{_{\@\perp(\@a\@)\@,(\.\!b\.\!)}}\@(w)-w_{_{\@\perp(\@a\@),(\.\!b\.\!)}}\big)\delta\big(l_b\,\@\big)=\pm1\,.
\end{align}
Here, the same coordinate system shifting method as in Appendix \ref{Appendix_derivative} has been applied, and the general property of the Dirac delta function, which has been recalled in \eqref{delta_integral}, has been used. In particular, in the middle term, the system has been shifted by the vector $(0,0,w_b)^{\T}$ for each direction $b\neq a$. The possible minus sign in the final result can appear due to the possibly reversed orientations of the face $\bar{S}^{l_{\@(\@a\@)}\@\pos{v}\@}_{\@(w)}$ and the link $l_a$. When these orientations agree, the outcome of \eqref{delta_surface_integral} is $1$. It is the case assumed in the derivation of the Poisson brackets in \eqref{holonomy_flux}.

Another step toward the replacement of the flux ${\bar{f}_i\@\big(\@\bar{S}^{l\.\pos{v}}_{\@(w)}\@\big)}$ in \eqref{holonomy_flux} with the functional $\mathbf{W}(R,n)$ (or $\mathbf{V}(R)$) is the repetition of the previous derivation for the distribution of the intersection point $w$ along the whole link $l$. Let ${\mathcal{E}_i\pos{v}}$ denotes the linear density of the gravitational momentum along the link ${l\.\pos{v}}$. Due to the link's linearity (the density becomes consequently labeled by ${\bar{\mathcal{E}}_i\pos{v}}$), the following relation holds,
\begin{align}
\label{density_momentum}
\bar{\mathtt{E}}_i^{l}\pos{v}:=
\!\int_{\@l\.\pos{v}}\!\!\!\!\!\!\bar{\mathcal{E}}_i\pos{v}
=\frac{\bar{E}^{l\.\pos{v}}_i\@\pos{v}\!}{\mathbb{L}_0\.\varepsilon\.\pos{v}\!}\int\!\!dl\.\pos{v}
=\bar{E}^{l\.\pos{v}}_i\@\pos{v}=:\bar{E}^{l}_i\pos{v}\,.
\end{align}
Then, one can derive the flux in \eqref{flux_momentum} of the momentum distribution $\bar{\mathtt{E}}_i^{l}\pos{v}$, 
\begin{align}
\label{density_flux}
\bar{\mathtt{f}}_i^{l\@}\big(R\@\,\pos{v}\big):=
\!\int_{\!\bar{S}^{l\pos{\@v\@}}_{\@(\@w\@)}}\!\!\!\!\bar{\mathtt{E}}_i^{l}\pos{v}
=l^2_{\!\perp}\pos{v}\.\bar{E}^{l}_i\pos{v}\,.
\end{align}

Finally, the derivation of the linear distribution of the Poisson brackets between the link holonomy $\bar{h}_l$ and the flux in \eqref{holonomy_flux} leads to the non-trivial result:
\begin{align}
\label{holonomy_density_flux}
\begin{split}
\frac{1}{\mathbb{L}_0\.\varepsilon\.\pos{v}\!}
\int_{\@l\.\pos{v}}\!\!\!\!\!\!dw
\Big\{\bar{h}_l\.\pos{v},\bar{f}_i\@\Big(\@\bar{S}^{l\.\pos{v}}_{\@(w)}\@\Big)\@\Big\}
=&\,-\frac{\gamma\kappa}{4}\!\lim_{N\to\infty}\frac{1}{N}
\Bigg(\!
\tau^i\bar{h}_l\.\pos{v}\@+\bar{h}_l\.\pos{v}\tau^i
+2\!\sum_{n=1}^{N-1}\!\bar{h}_{l\.\pos{v,v+n\.l/N}}\tau^i\bar{h}_{l\.\pos{v+n\.l/N,v+l}}
\!\Bigg)
\\
\neq&\;\Big\{\bar{h}_l\.\pos{v},\bar{\mathtt{f}}_i^{l\@}\big(R\@\,\pos{v}\big)\@\Big\}\,.
\end{split}
\end{align}
In the upper line, the trapezoidal rule was applied to the integration analogous to the one in \eqref{density_momentum}. The integral along the link has been replaced by an infinite sum of the differences regarding infinitesimally short linear intervals $\mathbb{L}_0\.\varepsilon^l\pos{v}/N$. The link with the contributing gauge-invariant holonomy became decomposed accordingly to the intervals, which altogether compose this link, namely,
\begin{align}
\label{linear_link}
\!l\.\pos{v}:=l\.\pos{v,v\@+\@l}:=l\.\pos{v,v\@+\@l/N}\@\circ\pos{v\@+\@l/N,v\@+\@2l/N}\@\circ...\circ\pos{v\@+\@(N\@-\@1)/N,v\@+\@l}\,.
\end{align}
Finally, the quantity in the lower line of \eqref{holonomy_density_flux} has been inserted to indicate the resulting inequality. This quantity is not precisely defined due to the indeterminacy of the location of the intersection between the link ${l\.\pos{v}}$ and the area $\bar{S}^{l\.\pos{v}}_{\@(w)}$.

A closer look at the middle element in equation \eqref{holonomy_density_flux} reveals a problem. The quantity in the brackets is not gauge-invariant. One can demonstrate this issue explicitly on the following example,
\begin{align}
\label{gauge_variant}
h_{l\.\pos{v,v+l/2}}\tau^ih_{l\.\pos{v+l/2,v+l}}\to g\.h_{l\.\pos{v,v+l/2}}\.g^{-1}g\.\tau^ig^{-1}g\.h_{l\.\pos{v+l/2,v+l}}g^{-1}\@
=h_{l\.\pos{v,v+l/2}}\.g\.\tau^ig^{-1}h_{l\.\pos{v+l/2,v+l}}\,,
\end{align}
where $g,h\in\text{SU}(2)$, $\tau\in\mathfrak{su}(2)$, and, for simplicity, $N=2$. Due to this problem, the definition of the gauge-invariant functional $\mathbf{W}(R,n)$ (or $\mathbf{V}(R)$) cannot be correctly constructed as simply as in the form of a (fractional) power of the contracted fluxes ${\bar{f}_i\@\big(\@\bar{S}^{l\.\pos{v}}_{\@(w)}\@\big)}$. It is worth noting that a similar problem occurs in CLQG \cite{Dittrich:2007th}. There, despite the name `volume operator', the operator related to $\mathbf{V}(R)$ is not a quantized volume, neither the square root of the quantized fluxes ${f_i\@\big(\@S^{l\.\pos{v}}_{\@(w)}\@\big)}$. Instead, an \textit{ad hoc} introduced quantum construction is assumed \cite{Lewandowski:1996gk,Ashtekar:1997fb}, which approximately reproduces the volume in the (semi)classical limit \cite{Thiemann:1996au,Giesel:2005bk}. In the methodology of this article, any methods of an \textit{ad hoc} introduction of quantum constructions are disfavored. Therefore, a direct resolution of the indicated problem is proposed, analogously to the one regarding a more general piecewise analytical link in \cite{Bilski:2021ysc}.

One can introduce the following quantity,
\begin{align}
\label{gauge_flux}
\bar{G}_i^{l\@}\big(R\@\,\pos{v}\big):=
\!\int_{\!\bar{S}^{l\pos{\@v\@}}_{\@(\@w\@)}}\!
\int_{\@l\.\pos{v}}\!\!\!\!\!\!dw\.g^{-1}\@\big(\!\.l(\!\.w\!\.)\!\.\big)\bar{\mathcal{E}}_i(\!\.l(\!\.w\!\.)\!\.\big)g(\!\.l(\!\.w\!\.)\!\.\big)
=\frac{1}{\mathbb{L}_0\.\varepsilon\.\pos{v}\!}
\int_{\@l\.\pos{v}}\!\!\!\!\!\!dw\.g^{-1}\@\big(\!\.l(\!\.w\!\.)\!\.\big)\bar{f}_i\@\Big(\@\bar{S}^{l\.\pos{v}}_{\@(w)}\@\Big)g(\!\.l(\!\.w\!\.)\!\.\big)\,,
\end{align}
which can be viewed either as the flux in \eqref{density_flux} of the distribution of the gauge-invariant momentum density \eqref{density_momentum} or the distribution of the gauge-invariant flux in \eqref{flux_momentum} --- see \cite{Bilski:2021ysc}. By using this quantity to derive the Poisson brackets analogous to \eqref{holonomy_density_flux}, the obtained result still appears to be problematic,
\begin{align}
\label{holonomy_gauge_flux}
\Big\{\bar{h}_l\.\pos{v},\bar{G}_i^{\,\@l\!}\big(R\@\,\pos{v}\big)\@\Big\}
=-\frac{\gamma\kappa}{4}\!\lim_{N\to\infty}\frac{1}{N}
\Bigg(\!
\tau^i\bar{h}_l\.\pos{v}\@+\bar{h}_l\.\pos{v}\tau^i
+2\@\sum_{n=1}^{N-1}\prod_{j=1}^3\bar{h}^{(j)}_{l\pos{v,v+n\.l/N}}\tau^i\bar{h}^{(j)}_{l\pos{v+n\.l/N,v+l}}
\!\Bigg).
\end{align}
The product in the last element determines the transfer of the $\mathfrak{su}(2)$ generators between the exponential maps corresponding to the intervals ${l\.\pos{v,v\@+\@n\.l/N}}$ and ${l\.\pos{v\@+\@n\.l/N,v\@+\@l}}$. These exponential maps from representations to group elements are defined as follows,
\begin{align}
\label{exponential_map}
h_{l\pos{v}}^{(j)}\@
:=\mathcal{P}\exp\!\bigg(\!\int_{\@l\pos{v}}\!\!\!\!\!\!ds\,\dot{\ell}^a\pos{v}\@(s)\.A_a^{(j)}\@\big(l(s)\big)\tau^{(j)}\!\bigg),
\end{align}
where the product ${\prod_{j=1}^3\@h_{l\pos{v}\@}^{\@(j\!\.)}}=h_{l\pos{v}\@}=h_{l\pos{v\!\.,v+l}\@}$ reproduces the definition of the holonomy in \eqref{holonomy_general}. It should be emphasized that the expression in \eqref{holonomy_gauge_flux} is gauge-invariant along the link, except the locations of the generator $\tau^i$. Each element of the summation in this expression transforms as follows,
\begin{align}
\label{gauge_invariant_maps}
g\.\bar{h}_{l\pos{v,w}}g^{-1}\tau^ig\.\bar{h}_{l\pos{w,v+l}}g^{-1}=\prod_{j=1}^3\bar{h}^{(j)}_{l\pos{v,w}}\tau^i\bar{h}^{(j)}_{l\pos{w,v+l}}\,.
\end{align}
Here, the gauge symmetry is preserved along the whole linear link ${l\.\pos{v}}$, except the point ${v\@+\@n\.l/N}$, where the generator has been added. 

The derivation of the right-hand side of formula \eqref{holonomy_gauge_flux} is presented in detail in Appendix~\ref{Appendix_continuous_flux}. The result of this calculation in the limit $N\to\infty$ leads to the expression
\begin{align}
\label{holonomy_flux_final}
\Big\{\bar{h}_l\.\pos{v},\bar{G}_i^{\,\@l\!}\big(R\@\,\pos{v}\big)\@\Big\}
=-\frac{\gamma\kappa}{4}
\Bigg(\!
\tau^i\bar{h}_l\.\pos{v}\@+\bar{h}_l\.\pos{v}\tau^i
+2\sin^2\!\bigg(\frac{\mathbb{L}_0\.\varepsilon_l\pos{v}\!}{4}\sum_{j=1}^3\bar{A}_l^{(j)}\pos{v}\!\bigg)\@\tau^i
\!\Bigg).
\end{align}
If the $\mathcal{O}(\varepsilon^2)$ corrections are neglected, this expression coincides with the sum of the Poisson brackets in \eqref{holonomy_flux} at the link's endpoints,
\begin{align}
\label{holonomy_flux_endpoints}
\bigg\{\bar{h}_l\.\pos{v},
\frac{1}{2}\bigg(\@
\bar{f}_i\@\Big(\@S^{l\.\pos{v}}_{\@\pos{v}}\@\Big)+\bar{f}_i\@\Big(\@S^{l\.\pos{v}}_{\@\pos{v+l}\@}\@\Big)
\!\bigg)
\@\bigg\}\,.
\end{align}

Finally instead of the holonomy on the left-hand side of the formula in \eqref{holonomy_flux_final}, one should concern the functional of holonomies that regularizes the gravitational connection. This leads to the following result,
\begin{align}
\label{holonomy_flux_regularized}
\Big\{\Big(\bar{h}_l^{\mathstrut}\,\@\pos{v}-\bar{h}_l^{-1}\pos{v}\Big),\bar{G}_i^{\,\@l\!}\big(R\@\,\pos{v}\big)\@\Big\}
=-\frac{\gamma\kappa}{4}
\Bigg(\!
\tau^i\Big(\bar{h}_l^{\mathstrut}\,\@\pos{v}-\bar{h}_l^{-1}\pos{v}\Big)
+\Big(\bar{h}_l^{\mathstrut}\,\@\pos{v}-\bar{h}_l^{-1}\pos{v}\Big)\tau^i
\!\Bigg).
\end{align}
This outcome equals exactly to the Poisson brackets in \eqref{holonomy_flux_endpoints} in which, as in \eqref{holonomy_flux_regularized}, the holonomy along a link has been replaced with the pair of reciprocal holonomies. It is worth emphasizing that concerning a piecewise linear approximation of an almost analytical link ${l\.\pos{v}}$, in which the linearity of segments is revealed only at the scale of infinitely many intervals in \eqref{linear_link}, the Poisson brackets in \eqref{holonomy_flux_regularized} lead to the same result \cite{Bilski:2021ysc}.

	%%%%%%%%%	%%%%%%%%%	%%%%%%%%%	%%%%%%%%%
\subsection{Discrete representation of geometry}\label{Sec_Regularization_/_geometry}

\noindent
The last step in the construction of the regularization of the connection-momentum algebra in \eqref{Poisson_qp} is the replacement of the distribution ${\bar{G}_i^{l\@}\big(R\@\,\pos{v}\big)}$ in \eqref{holonomy_flux_regularized} with the functional $\mathbf{W}(C^{\textsc{q\@h}}_{\alpha}\@,n)$ (or $\mathbf{V}(C^{\textsc{q\@h}}_{\alpha})$). In this step, the connection smearing in \eqref{holonomy_regularized} is going to be linked with the momentum regularization in \eqref{e-regularization} (or \eqref{v-regularization}).

One should first determine how the region $R$, which defines the quantities $\mathbf{W}(R,n)$ and $\mathbf{V}(R)$, becomes adjusted to an elementary cell in formulas \eqref{e-regularization} and \eqref{v-regularization}, respectively. The definition in \eqref{W_volume} of the former quantity includes the latter and, in this case, the adjustment reads
\begin{align}
\label{W_cell}
\begin{split}
\!\mathbf{W}\big(C^{\textsc{q\@h}}_{\alpha},n\big)
=&\int_{\!C^{\textsc{q\!\.h}}_{\alpha}}\!\!\!\!\!\!d^3s
\bigg(\bigg|\frac{1}{3!}\epsilon_{abc}^{\scriptscriptstyle-}\epsilon^{ijk}\bar{E}^a_i\@(s)\bar{E}^b_j\@(s)\bar{E}^c_k\@(s)\bigg|\bigg)^{\!\!n}
=
\bigg(\bigg|-\frac{2}{3}\.\text{tr}\@\big[\bar{E}(s)\@\wedge\@\bar{E}(s)\@\wedge\@\bar{E}(s)\big]\bigg|\bigg)^{\!\!n}
\bigg|_{s\in C^{\textsc{q\!\.h}}_{\alpha}}
\int_{\!C^{\textsc{q\!\.h}}_{\alpha}}\!\!\!\!\!\!d^3s
\\
=&\:
\bigg(\bigg|\frac{2}{3}\.\text{tr}\@\Big[
\bar{G}\big(C^{\textsc{q\@h}}_{\alpha}\big)\@\wedge\@
\bar{G}\big(C^{\textsc{q\@h}}_{\alpha}\big)\@\wedge\@
\bar{G}\big(C^{\textsc{q\@h}}_{\alpha}\big)
\Big]\bigg|\mathbf{V}^{-2}_{\!\textsc{\!e\.\!u\@c\.\!l\!}}\big(C^{\textsc{q\@h}}_{\alpha}\big)\!\bigg)^{\!\!n}\bigg|_{s\in C^{\textsc{q\!\.h}}_{\alpha}}
(\varepsilon_{\alpha}\.\mathbb{L}_0)^{3}
\\
=&\:
\bigg(\bigg|\frac{1}{36}\.\text{tr}\@\Big[
\Big(
\bar{G}^{\,\@b_{\textsc{i}\@}(\@F^{\textsc{iii}}_{\!\alpha}\@)}
\!\wedge\@\bar{G}^{\,\@b_{\textsc{ii}\@}(\@F^{\textsc{iii}}_{\!\alpha}\@)}
\!+\@\bar{G}^{\,\@b_{\textsc{i}\@}(\@F'^{\textsc{iii}}_{\!\alpha}\@)}
\!\wedge\@\bar{G}^{\,\@b_{\textsc{ii}\@}(\@F'^{\textsc{iii}}_{\!\alpha}\@)}
\Big)
\!\wedge\!
\Big(
\bar{G}^{\,\@b_{\textsc{iii}\@}(\@F^{\textsc{i}}_{\!\alpha}\@)}
\!+\@\bar{G}^{\,\@b_{\textsc{iii}\@}(\@F^{\textsc{ii}}_{\!\alpha}\@)}
\!+\@\bar{G}^{\,\@b_{\textsc{iii}\@}(\@F'^{\textsc{i}}_{\!\alpha}\@)}
\!+\@\bar{G}^{\,\@b_{\textsc{iii}\@}(\@F'^{\textsc{ii}}_{\!\alpha}\@)}
\Big)
\\
&+\@
\Big(
\bar{G}^{\,\@b_{\textsc{ii}\@}(\@F^{\textsc{i}}_{\!\alpha}\@)}
\!\wedge\@\bar{G}^{\,\@b_{\textsc{iii}\@}(\@F^{\textsc{i}}_{\!\alpha}\@)}
\!+\@\bar{G}^{\,\@b_{\textsc{ii}\@}(\@F'^{\textsc{i}}_{\!\alpha}\@)}
\!\wedge\@\bar{G}^{\,\@b_{\textsc{iii}\@}(\@F'^{\textsc{i}}_{\!\alpha}\@)}
\Big)
\!\wedge\!
\Big(
\bar{G}^{\,\@b_{\textsc{i}\@}(\@F^{\textsc{ii}}_{\!\alpha}\@)}
\!+\@\bar{G}^{\,\@b_{\textsc{i}\@}(\@F^{\textsc{iii}}_{\!\alpha}\@)}
\!+\@\bar{G}^{\,\@b_{\textsc{i}\@}(\@F'^{\textsc{ii}}_{\!\alpha}\@)}
\!+\@\bar{G}^{\,\@b_{\textsc{i}\@}(\@F'^{\textsc{iii}}_{\!\alpha}\@)}
\Big)
\\
&+\@
\Big(
\bar{G}^{\,\@b_{\textsc{iii}\@}(\@F^{\textsc{ii}}_{\!\alpha}\@)}
\!\wedge\@\bar{G}^{\,\@b_{\textsc{i}\@}(\@F^{\textsc{ii}}_{\!\alpha}\@)}
\!+\@\bar{G}^{\,\@b_{\textsc{iii}\@}(\@F'^{\textsc{ii}}_{\!\alpha}\@)}
\!\wedge\@\bar{G}^{\,\@b_{\textsc{i}\@}(\@F'^{\textsc{ii}}_{\!\alpha}\@)}
\Big)
\!\wedge\!
\Big(
\bar{G}^{\,\@b_{\textsc{ii}\@}(\@F^{\textsc{iii}}_{\!\alpha}\@)}
\!+\@\bar{G}^{\,\@b_{\textsc{ii}\@}(\@F^{\textsc{i}}_{\!\alpha}\@)}
\!+\@\bar{G}^{\,\@b_{\textsc{ii}\@}(\@F'^{\textsc{iii}}_{\!\alpha}\@)}
\!+\@\bar{G}^{\,\@b_{\textsc{ii}\@}(\@F'^{\textsc{i}}_{\!\alpha}\@)}
\Big)
\Big]\bigg|\bigg)^{\!\!n}
(\varepsilon_{\alpha}\.\mathbb{L}_0)^{(3-6n)}\!
\\
=&\;(\varepsilon_{\alpha}\.\mathbb{L}_0)^{(3-6n)}\.\mathbf{V}^{2n}\@\big(C^{\textsc{q\@h}}_{\alpha}\big)\,.
\end{split}
\end{align}
In the steps in the first two lines, the local constancy of elementary cells' geometry is applied. In the third line, the transition from the mosaic manifold $\Sigma_t^{\hash^{\@\textsc{qh}}}$ into the quadrilaterally hexahedral graph $\Gamma^{\textsc{q\@h}}$ is postulated. Let it be recalled that the canonical quantization of the theory, which is based on the methods introduced in this article (and will be described in detail in \cite{Bilski:2021_RCT_III}) requires the implementation of propagating degrees of freedom on a graph --- this issue has been explained in Sec.~\ref{Sec_Regularization_/_Wigner}. Thus, both $A_a$ and $E^a$ fields need to be lattice-smeared along some line segments. The structure of this smearing in \eqref{W_cell} reveals the maximally symmetric distribution of all different compositions of the bimedian exterior triple products that equal the geometric definition of the oriented volume of $C^{\textsc{q\@h}}_{\alpha}$. It is worth mentioning that the orientation is then removed by the norm (at the quantum level the norm can be determined by the metric-independent inner product \cite{Ashtekar:1994wa,Ashtekar:1995zh,Baez:1995md,Thiemann:1996hw,Thiemann:1997rv}). In the last step in \eqref{W_cell}, the local constancy of the cells' geometry is used again. Regarding this feature, the approximated relation in \eqref{W_trick_matter_corrections} between $\mathbf{W}\big(C^{\textsc{q\@h}}_{\alpha},n\big)$ and $\mathbf{V}\big(C^{\textsc{q\@h}}_{\alpha}\big)$ becomes equality.

%
%%%	FIGURE	%%%
%
\begin{figure}[h]
\vspace{5pt}%
\begin{center}
\begin{tikzpicture}[scale=0.5]
%	secondary quadrilateral
\draw[cl05]{(1.0,11.4) -- (8.4,10.4)};
\draw[cl05]{(3.0,4.0) -- (9.0,3.0)};
\draw[cl05]{(1.0,11.4) -- (3.0,4.0)};
\draw[cl05]{(8.4,10.4) -- (9.0,3.0)};
%	lower diagonals
\draw[cl15]{(1.0,0.4) -- (3.0,4.0)};
\draw[cl15]{(6.4,0.0) -- (9.0,3.0)};
\node at (5.55,2.15) {$c^{\textsc{iii}}_{\alpha}$};
%	altitude
\draw[cb15]{(4.8,3.0) -- (4.1,9.7)};
\draw[cb05]{(4.35,3.35) -- (4.75,3.4)};
\draw[cb05]{(4.35,3.35) -- (4.4,2.95)};
\node at (5.1,6.0) {$\mathbf{h}^{\textsc{iii}}_{\alpha}$};
%	centr./bim.
\draw[cg15]{(2.0,2.2) -- (7.7,1.5)};
\node at (3.3,2.45) {$b_{\textsc{ii}}\@(\@F^{\textsc{iii}}_{\@\alpha}\@)$};
\draw[cg15]{(6.0,3.5) -- (3.7,0.2)};
\node at (5.85,3.95) {$b_{\textsc{i}}\@(\@F^{\textsc{iii}}_{\@\alpha}\@)$};
\node at (4.85,1.85) {\textbullet};
%	upper diagonals
\draw[cl15]{(0.0,10.0) -- (1.0,11.4)};
\draw[cl15]{(7.0,7.0) -- (8.4,10.4)};
\node at (4.75,9.95) {$c'^{\textsc{iii}}_{\alpha}$};
%	centr./bim.
\draw[cg15]{(0.5,10.7) -- (7.7,8.7)};
\node at (2.4,10.7) {$b_{\textsc{ii}}\@(\@F'^{\textsc{iii}}_{\@\alpha}\@)$};
\draw[cg15]{(4.7,10.9) -- (3.5,8.5)};
\node at (4.9,11.3) {$b_{\textsc{i}}\@(\@F'^{\textsc{iii}}_{\@\alpha}\@)$};
\node at (4.1,9.7) {\textbullet};
%	primary quadrilateral
\draw[cl25]{(0.0,10.0) -- (7.0,7.0)};
\draw[cl25]{(1.0,0.4) -- (6.4,0.0)};
\draw[cl25]{(0.0,10.0) -- (1.0,0.4)};
\draw[cl25]{(7.0,7.0) -- (6.4,0.0)};
\end{tikzpicture}
\end{center}
\vspace{-10pt}%
\caption{Altitude $\mathbf{h}^{\textsc{iii}}_{\alpha}$ of the quadrilateral hexahedron $C^{\textsc{q\@h}}_{\alpha}$ dropped to the base base $F_{\@\alpha}^{\textsc{iii}}$ from the centroid $c'^{\textsc{iii}}_{\alpha}$}
\label{QH_altitude}
\end{figure}
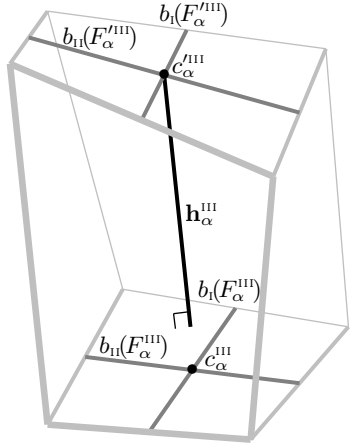
%
%%%	FIGURE	%%%
%
The structure of the fields' smearing on the bimedian exterior triple products represents the symmetrization of the cell's Euclidean volume formula. This formula has already been introduced in the abstract form in \eqref{averaged link}. Its specific geometric form equals
\begin{align}
\label{volume_area-altitude}
\mathbf{V}_{\!\textsc{\!e\.\!u\@c\.\!l\!}}\big(C^{\textsc{q\@h}}_{\alpha}\big)
=\mathbf{A}\big(F_{\@\alpha}^{\texttt{\#}}\big)\.\mathbf{h}^{\texttt{\#}}_{\alpha}\,,
\end{align}
where $\mathbf{A}(F_{\@\alpha}^{\texttt{\#}})$ is the area of the face $F_{\@\alpha}^{\texttt{\#}}$, and $\mathbf{h}^{\texttt{\#}}_{\alpha}$ is the altitude dropped to the face $F_{\@\alpha}^{\texttt{\#}}$ from the vertex centroid $c'^{\texttt{\#}}_{\alpha}:=c'^{\texttt{\#}}_{\alpha}(F'^{\texttt{\#}}_{\alpha})$ of the opposite face $F'^{\texttt{\#}}_{\alpha}$ --- see FIG.~\ref{QH_altitude}. For instance, concerning the area $\mathbf{A}(F_{\@\alpha}^{\textsc{iii}})=|{b_{\textsc{i}}(F^{\textsc{iii}}_{\@\alpha})}\@\wedge\@{b_{\textsc{ii}}(F^{\textsc{iii}}_{\@\alpha})}|$, the volume in \eqref{volume_area-altitude} reads
\begin{align}
\label{area-altitude}
\Big|b_{\textsc{i}}(F^{\textsc{iii}}_{\@\alpha})\@\wedge\@b_{\textsc{ii}}(F^{\textsc{iii}}_{\@\alpha})\@\wedge\@
\frac{1}{4}\big(
b_{\textsc{iii}}(F^{\textsc{i}}_{\@\alpha})
+b_{\textsc{iii}}(F^{\textsc{ii}}_{\@\alpha})
+b_{\textsc{iii}}(F'^{\textsc{i}}_{\@\alpha})
+b_{\textsc{iii}}(F'^{\textsc{ii}}_{\@\alpha})
\big)\Big|\,.
\end{align}
The third factor in this expression, constructed as a mean of four bimedians, is the oriented line segment emanated from the centroid $c^{\textsc{iii}}_{\alpha}$ and going to the centroid $c'^{\textsc{iii}}_{\alpha}$ (of the opposite face). The altitude $\mathbf{h}^{\textsc{iii}}_{\alpha}$, sketched in FIG.~\ref{QH_altitude}, is the norm of the projection of this segment into the normal of $\mathbf{A}\big(F_{\@\alpha}^{\textsc{iii}}\big)$. This explains the bimedian exterior triple product in \eqref{area-altitude}. Then, the structure of these products in \eqref{W_cell} is their mean, averaged over six faces of $C^{\textsc{q\@h}}_{\alpha}$. This arithmetic mean represents the symmetrization of the six face-related locations of the same local coordinate system.

The last statement is worth a longer comment. Each flat face connects two elementary cells. For instance, let the face $F^{\textsc{iii}}_{\@1}\@=F'^{\textsc{iii}}_{\@2}$ connects $C^{\textsc{q\@h}}_{1}$ and $C^{\textsc{q\@h}}_{2}$. A possibly different local coordinate system is related to each of these cells. However, independently of choosing the first or the second local system, the following pairs of holonomies $\bar{h}_{\,\@b_{\textsc{i}\@}(\@F^{\textsc{iii}}_{\!1}\@)}^{\mathstrut}\@=\bar{h}_{\,\@b_{\textsc{i}\@}(\@F'^{\textsc{iii}}_{\!2}\@)}^{\mathstrut}$, $\bar{h}_{\,\@b_{\textsc{ii}\@}(\@F^{\textsc{iii}}_{\!1}\@)}^{\mathstrut}\@=\bar{h}_{\,\@b_{\textsc{ii}\@}(\@F'^{\textsc{iii}}_{\!2}\@)}^{\mathstrut}$, and the pair of areas $\mathbf{A}(F^{\textsc{iii}}_{\@1})=\mathbf{A}(F'^{\textsc{iii}}_{\@2})$ indicate the same objects. These equalities are a consequence of the construction of the mosaic manifold $\Sigma_t^{\hash}$, introduced in Sec.~\ref{Sec_Regularization_/_volume}. They are a result of fixing the orientations of diffeomorphism transformations through the face $F^{\textsc{iii}}_{\@1}\@=F'^{\textsc{iii}}_{\@2}$.

In this fixing, the two-dimensional coordinate systems at both sides of the face are set to be equal. The third dimensions are related as follows. The ray $r_1$ emanated from a point ${p_{\@F}}$ on $F^{\textsc{iii}}_{\@1}$ (inside $C^{\textsc{q\@h}}_{1}$) corresponds to a unique point $p_{1}$ at the boundary $\partial C^{\textsc{q\@h}}_{1}$, i.e. $p_{1}$ is the intersection of the ray and the boundary. There is a $1\@:\@1$ map from the boundary $\partial C^{\textsc{q\@h}}_{1}$ into the boundary ${\partial C^{\textsc{c\@u\@b\@e}}_{\@\@F^{\textsc{iii}}_{\@1}\@}}$ of the cube based on the face $F^{\textsc{iii}}_{\@1}$. Let $p_1^{\textsc{c\@u\@b\@e}}$ denote the mapped intersection $p_1$ into the boundary of the cube, and $r_1^{\textsc{c\@u\@b\@e}}$ denote the corresponding ray (emanated from the same point ${p_{\@F}}$ and going through $p_1^{\textsc{c\@u\@b\@e}}$). Let ${\partial C^{\textsc{c\@u\@b\@e}}_{\@\@F'^{\textsc{iii}}_{\@2}\@}}$ denotes the boundary of the cube reflected through the face $F^{\textsc{iii}}_{\@1}\@=F'^{\textsc{iii}}_{\@2}$. The diffeomorphic continuation of the direction indicated by the ray $r_1^{\textsc{c\@u\@b\@e}}$ in ${C^{\textsc{c\@u\@b\@e}}_{\@\@F^{\textsc{iii}}_{\@1}\@}}$ through the face $F^{\textsc{iii}}_{\@1}\@=F'^{\textsc{iii}}_{\@2}$ into the cube ${C^{\textsc{c\@u\@b\@e}}_{\@\@F'^{\textsc{iii}}_{\@2}\@}}$ is going to be denoted by $r_2^{\textsc{c\@u\@b\@e}}$. Another $1\@:\@1$ map from the boundary ${\partial C^{\textsc{c\@u\@b\@e}}_{\@\@F'^{\textsc{iii}}_{\@2}\@}}$ into the boundary $\partial C^{\textsc{q\@h}}_{2}$ transforms the position of the intersection $p_2^{\textsc{c\@u\@b\@e}}$ between the ray $r_2^{\textsc{c\@u\@b\@e}}$ and the boundary ${\partial C^{\textsc{c\@u\@b\@e}}_{\@\@F'^{\textsc{iii}}_{\@2}\@}}$ into $p_{2}$ on the boundary $\partial C^{\textsc{q\@h}}_{2}$. The ray $r_2$, emanated from ${p_{\@F}}$ and going through $p_{2}$, defines the diffeomorphic continuation of the direction indicated by the ray $r_1$ through the face $F^{\textsc{iii}}_{\@1}\@=F'^{\textsc{iii}}_{\@2}$. One can easily demonstrate that the same construction determines $r_1$ as the diffeomorphic continuation of $r_2$.

%
%%%	FIGURE	%%%
%
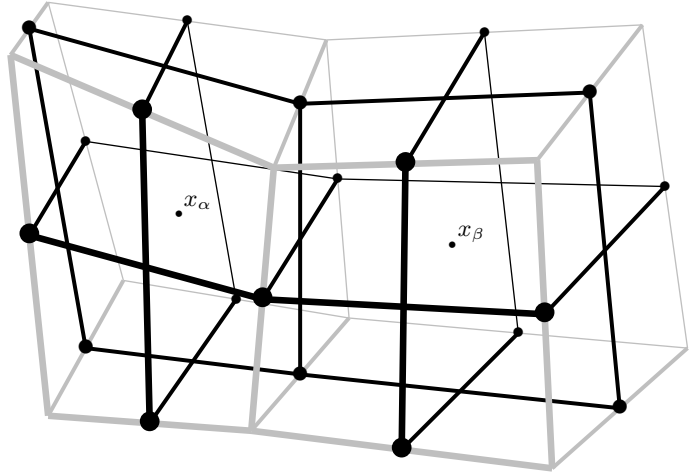
\begin{figure}[h]
\vspace{5pt}%
\begin{center}
\begin{tikzpicture}[scale=0.5]
%	secondary quadrilateral
\draw[cl05]{(1.0,11.4) -- (8.4,10.4)};
\draw[cl05]{(3.0,4.0) -- (9.0,3.0)};
\draw[cl05]{(1.0,11.4) -- (3.0,4.0)};
\draw[cl05]{(8.4,10.4) -- (9.0,3.0)};
\draw[cl05]{(8.4,10.4) -- (16.8,10.8)};
\draw[cl05]{(9.0,3.0) -- (18.0,2.2)};
\draw[cl05]{(16.8,10.8) -- (18.0,2.2)};
%	centr./bim.
\node at (2.0,7.7) {\textbullet};
\node at (8.7,6.7) {\textbullet};
\node at (17.4,6.5) {\textbullet};
\node at (4.7,10.9) {\textbullet};
\node at (6.0,3.5) {\textbullet};
\node at (12.6,10.6) {\textbullet};
\node at (13.5,2.6) {\textbullet};
\draw[cb05]{(4.7,10.9) -- (6.0,3.5)};
\draw[cb05]{(2.0,7.7) -- (8.7,6.7)};
\draw[cb05]{(12.6,10.6) -- (13.5,2.6)};
\draw[cb05]{(8.7,6.7) -- (17.4,6.5)};
%	lower diagonals
\draw[cl15]{(1.0,0.4) -- (3.0,4.0)};
\draw[cl15]{(6.4,0.0) -- (9.0,3.0)};
\draw[cl15]{(14.4,-1.0) -- (18.0,2.2)};
%	centr./bim.
\node at (2.0,2.2) {\Large\textbullet};
\node at (7.7,1.5) {\Large\textbullet};
\node at (16.2,0.6) {\Large\textbullet};
\draw[cb15]{(0.5,10.7) -- (2.0,2.2)};
\draw[cb15]{(2.0,7.7) -- (0.5,5.2)};
%
%%%	%%%	%%%	%%%
\node at (4.475,5.775) {\tiny\textbullet};
\node at (4.975,6.075) {$x_{\alpha}$};
%%%	%%%	%%%	%%%
%
\draw[cb15]{(2.0,2.2) -- (7.7,1.5)};
\draw[cb15]{(6.0,3.5) -- (3.7,0.2)};
%
%%%	%%%	%%%	%%%
\node at (11.75,4.95) {\tiny\textbullet};
\node at (12.25,5.25) {$x_{\beta}$};
%%%	%%%	%%%	%%%
%
\draw[cb15]{(7.7,1.5) -- (16.2,0.6)};
\draw[cb15]{(13.5,2.6) -- (10.4,-0.5)};
%	upper diagonals
\draw[cl15]{(0.0,10.0) -- (1.0,11.4)};
\draw[cl15]{(7.0,7.0) -- (8.4,10.4)};
\draw[cl15]{(14.0,7.2) -- (16.8,10.8)};
%	centr./bim.
\node at (0.5,10.7) {\Large\textbullet};
\node at (7.7,8.7) {\Large\textbullet};
\node at (15.4,9.0) {\Large\textbullet};
\draw[cb15]{(0.5,10.7) -- (7.7,8.7)};
\draw[cb15]{(4.7,10.9) -- (3.5,8.5)};
\draw[cb15]{(7.7,8.7) -- (7.7,1.5)};
\draw[cb15]{(8.7,6.7) -- (6.7,3.5)};
\draw[cb15]{(7.7,8.7) -- (15.4,9.0)};
\draw[cb15]{(12.6,10.6) -- (10.5,7.1)};
\draw[cb15]{(15.4,9.0) -- (16.2,0.6)};
\draw[cb15]{(17.4,6.5) -- (14.2,3.1)};
%	primary quadrilateral
\draw[cl25]{(0.0,10.0) -- (7.0,7.0)};
\draw[cl25]{(1.0,0.4) -- (6.4,0.0)};
\draw[cl25]{(0.0,10.0) -- (1.0,0.4)};
\draw[cl25]{(7.0,7.0) -- (6.4,0.0)};
\draw[cl25]{(7.0,7.0) -- (14.0,7.2)};
\draw[cl25]{(6.4,0.0) -- (14.4,-1.0)};
\draw[cl25]{(14.0,7.2) -- (14.4,-1.0)};
%	centr./bim.
\node at (0.5,5.2) {\huge\textbullet};
\node at (6.7,3.5) {\huge\textbullet};
\node at (14.2,3.1) {\huge\textbullet};
\node at (3.5,8.5) {\huge\textbullet};
\node at (3.7,0.2) {\huge\textbullet};
\node at (10.5,7.1) {\huge\textbullet};
\node at (10.4,-0.5) {\huge\textbullet};
\draw[cb25]{(3.5,8.5) -- (3.7,0.2)};
\draw[cb25]{(0.5,5.2) -- (6.7,3.5)};
\draw[cb25]{(10.5,7.1) -- (10.4,-0.5)};
\draw[cb25]{(6.7,3.5) -- (14.2,3.1)};
\end{tikzpicture}
\end{center}
\vspace{-10pt}%
\caption{Cells $C^{\textsc{q\@h}}_{\alpha}$ and $C^{\textsc{q\@h}}_{\beta}$ (gray), corresponding elementary bimedian graphs ($B$-cells) $B_{\alpha}$ and $B_{\beta}$ (black),\\
with their related centroids $x_{\alpha}$ and $x_{\beta}$, respectively}
\label{B_graph}
\end{figure}
%
%%%	FIGURE	%%%
%
Each face is a location of a pair of linear and gauge-invariant holonomies along bimedians that formally do not intersect at the face's centroid (positioned at the intersection of these bimedians). This apparently contradictory structure is sketched in FIG~\ref{B_graph}, where the holonomies' intersections are labeled by black dots. This structure is a consequence of the simplification of pairs of opposite links related to each flat face. These pairs of links are replaced by the corresponding bimedians (equidistant from both links). Hence, during quantization, a pair of Hilbert spaces can be assigned to bimedians of each face as a tensor product of these spaces (holonomies formally do not intersect at faces). Diffeomorphism transformations through faces are well-defined without using a metric tensor. All the non-trivial transformations occur only at edges, where gauge invariance has to be implemented by solving the first-class constraints corresponding to SU$(2)$ and to spatial diffeomorphism transformations. This procedure, based directly on the methods known from CLQG \cite{Ashtekar:1995zh,Ashtekar:2004eh,Zapata:1997db,Zapata:1997da,Fleischhack:2004jc,Lewandowski:2005jk,Thiemann:2007zz}, will be described in \cite{Bilski:2021_RCT_III}.

Thus, in the transition from the mosaic manifold $\Sigma_t^{\hash^{\@\textsc{qh}}}$ into a graph, faces are replaced by their corresponding pairs of bimedians (with non-intersecting holonomies). Consequently, the cell's edges correspond to nodes, where the projectors into the equivalence classes of gauge transformations will be located. The newly defined elementary structures $B_{\alpha}$, sketched in FIG~\ref{B_graph}, are going to be called elementary bimedian graphs, elementary bimedian cells, or simply $B$-cells\footnote{The abbreviation $B$-cell can be explained as a bimedian cell. If one prefers to use the graphical connotation regarding $B^{\textsc{q\@h}}_{\alpha}$ and $B^{\textsc{q\@h}}_{\beta}$ in FIG~\ref{B_graph}, the author also suggests the names ``box cell'' or ``box region''. Consequently, after the quantization, $B$-states will be known as ``bimedian sates'' or ``box sates''.}. Each bimedian cell $B_{\alpha}$ is uniquely related to a cell $C^{\textsc{q\@h}}_{\alpha}$ by a reversal map. The entire graph is going to be denoted by $\Gamma^{\textsc{b}}$ and called the bimedian graph (or the ``box graph''). It should be also emphasized that the whole map between $\Gamma^{\textsc{b}}$ and the quadrilaterally hexahedral graph $\Gamma^{\textsc{q\@h}}$ is analogously construable, and it is also a unique reversal relation.

%
%%%	FIGURE	%%%
%
\begin{figure}[h]
\vspace{5pt}%
\begin{center}
\begin{tikzpicture}[scale=0.75]
\draw[cb15]{(7.7,8.7) -- (7.7,1.5)};
\node at (7.7,8.7) {\large\textbullet};
\node at (7.7,1.5) {\large\textbullet};
%	secondary quadrilateral
\draw[cl05]{(1.0,11.4) -- (8.4,10.4)};
\draw[cl05]{(3.0,4.0) -- (9.0,3.0)};
\draw[cl05]{(1.0,11.4) -- (3.0,4.0)};
\draw[cl05]{(8.4,10.4) -- (9.0,3.0)};
%	centr./bim.
%\draw[cg05]{(4.7,10.9) -- (6.0,3.5)};
%\node at (4.9,11.1) {$b_{\textsc{iii}}(F^{\textsc{i}}_{\@\alpha})$};
\draw[cg05]{(1.2,10.66) -- (8.46,9.66)};	%%%	%%%	%%%	%%%
\draw[cg05]{(1.4,9.92) -- (8.52,8.92)};	%%%	%%%	%%%	%%%
\draw[cg05]{(1.6,9.18) -- (8.58,8.18)};	%%%	%%%	%%%	%%%
\draw[cg05]{(1.8,8.44) -- (8.64,7.44)};	%%%	%%%	%%%	%%%
\draw[cg05]{(2.0,7.7) -- (8.7,6.7)};		%%%	%%%	%%%	%%%
\draw[cg05]{(2.2,6.96) -- (8.76,5.96)};	%%%	%%%	%%%	%%%
\draw[cg05]{(2.4,6.22) -- (8.82,5.22)};	%%%	%%%	%%%	%%%
\draw[cg05]{(2.6,5.48) -- (8.88,4.48)};	%%%	%%%	%%%	%%%
\draw[cg05]{(2.8,4.74) -- (8.94,3.74)};	%%%	%%%	%%%	%%%
%	lower diagonals
\draw[cl15]{(1.0,0.4) -- (3.0,4.0)};
\draw[cl15]{(6.4,0.0) -- (9.0,3.0)};
%	centr./bim.
%\draw[cg15]{(0.5,10.7) -- (2.0,2.2)};
%\node at (-0.4,10.5) {$b_{\textsc{iii}}(F^{\textsc{ii}}_{\@\alpha})$};
\draw[cg15]{(1.2,10.66) -- (0.1,9.04)};	%%%	%%%	%%%	%%%
\draw[cg15]{(1.4,9.92) -- (0.2,8.08)};		%%%	%%%	%%%	%%%
\draw[cg15]{(1.6,9.18) -- (0.3,7.12)};		%%%	%%%	%%%	%%%
\draw[cg15]{(1.8,8.44) -- (0.4,6.16)};		%%%	%%%	%%%	%%%
\draw[cg15]{(2.0,7.7) -- (0.5,5.2)};		%%%	%%%	%%%	%%%
\draw[cg15]{(2.2,6.96) -- (0.6,4.24)};		%%%	%%%	%%%	%%%
\draw[cg15]{(2.4,6.22) -- (0.7,3.28)};		%%%	%%%	%%%	%%%
\draw[cg15]{(2.6,5.48) -- (0.8,2.32)};		%%%	%%%	%%%	%%%
\draw[cg15]{(2.8,4.74) -- (0.9,1.36)};		%%%	%%%	%%%	%%%
%%%	%%%	%%%	%%%
\node at (4.475,5.775) {\scriptsize\textbullet};
\node at (4.8,6.0) {$x_{\alpha}$};
%%%	%%%	%%%	%%%
%\draw[cg15]{(2.0,2.2) -- (7.7,1.5)};
%\draw[cg15]{(6.0,3.5) -- (3.7,0.2)};
%	upper diagonals
\draw[cl15]{(0.0,10.0) -- (1.0,11.4)};
\draw[cl15]{(7.0,7.0) -- (8.4,10.4)};
%	centr./bim.
%\draw[cg15]{(0.5,10.7) -- (7.7,8.7)};
%\draw[cg15]{(4.7,10.9) -- (3.5,8.5)};
\draw[cg15]{(8.46,9.66) -- (6.94,6.3)};		%%%	%%%	%%%	%%%
\draw[cg15]{(8.52,8.92) -- (6.88,5.6)};		%%%	%%%	%%%	%%%
\draw[cg15]{(8.58,8.18) -- (6.82,4.9)};		%%%	%%%	%%%	%%%
\draw[cg15]{(8.64,7.44) -- (6.76,4.2)};		%%%	%%%	%%%	%%%
\draw[cg15]{(8.7,6.7) -- (6.7,3.5)};			%%%	%%%	%%%	%%%
\draw[cg15]{(8.76,5.96) -- (6.64,2.8)};		%%%	%%%	%%%	%%%
\draw[cg15]{(8.82,5.22) -- (6.58,2.1)};		%%%	%%%	%%%	%%%
\draw[cg15]{(8.88,4.48) -- (6.52,1.4)};		%%%	%%%	%%%	%%%
\draw[cg15]{(8.94,3.74) -- (6.46,0.7)};		%%%	%%%	%%%	%%%
%	primary quadrilateral
\draw[cl25]{(0.0,10.0) -- (7.0,7.0)};
\draw[cl25]{(1.0,0.4) -- (6.4,0.0)};
\draw[cl25]{(0.0,10.0) -- (1.0,0.4)};
\draw[cl25]{(7.0,7.0) -- (6.4,0.0)};
%	centr./bim.
%\draw[cg25]{(3.5,8.5) -- (3.7,0.2)};
%\node at (3.7,-0.1) {$b_{\textsc{iii}}(F'^{\textsc{i}}_{\@\alpha})$};
\draw[cg25]{(0.1,9.04) -- (6.94,6.3)};		%%%	%%%	%%%	%%%
\draw[cg25]{(0.2,8.08) -- (6.88,5.6)};		%%%	%%%	%%%	%%%
\draw[cg25]{(0.3,7.12) -- (6.82,4.9)};		%%%	%%%	%%%	%%%
\draw[cg25]{(0.4,6.16) -- (6.76,4.2)};		%%%	%%%	%%%	%%%
\draw[cg25]{(0.5,5.2) -- (6.7,3.5)};		%%%	%%%	%%%	%%%
\draw[cg25]{(0.6,4.24) -- (6.64,2.8)};		%%%	%%%	%%%	%%%
\draw[cg25]{(0.7,3.28) -- (6.58,2.1)};		%%%	%%%	%%%	%%%
\draw[cg25]{(0.8,2.32) -- (6.52,1.4)};		%%%	%%%	%%%	%%%
\draw[cg25]{(0.9,1.36) -- (6.46,0.7)};		%%%	%%%	%%%	%%%
\node at (8.56,1.9) {$b_{\,\@\textsc{iii}}(F'^{\textsc{ii}}_{\@\alpha})$};
\end{tikzpicture}
\end{center}
\vspace{-10pt}%
\caption{Slicing of $C^{\textsc{q\@h}}_{\alpha}$ (or $B_{\alpha}$) along direction \textsc{III}}
\label{QH_Slicing}
\end{figure}
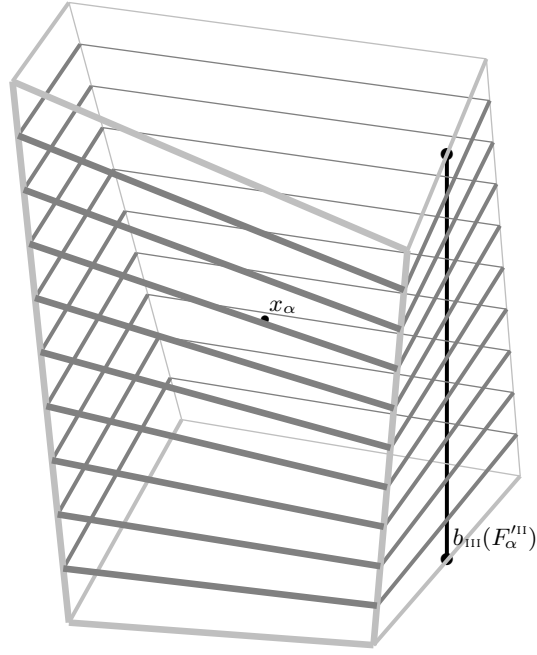
%
%%%	FIGURE	%%%
%
Finally, the Poisson algebra in \eqref{holonomy_flux_regularized} adjusted to bimedian line segments that indicate directions in equation \eqref{W_cell} takes the following form:
\begin{align}
\label{holonomy_flux_bimedian}
\Big\{\Big(\bar{h}_{\,\@b_{\textsc{iii}\@}(\@F'^{\textsc{ii}}_{\!\alpha}\@)}^{\mathstrut}
\!-\bar{h}_{\,\@b_{\textsc{iii}\@}(\@F'^{\textsc{ii}}_{\!\alpha}\@)}^{-1}\Big),
\bar{G}_i^{\,\@b_{\textsc{iii}\@}(\@F'^{\textsc{ii}}_{\!\alpha}\@)}\@\Big\}
=-\frac{\gamma\kappa}{4}
\Bigg(\!
\tau^i\Big(
\bar{h}_{\,\@b_{\textsc{iii}\@}(\@F'^{\textsc{ii}}_{\!\alpha}\@)}^{\mathstrut}\!-\bar{h}_{\,\@b_{\textsc{iii}\@}(\@F'^{\textsc{ii}}_{\!\alpha}\@)}^{-1}
\Big)
+\Big(
\bar{h}_{\,\@b_{\textsc{iii}\@}(\@F'^{\textsc{ii}}_{\!\alpha}\@)}^{\mathstrut}\!-\bar{h}_{\,\@b_{\textsc{iii}\@}(\@F'^{\textsc{ii}}_{\!\alpha}\@)}^{-1}
\Big)\tau^i
\!\Bigg).
\end{align}
In this example, sketched in FIG.~\ref{QH_Slicing}, the gauge-invariant flux distribution (defined in \eqref{gauge_flux}) became replaced by
\begin{align}
\label{gauge_flux_bimedian}
\bar{G}_i^{\,\@b_{\textsc{iii}\@}(\@F'^{\textsc{ii}}_{\!\alpha}\@)}
=\frac{1}{\mathbb{L}_0\.\varepsilon^{\mathstrut}_{\!\.b_{\textsc{iii}\@}(\@F'^{\textsc{ii}}_{\!\alpha}\@)}\!}
\int_{\!b_{\textsc{iii}\@}(\@F'^{\textsc{ii}}_{\!\alpha}\@)}\!\!\!\!\!\!\!\!\!\!\!\!dw
\.g^{-1}\@\big(b(\!\.w\!\.)\big)
\bar{f}_i\big(\bar{S}^{\textsc{iii}}_{\@(w_{\alpha\@})\@}\big)g\big(b(\!\.w\!\.)\big)\,.
\end{align}
It should be emphasized that the intervals in \eqref{linear_link} were replaced by the infinitesimally short segments $dw$ of the bimedian corresponding to the slices determined as in FIG.~\ref{QH_Slicing}. More precisely, these slices were constructed by decomposing each of four bimedians $b_{\.\textsc{iii}}(F'^{\textsc{ii}}_{\@\alpha})$, $b_{\textsc{iii}}(F^{\textsc{i}}_{\@\alpha})$, $b_{\textsc{iii}}(F^{\textsc{ii}}_{\@\alpha})$, and $b_{\textsc{iii}}(F'^{\textsc{i}}_{\@\alpha})$ into the same number $N$ of equal intervals $dw$. Next, the surfaces $\bar{S}^{\textsc{iii}}_{\@(w_{\alpha\@})\@}$ were defined as the flat surfaces constructed between the endpoints of the corresponding intervals of the four mentioned bimedians. This construction entailed the relation
\begin{align}
\label{gauge_flux_surface}
\bar{S}^{\textsc{iii}}_{\@(w_{\alpha\@})\@}
\@:=\bar{S}^{\,\@b_{\textsc{iii}\@}(\@F^{\textsc{i}}_{\!\alpha}\@)}_{\@(w)}
\@=\bar{S}^{\,\@b_{\textsc{iii}\@}(\@F^{\textsc{ii}}_{\!\alpha}\@)}_{\@(w)}
\@=\bar{S}^{\,\@b_{\textsc{iii}\@}(\@F'^{\textsc{i}}_{\!\alpha}\@)}_{\@(w)}
\@=\bar{S}^{\,\@b_{\textsc{iii}\@}(\@F'^{\textsc{ii}}_{\!\alpha}\@)}_{\@(w)}\,.
\end{align}
It is also worth noting that the result in \eqref{holonomy_flux_bimedian} holds due to the implicit assumption of the finite value of the angle between each surface $\bar{S}^{\textsc{iii}}_{\@(w_{\alpha\@})\@}$ and the bimedian $b_{\.\textsc{iii}}(F'^{\textsc{ii}}_{\@\alpha})$. This finiteness guarantees the outcome of \eqref{delta_surface_integral} concerning the derivation of a Gaussian integral\footnote{For details, see \cite{Thiemann:2007zz}. The procedure in the cited textbook is rigorously described regarding an analytical link intersecting a surface. Thus, it instantly holds for a linear link.}. The infinitesimal or zero values of the angle are excluded by the convexity of cells.

A few comments are needed at the end. The formula in \eqref{holonomy_flux_bimedian} is written concerning a single bimedian. The expression in \eqref{W_cell} contains twelve of these objects. Therefore, the Poisson brackets in \eqref{Poisson_AE} have to be accordingly specified regarding the lattice structure. This specification is formulated in two steps. First, one uses the separability of the mosaic manifold $\Sigma_t^{\hash^{\@\textsc{qh}}}$. Then, one adjusts the locally constant spatial coordinates to the bimedians of flat cell's faces. In the first step, one obtains the Poisson structure regarding the graph $\Gamma^{\textsc{q\@h}}$,
\begin{align}
\label{Poisson_mosaic}
\!\{X,Y\}_{\!_{\@\bar{A}\@,\bar{E}}\!}:=\frac{\gamma\kappa}{2}\!\int_{\!\Sigma_t^{\hash^{\@\textsc{qh}}}}\!\!\!\!\!\!\!\!\!d^3x\bigg(\!
\frac{\delta X}{\delta\bar{E}^a_i\!\.(\!\.x\!\.)\!}\.\frac{\delta Y}{\delta\bar{A}_a^i\!\.(\!\.x\!\.)\!}
-\frac{\delta X}{\delta\bar{A}_a^i\!\.(\!\.x\!\.)\!}\.\frac{\delta Y}{\delta\bar{E}^a_i\!\.(\!\.x\!\.)\!}\.
\!\bigg)\@
=\frac{\gamma\kappa}{2}\@\sum_{\alpha}\!\int_{\!C^{\textsc{q\@h}}_{\alpha}}\!\!\!\!\!\!d^3x
\bigg(\!
\frac{\delta X}{\delta\bar{E}^a_i\!\.(\@B_{\@\alpha}\@)\!}\.\frac{\delta Y}{\delta\bar{A}_a^i\!\.(\@B_{\@\alpha}\@)\!}
-\frac{\delta X}{\delta\bar{A}_a^i\!\.(\@B_{\@\alpha}\@)\!}\.\frac{\delta Y}{\delta\bar{E}^a_i\!\.(\@B_{\@\alpha}\@)\!}\.
\!\bigg),\!
\end{align}
where the summation runs over elementary cells. Precise implementation of the second step requires analysis of the symplectic form.

The splitting of the integral in \eqref{Poisson_mosaic} into elementary cells allows one to restrict the analysis to a single cell. The corresponding symplectic form of the gravitational canonical variables, located regarding the cell's linear edges, reads
\begin{align}
\begin{split}
\label{symplectic_mosaic}
\!\Omega_{\!_{\bar{A}\@,\bar{E}}\!}=&-\@\frac{2}{\gamma\kappa}\!\int_{\!C^{\textsc{q\@h}}_{\alpha}}\!\!\!\!\!\!d^3x
\Big(\@
\delta_1\bar{A}_a^i\!\.(\!\.x\!\.)\.\delta_2\bar{E}^a_i\!\.(\!\.x\!\.)
-\delta_2\bar{A}_a^i\!\.(\!\.x\!\.)\.\delta_1\bar{E}^a_i\!\.(\!\.x\!\.)
\@\Big)
=-\frac{2\.\mathbb{L}_0^3}{\gamma\kappa}\varepsilon_{\alpha}^3
\Big(\@
\delta_1\bar{A}_a^i\!\.(\@B_{\@\alpha}\@)\.\delta_2\bar{E}^a_i\!\.(\@B_{\@\alpha}\@)
-\delta_2\bar{A}_a^i\!\.(\@B_{\@\alpha}\@)\.\delta_1\bar{E}^a_i\!\.(\@B_{\@\alpha}\@)
\@\Big)\!
\\
\equiv&-\@\frac{\mathbb{L}_0}{2\gamma\kappa}\sum_{b_{\@(\@a\@)}\!\in\@B_{\@\alpha}\!}
\!\!\!\varepsilon_{\alpha}^{b_{\@(\@a\@)}}\@
\Big(\@
\delta_1\bar{A}_{b_{\@(\@a\@)}}^i\!(\@B_{\@\alpha}\@)\.\delta_2\bar{G}^a_i\!\.(\@B_{\@\alpha}\@)
-\delta_2\bar{A}_{b_{\@(\@a\@)}}^i\!(\@B_{\@\alpha}\@)\.\delta_1\bar{G}^a_i\!\.(\@B_{\@\alpha}\@)
\@\Big)
=
\delta_1\bar{C}_a^i\!\.(\@B_{\@\alpha}\@)\.\delta_2\bar{P}_a^i\!\.(\@B_{\@\alpha}\@)
-\delta_2\bar{C}_a^i\!\.(\@B_{\@\alpha}\@)\.\delta_1\bar{P}_a^i\!\.(\@B_{\@\alpha}\@)\,.
\end{split}
\end{align}
Here, the transition from the first line is realized by using a unique relation between elementary cells and $B$-cells. This relation is illustrated in FIG~\ref{B_graph}. The summation in the second line runs over fours of bimedians $b_a$, and the mean of these bimedians forms the line segment indicating the direction $x^a$. It is also worth noting that the group elements $g$ and $g^{-1}$ in $\bar{G}^a_i$ (see \eqref{gauge_flux_bimedian} or the definition in \eqref{gauge_flux}) vanish due to the contraction of the internal indices, each representing a different orientation of $g$ and $g^{-1}$ in the internal space. More precisely, a triple of the orthogonal SU$(2)$ elements at any link's starting point is proportional to the unit element. The constant of proportionality is then canceled by the inverse constant from the reciprocal triple of these elements at the link's endpoint. Finally, the pair of the new variables corresponding to the graph $\Gamma^{\textsc{b}}$, introduced in the last step in \eqref{symplectic_mosaic}, is defined as follows:
\begin{align}
\label{symplectic_C}
\bar{C}^i_a\!\.(\@B_{\@\alpha}\@):=&\;
\frac{\mathbb{L}_0\@}{4}\!\!\sum_{b_{\@(\@a\@)}\!\in\@B_{\@\alpha}\!\!}
\!\!\varepsilon_{\alpha}^{b_{\@(\@a\@)}}\@\bar{A}_{b_{\@a}}^i\!(\@B_{\@\alpha}\@)
=
\frac{1}{8}\!\!\sum_{b_{\@(\@a\@)}\!\in\@B_{\@\alpha}\!\!}
\!\!\Big(\bar{h}_{b_{\@a}}^{\mathstrut}\pos{v}-\bar{h}_{b_{\@a}}^{-1}\pos{v}\Big)\,,
\\
\label{symplectic_P}
\bar{P}_i^a\!\.(\@B_{\@\alpha}\@):=&
-\frac{2}{\gamma\kappa}\bar{G}^a_i\!\.(\@B_{\@\alpha}\@)
=-\frac{1}{\gamma\kappa}\Big(\@
\bar{f}_i\big(\bar{F}^{p}_{\@\alpha}\big)+\bar{f}_i\big(\bar{F}'^{p}_{\@\alpha}\big)
\!\Big)\,.
\end{align}

The description of the symplectic structure complements the specification of the holonomy-flux algebra adjusted to the bimedian graph.

	%%%%%%%%%	%%%%%%%%%	%%%%%%%%%	%%%%%%%%%
\subsection{Graph representation of the Lorentz connection's rapidity}\label{Sec_Regularization_/_rapidity}

\noindent
In the complete analysis concerning the construction of the lattice representation of gravitational kinematics, one more element has to be specified. As mentioned in Sec.~\ref{Sec_Regularization_/_Gauge_bosons}, the Lorentz connection decomposes into the axis-angle vector and rapidity. The Ashtekar-Barbero connection is constructed as a sum of both elements --- see \eqref{Ashtekar_connection}. However, the total Hamiltonian of the gravitational field is not formulated directly in terms of the $A_a$ and $E^a$ fields. Both gauge-generating constraints contributing to the Hamiltonian depend only on these fields, but the scalar constraint is also a functional of the rapidity
\begin{align}
\label{rapidity}
K^i_a:=\Gamma^i_{\ 0a}\,.
\end{align}
Therefore, a method of expressing this quantity as a functional of canonical fields is needed.

As demonstrated in \cite{Bojowald:2007nu}, the needed method, introduced by Thiemann \cite{Thiemann:1996ay,Thiemann:1996aw,Thiemann:2007zz} in the absence of fermions\footnote{In the absence of the torsion tensor, the rapidity in \eqref{rapidity} coincide with the dreiben-contracted extrinsic curvature.}, holds also in the presence of the fermionic degrees of freedom. In this construction, the following identity is applied,
\begin{align}
\label{curvature_trick}
K^i_a=\frac{\delta\mathbf{K}}{\delta E^a_i}=-\frac{2}{\gamma\kappa}\big\{ A^i_a,\mathbf{K}\big\}\,,
\end{align}
where $\mathbf{K}=\int\!d^3xK^i_aE^a_i\.$. This integrated densitized trace of the rapidity is then expressible by the Poisson brackets of the functionals of only canonical variables, namely,
\begin{align}
\label{curvature_Hamiltonian}
\mathbf{K}
=\frac{1}{\gamma^2\kappa}
\bigg\{
\!\int\!\!d^3x|E|^{-\frac{1}{2}}\epsilon^{ijk}F^i_{ab}E_j^aE_k^b
,\mathbf{V}
\bigg\}.
\end{align}

Next, by applying equation \eqref{curvature_Hamiltonian} into \eqref{curvature_trick} and postulating the distribution of the rapidity along a linear link, one finds
\begin{align}
\begin{split}
\label{curvature_linear}
\bar{K}^i_a\pos{v}
=&\;\frac{4}{\gamma^3\kappa^2}\Bigg\{\!
\.\text{tr}\big(\bar{A}_a\pos{v}\tau^i\big),
\bigg\{\!\sum_{\alpha}\!\int_{\!C^{\textsc{q\@h}}_{\alpha}}\!\!\!\!\!\!d^3x
|\bar{E}|^{-\frac{1}{2}}\epsilon^{jkl}\bar{F}^j_{bc}\bar{E}_k^b\bar{E}_l^c,\mathbf{V}\@\big(\@C^{\textsc{q\@h}}_{\alpha}\@\big)\bigg\}
\!\Bigg\}
\\
=&\;\frac{4\,\text{sgn}\big(\text{det}(E^b_j)\big)\!}{\gamma^4\kappa^3\.\mathbb{L}_0\.\varepsilon_{(\!\.a\!\.)}\pos{v}}
\Bigg\{
\text{tr}\bigg(\!\Big(\@\bar{h}_{a}^{\mathstrut}\pos{v}-\bar{h}_{a}^{-1}\pos{v}\@\Big)\tau^i\!\bigg),
\\
&\;\bigg\{\!
\sum_{\alpha}\!\frac{1}{\varepsilon_{\alpha}^3\.\mathbb{L}_0^3\!}\int_{\!C^{\textsc{q\@h}}_{\alpha}}\!\!\!\!\!\!d^3x
\.\epsilon^{cde}_{\scriptscriptstyle+}
\.\text{tr}\bigg(\!
\Big(\@\bar{h}_{de}\pos{\!\.v_{\@\alpha\@}}\!-\@\bar{h}_{de}^{-1}\pos{\!\.v_{\@\alpha\@}}\@\Big)
\Big\{\@
\Big(\@\bar{h}_{c}^{\mathstrut}\pos{\!\.v_{\@\alpha\@}}\!-\@\bar{h}_{c}^{-1}\pos{\!\.v_{\@\alpha\@}}\@\Big)
,\@\mathbf{V}\@\big(\@C^{\textsc{q\@h}}_{\alpha}\@\big)
\@\Big\}
\@\@\bigg)\@
,\@\mathbf{V}\@\big(\@C^{\textsc{q\@h}}_{\alpha}\@\big)
\!\bigg\}
\!\Bigg\}
\@+\mathcal{O}\big(\varepsilon^3\big)
\\
=&\;\frac{4\,\text{sgn}\big(\text{det}(E^b_j)\big)\!}{\gamma^4\kappa^3\.\mathbb{L}_0\.\varepsilon_{(\!\.a\!\.)}\pos{v}}
\Bigg\{
\text{tr}\bigg(\!\Big(\@\bar{h}_{a}^{\mathstrut}\pos{v}-\bar{h}_{a}^{-1}\pos{v}\@\Big)\tau^i\!\bigg),
\\
&\;\bigg\{\!\sum_{\alpha}\!
\epsilon^{pqr}_{\scriptscriptstyle+}
\.\text{tr}\bigg(
\Big(\@
\bar{h}_{qr}\big(\!F^{(\@p\@)}_{\@\alpha}\@\big)\!+\@\bar{h}_{qr}\big(\!F'^{(\@p\@)}_{\@\alpha}\@\big)
\!-\@\bar{h}_{qr}^{-1}\big(\!F^{(\@p\@)}_{\@\alpha}\@\big)\!-\@\bar{h}_{qr}^{-1}\big(\!F'^{(\@p\@)}_{\@\alpha}\@\big)
\@\Big)
\Big\{
\bar{C}^i_p\!\.(\@B_{\@\alpha}\@),
\mathbf{V}\@\big(\@C^{\textsc{q\@h}}_{\alpha}\@\big)\@\Big\}
\bigg)
,\mathbf{V}\@\big(\@C^{\textsc{q\@h}}_{\alpha}\@\big)
\!\bigg\}
\!\Bigg\}
\@+\mathcal{O}\big(\varepsilon^3\big)\,.
\end{split}
\end{align}
In the first step, the gravitational momenta have been regularized by using the identity in \eqref{trick_gravity_TT}. Then, imposing relations \eqref{holonomy_regularized}, \eqref{loop_regularized}, and \eqref{curvature_distribution_linear}, the gravitational connection and its curvature have been replaced by holonomies. In the second step, the variables defined along the spatial directions, which are oriented along piecewise linear curves, have been adjusted to the structure of the bimedian graph --- compare \eqref{symplectic_mosaic}. It is worth mentioning that the corresponding adjustment of the difference $\bar{h}_{a}^{\phantom{1}}\pos{v}-\bar{h}_{a}^{-1}\pos{v}$ requires an additional definition of the $B$-cell smearing of $\bar{K}^i_a\pos{v}$.

The natural method of smearing the rapidity is the application of the same procedure that has been used for the gravitational connection smearing in \eqref{bimedian_connection}. Thus, one should define the bimedian adjustment of the locally constant spatial direction by the averaging procedure analogous to \eqref{symplectic_C}. The resulting bimedian-smeared rapidity reads
\begin{align}
\label{symplectic_R}
\bar{R}^i_a\!\.(\@B_{\@\alpha}\@):=&\;
\frac{\mathbb{L}_0\@}{4}\!\!\sum_{b_{\@(\@a\@)}\!\in\@B_{\@\alpha}\!\!}
\!\!\varepsilon_{\alpha}^{b_{\@(\@a\@)}}\@\bar{K}_{b_{\@a}}^i\!(\@B_{\@\alpha}\@)\,.
\end{align}
This expression supplements the canonical pair smearing in \eqref{symplectic_C} and \eqref{symplectic_P}. It completes the set of tools that determines the smeared form of the gravitational Hamiltonian on the $\Gamma^{\textsc{b}}$ graph.

This section closes the analysis regarding the construction and inspection of the lattice regularization of gravitational degrees of freedom. The obtained procedure is as precise as standard formulations of classical theories (mechanics, electrodynamics) and QFTs for gauge fields. In these formulations, the conventionally assumed approximations of group symmetries by their representations neglect cubic and higher powers of distance-related expansion parameters. Therefore, the lattice gravity framework introduced in this article is a {(i)} sufficiently accurate {(ii)} nonperturbative {(iii)} canonically quantizable gauge theory that {(iv\!\.)} preserves the general postulate of relativity. Up to the author's knowledge, this is the only model of gravitation that satisfies all of these restrictions\footnote{The standard construction of the classical Hamiltonian corresponding to CLQG \cite{Ashtekar:2004eh,Thiemann:2007zz} fulfills conditions {(ii)} and {(iv\!\.)}. The most popular formulations (for instance, the dressed metric approach \cite{Ashtekar:2011ni}) of the related effective cosmological model, called loop quantum cosmology \cite{Ashtekar:2003hd,Bojowald:2008zzb,Ashtekar:2009vc,Ashtekar:2011ni}, do not satisfy any of the four listed restrictions, \textit{cf.} \cite{Bilski:2019tji}.}.

	%%%%%%%%%	%%%%%%%%%	%%%%%%%%%	%%%%%%%%%

	%%%%%%%%%	%%%%%%%%%	%%%%%%%%%	%%%%%%%%%
\section{Summary}\label{Sec_Summary}

\noindent
This manuscript has been written to demonstrate that the lattice smearing of the Ashtekar variables can be defined with the quadratic precision in the expansion parameter. Moreover, the same smearing holds for the extension of these variables, which applies to coupling with the fermionic matter. The introduced method determines the classical lattice theory of gravity, expressed in terms of holonomies of gravitational connections and fluxes of densitized dreibeins. This theory is generally relativistic, i.e. it has SE and BI, but it is not formulated in GCA. Instead, the Hamiltonian-Dirac formalism \cite{Hamilton:1834,Hamilton:1835,Dirac:1950pj} determines TGA.

The lattice theory of gravity can be then canonically quantized by explicitly using the standard Dirac-Heisenberg-DeWitt \cite{Dirac:1925jy,Heisenberg:1929xj,DeWitt:1967yk} procedure for gauge fields. In this procedure, the states space is a fiber sum of cell's states, each defined as a sum of $24$ combinations of tensor products of triples of cylindrical functions. These triples are selected as in expression \eqref{W_cell}, and each cylindrical function, defined in the same manner as in CLQG \cite{Ashtekar:1993wf,Marolf:1994cj,Ashtekar:1994mh,Ashtekar:1994wa,Ashtekar:1995zh}, depends on the connection only through a linear holonomy. Details regarding the construction of this model are going to be presented in the forthcoming article \cite{Bilski:2021_RCT_III}.

All the introduced framework does not concern only gravitational selfinteractions. The gravitational degrees of freedom minimally coupled with any bosonic field, i.e. coupled only by the metric tensor contraction, are smeared in the same manner. Moreover, the analysis regarding fermions \cite{Mercuri:2006um,Bojowald:2007nu} has allowed identifying the modification of the gravitational connection that removes its presence from the Dirac action coupled with the gravitational field. This modification does not change the symplectic form concerning the canonical gravitational variables. Therefore, after the generalization of the connection to a torsion-full object \cite{Hehl:1976kj,Hehl:1994ue}, the lattice smearing of the modified gravitational variables establishes the minimal coupling between gravity and all the matter fields in the Standard Model. Finally, all these fields can also be smeared all over the same lattice by selecting their continuous Yang-Mills representation and using then the same smearing procedure, which has been formulated for the gravitational variables. The resulting lattice theory of fundamental fields' interactions will approximate the classical continues field theory as precisely, as any gauge group representation approximates its corresponding Lie group in a standard canonical formulation of QFT with internal degrees of freedom.

The analysis regarding all fundamental fields, which are possibly minimally coupled one to another, requires a precise introduction of the whole fundamental methodology of physics concerning the gauge field's representation of the dynamics of nature. This philosophical introduction is in preparation \cite{Bilski:2021_RCT_I}. In general, it demonstrates how to describe all fundamental interactions by using the same mathematical object --- a gauge field and the same physical method --- the Dirac-Heisenberg-DeWitt quantization of a lattice gauge theory. The SE of this framework is established in the Einsteinian procedure, namely, by the minimal coupling with the dynamical metric tensor. Its BI is determined by the diffeomorphisms independent lattice' structure, where this independence decouples from the gravitational dynamics after solving first-class constraints. Moreover, by defining a quantum states space over the same lattice' structure, BI will be preserved after the quantization and then after the derivation of the (semi)classical limit of the so-constructed quantum theory.

This article also describes the improvement of the lattice regularization in CLQG, which leads to a reliable model. As a result, all constructional approximations reach the precision of the ones concerning gauge invariance. The postulated piecewise linearity of a lattice is not mandatory, and the theory can be founded over the piecewise analytical lattice (exactly like in CLQG \cite{Thiemann:2007zz}). However, the improvement of the approximation in the smearing procedure in CLQG is mandatory to receive the precision comparable with phenomenologically applicable theories.

	%%%%%%%%%	%%%%%%%%%	%%%%%%%%%	%%%%%%%%%

	%%%%%%%%%	%%%%%%%%%	%%%%%%%%%	%%%%%%%%%
\appendix

	%%%%%%%%%	%%%%%%%%%	%%%%%%%%%	%%%%%%%%%
\section{Short path holonomy expansion}\label{Appendix_holonomy_path}

\noindent
Let the object $h_{\ell}\pos{v}$ represents a reciprocal holonomy along the path $\ell(s)$, where $s\in[0,l]$, and the path is embedded in the spatial manifold as follows, $v:=\ell(0)$ and ${v\@+\@l}:=\ell(l)$. By expanding it at the point $v$ around the short length $l$, one obtains:
\begin{align}
\begin{split}
\label{holonomy_expansion_abstract}
h_{\ell}\pos{v}=&\;\mathcal{P}\exp\!\bigg(\!\int_{\@0}^{l}\!\!\!ds\.A\!\.\big(\ell(s)\big)\!\bigg)
=\mathds{1}+lA\!\.\big(\ell(0)\big)+\frac{1}{2}l^2\dot{A}\!\.\big(\ell(0)\big)+\frac{1}{2}l^2A^2\!\.\big(\ell(0)\big)+\mathcal{O}(l^3)
\\
=&\;\mathds{1}+lA^i_a\!\.(v)\.\dot{\ell}^a\!\.(v)\.\tau^i
+\frac{1}{2}l^2\partial_aA_b^i\!\.(v)\.\dot{\ell}^a\!\.(v)\.\dot{\ell}^b\!\.(v)\.\tau^i
+\frac{1}{2}l^2A^i_a\!\.(v)\.A_b^j\!\.(v)\.\dot{\ell}^a\!\.(v)\.\dot{\ell}^b\!\.(v)\.\tau^i\tau^j
+\mathcal{O}(l^3)\,.
\end{split}
\end{align}
The basis vector $\dot{\ell}(0)$ is the tangent to the path at point $v$. From the perspective of the path, the whole system in \eqref{holonomy_expansion_abstract} is one-dimensional and the position along the path is labeled by $s$. However, the geometry of this path's embedding can be non-Euclidean in general.

Let $p$ defines the direction of an analytical link $l_{(p)}$ that starts at $v$ and ends at $v\@+\@l_{(p)}$. Let the link's length be determined along its direction via the small parameter $\varepsilon_{(p)}\pos{v}\ll1$ and length's unit be set to $\mathbb{L}_0$. The latter quantity is a globally defined constant. Thus, the parameter $\varepsilon_{(p)}\pos{v}$ determines the link's embedding by setting the position, direction, and length of the path. However, the geometry measured along the path (along the $p$-direction) is set to be Euclidean. The holonomy defined along this path reads
\begin{align}
\begin{split}
\label{holonomy_expansion_path}
h_{(p)}\pos{v}:=h_{l_{(p)}\pos{v}}=&\;\mathds{1}+\mathbb{L}_0\.\varepsilon_{(p)}\pos{v}A_{(p)}\@(v)
+\frac{1}{2}\mathbb{L}_0^2\.\varepsilon_{(p)}^2\pos{v}\.\partial_{(p)}A_{(p)}\@(v)
+\frac{1}{2}\mathbb{L}_0^2\.\varepsilon_{(p)}^2\pos{v}A^2_{(p)}\@(v)
+\mathcal{O}\big(\varepsilon_{(p)}^3\big)\,.
\end{split}
\end{align}
It is worth emphasizing that all the elements have been decomposed along the only spatial direction in the link's one-dimensional system, namely $p$.

	%%%%%%%%%	%%%%%%%%%	%%%%%%%%%	%%%%%%%%%
\section{Functional derivative of a holonomy}\label{Appendix_derivative}

\noindent

The functional derivative of a holonomy regarding the gravitational connection reads
\begin{align}
\label{variation_holonomy}
\frac{\delta h_{l\.\pos{v}}[A]}{\delta A^i_a(w)\@}=\left\{\begin{array}{ll}
\frac{1}{2}\!\int_{\@v}^{v\@+\@l}\!\!ds\.\dot{l}^a\@(s)\.\tau^ih_{l\.\pos{v}}\delta^3\@\big(l(s)-w\big)
\quad&\text{ for }w=v\text{, being the source of } l
\\
\int_{\@v}^{v\@+\@l}\!\!ds\.\dot{l}^a\@(s)\.h_{l\.\pos{v,s}}\tau^ih_{l\.\pos{s,v+l}}\delta^3\@\big(l(s)-w\big)
\quad&\text{ for }w\in(v,v+l)\text{, being inside } l
\\
\frac{1}{2}\!\int_{\@v}^{v\@+\@l}\!\!ds\.\dot{l}^a\@(s)\.h_{l\.\pos{v}}\tau^i\.\delta^3\@\big(l(s)-w\big)
\quad&\text{ for }w=v+l\text{, being the target of } l
\\
0
\quad&\text{ for }\forall_s\,w\neq l(s)\text{, being disjoint from } l\,.
\end{array}\right.
\end{align}
This result, which depends on the position of the point where the derivative is implemented, was first derived in \cite{Lewandowski:1993zq}. A simple description of this derivation is written in \cite{Dona:2010hm}. Here, the second case, i.e. $w\in(v,v+l)$, will be discussed in detail. For a simpler notation, one can assume that the vector field is oriented towards the $z$-direction. One can then calculate the resulting path integral,
\begin{align}
\label{variation_trick_z}
\frac{\delta h_{l\.\pos{v}}[A]}{\delta A^i_z(w)\@}\bigg|_{w\in(v,v+l)}\!\!
=\!\int_{\@v}^{v\@+\@l}\!\!\!\!\!\!ds\.\dot{l}^z\@(s)\.h_{l\.\pos{v,s}}\tau^ih_{l\.\pos{s,v+l}}
\delta\big(l^x\@(s)-w^x\big)\delta\big(l^y\@(s)-w^y\big)\delta\big(l^z\@(s)-w^z\big)\,,
\end{align}
in the following steps.

One should first recognize that the $z$-coordinate of the intersection point $w$ satisfies the relation $w^z\in(v,v+l)$. Next, one can shift the entire system by the vector $(0,0,-w^z)^{\T}$, getting
\begin{align}
\label{variation_trick_shift}
\begin{split}
&\int_{\@v\@-\@w^z}^{v\@+\@l\@-\@w^z}\!\!\!\!\!\!\!ds\.\dot{l}^z\@(s)\.h_{l\.\pos{v-w^z,s}}\tau^ih_{l\.\pos{s,v+l-w^z}}
\delta\big(l^x\@(s)-w^x\big)\delta\big(l^y\@(s)-w^y\big)\delta\big(l^z\@(s)\big)
\\
=&\Bigg(\!\int_{\@v\@-\@w^z}^0\!+\!\int_0^{v\@+\@l\@-\@w^z}\!\Bigg)ds\.\dot{l}^z\@(s)\.h_{l\.\pos{v-w^z,s}}\tau^ih_{l\.\pos{s,v+l-w^z}}
\delta\big(l^x\@(s)-w^x\big)\delta\big(l^y\@(s)-w^y\big)\delta\big(l^z\@(s)\big)\,,
\end{split}
\end{align}
where $s-w^z$ has been changed into a new variable $s$, and the fact that $v<w^z<v+l$, hence $0\in(v-w^z,v+l-w^z)$, has been used. The integrals can be regularized be applying the following identity,
\begin{align}
\label{delta_integral}
\int_0^{\infty}\!\!\!\!\!ds\.f(s)\.\delta(s)=\!\int_{\@-\infty}^0\!\!\!\!\!\!ds\.f(s)\.\delta(s)=\frac{1}{2}f(0)
\,,
\end{align}
which is based on the general property of the Dirac delta function that is symmetric around $0$. Consequently, one obtains a much simpler form of the expression in \eqref{variation_trick_shift}, namely
\begin{align}
\label{variation_trick_regularized}
h_{l\.\pos{v-w^z,0}}\tau^ih_{l\.\pos{0,v+l-w^z}}\delta\big(l^x\@(0)-w^x\big)\delta\big(l^y\@(0)-w^y\big)
=:h_{l\.\pos{v-w^z,0}}\tau^ih_{l\.\pos{0,v+l-w^z}}\delta^2\@\big(l\big.^{_{\perp z}}\@(0)-w\big.^{_{\perp z}}\big)\,.
\end{align}
Finally, by shifting the system back by the vector $(0,0,w^z)^{\T}$, one finds the result of the derivative in \eqref{variation_trick_z}. After rewriting it in the manner independent of any reference frame's orientation, one gets
\begin{align}
\label{variation_trick_inside}
\frac{\delta h_{l\.\pos{v}}[A]}{\delta A^i_a(w)\@}\bigg|_{w\in(v,v+l)}\!\!
=h_{l\.\pos{v,w}}\tau^ih_{l\.\pos{w,v+l}}\delta^2\@\big(l\big.^{_{\perp a}}\@(w)-w\big.^{_{\perp a}}\big)\,.
\end{align}
Analogous derivations provide the following outcomes,
\begin{subequations}
\begin{align}
\label{variation_trick_source}
\frac{\delta h_{l\.\pos{v}}[A]}{\delta A^i_a(w)\@}\bigg|_{w=v}\!\!
=&\;\frac{1}{2}\tau^ih_{l\.\pos{v}}\delta^2\@\big(l\big.^{_{\perp a}}\@(w)-w\big.^{_{\perp a}}\big)\,,
\\
\label{variation_trick_target}
\frac{\delta h_{l\.\pos{v}}[A]}{\delta A^i_a(w)\@}\bigg|_{w=v+l}\!\!
=&\;\frac{1}{2}h_{l\.\pos{v}}\tau^i\delta^2\@\big(l\big.^{_{\perp a}}\@(w)-w\big.^{_{\perp a}}\big)\,.
\end{align}
\end{subequations}

	%%%%%%%%%	%%%%%%%%%	%%%%%%%%%	%%%%%%%%%
\section{Elements of the gauge invariant algebra derivation on a link}\label{Appendix_continuous_flux}

\noindent
To derive the last element on the right-hand side of equation \eqref{holonomy_gauge_flux} it is convenient to introduce the following labeling of intervals, $\Delta l\.\pos{v_q}:=l\.\pos{v+(q-1)l/N,v+q\.l/N}=l\.\pos{\bar{v}_q-l/(2N),\bar{v}_q+l/(2N)}$, where $\bar{v}_q:=v+(q-1/2)l/N$ denotes the center of each interval. The summation contributing to this element can be rewritten as follows:
\begin{align}
\label{h,F_linear_density_sum}
\sum_{p=1}^{N-1}\!\prod_{j=1}^3\@h^{(j)}_{l\.\pos{v,v+p\.l/N}}\tau^ih^{(j)}_{l\.\pos{v+p\.l/N,v+l}}
=\sum_{p=1}^{N-1}\!\prod_{j=1}^3\prod_{q=1}^p\,\prod_{r=p+1}^N\!\!h^{(j)}_{\Delta l\.\pos{v_{\@q\@}}}\tau^ih^{(j)}_{\Delta l\.\pos{v_{\@r\@}}}\,.
\end{align}
In the limit $N\to\infty$, the connection smeared along each $\varepsilon^l/N$-short interval $\Delta l\.\pos{v_q}$ is arbitrarily well-approximated by the constant connection $\bar{A}^j_l\!\.(\bar{v}_q\!\.)$, which value equals the value of the continues one located at the interval's center, namely $A^j_l\!\.(\bar{v}_q\!\.)$. The corresponding holonomy reads
\begin{align}
\label{h,F_holonomy_const}
h^{(j)}_{\Delta l\.\pos{v_{\@q\@}}}=\exp\@\Big(\Delta l\.\pos{v_{(q)}}\bar{A}^{(j)}_l\!\.(\bar{v}_{(q)}\!\.)\.\tau^{(j)}\Big)
=\cos\@\bigg(\@\frac{1}{2}\Delta l\.\pos{v_{(q)}}\bar{A}^{(j)}_l\!\.(\bar{v}_{(q)}\!\.)\@\bigg)\mathds{1}
+2\sin\@\bigg(\@\frac{1}{2}\Delta l\.\pos{v_{(q)}}\bar{A}^{(j)}_l\!\.(\bar{v}_{(q)}\!\.)\@\bigg)\tau^{(j)}\,.
\end{align}

To simplify the notation, it is worth introducing an auxiliary variable $\alpha^j_q:=\frac{1}{2}{\Delta l\.\pos{v_{(q)}}}{\bar{A}^j_l\!\.(\bar{v}_{(q)}\!\.)}$. Consequently, the expression in \eqref{h,F_linear_density_sum} takes the form
\begin{align}
\label{h,F_linear_prod}
\begin{split}
&\sum_{p=1}^{N-1}\prod_{q=1}^p\prod_{r=p+1}^N\prod_{j=1}^3\!
\Big[
\cos\@\big(\alpha^{(j)}_q\big)\mathds{1}
+2\sin\@\big(\alpha^{(j)}_q\big)\tau^{(j)}
\Big]\tau^i\Big[
\cos\@\big(\alpha^{(j)}_r\big)\mathds{1}
+2\sin\@\big(\alpha^{(j)}_r\big)\tau^{(j)}
\Big]
\\
=&\sum_{p=1}^{N-1}\prod_{q=1}^p\prod_{r=p+1}^N\prod_{j=1}^3\!
\bigg[
\cos\@\big(\alpha^{(j)}_q\@+\alpha^{(j)}_r\big)\bigg(\@\frac{1}{2}\tau^i\@-2\.\tau^{(j)}\tau^i\tau^{(j)}\@\bigg)
+\cos\@\big(\alpha^{(j)}_q\@-\alpha^{(j)}_r\big)\bigg(\@\frac{1}{2}\tau^i\@+2\.\tau^{(j)}\tau^i\tau^{(j)}\@\bigg)
\\
&+
\sin\@\big(\alpha^{(j)}_q\@+\alpha^{(j)}_r\big)\Big(\tau^i\tau^{(j)}\@+\tau^{(j)}\tau^i\Big)
+\sin\@\big(\alpha^{(j)}_q\@-\alpha^{(j)}_r\big)\Big(\tau^i\tau^{(j)}\@+\tau^{(j)}\tau^i\Big)
\bigg]
\\
=&\;\frac{1}{2}\!\sum_{p=1}^{N-1}\prod_{q=1}^p\prod_{r=p+1}^N\prod_{j=1}^3\!
\bigg[
\tau^i\Big(\!
\cos\@\big(\alpha^{(j)}_q\@+\alpha^{(j)}_r\big)+2\sin\@\big(\alpha^{(j)}_q\@+\alpha^{(j)}_r\big)\tau^{(j)}
\Big)
\\
&+\Big(\!
\cos\@\big(\alpha^{(j)}_q\@+\alpha^{(j)}_r\big)+2\sin\@\big(\alpha^{(j)}_q\@+\alpha^{(j)}_r\big)\tau^{(j)}
\Big)\tau^i
+\tau^i-\cos\@\big(\alpha^{(j)}_q\@+\alpha^{(j)}_r\big)\tau^i
-\sin\@\big(\alpha^i_q\@-\alpha^i_r\big)
\bigg]\,,
\end{split}
\end{align}
where $\sum_{p=1}^{N-1}\@\prod_{q=1}^p\@\prod_{r=p+1}^N\@\cos\@\big(\alpha^{(j)}_q\@-\alpha^{(j)}_r\big)=1$. After performing the multiplications, which enter the trigonometric functions above as summations, this formula undergo further simplifications and leads to the following result:
\begin{align}
\label{h,F_linear_corrections}
\frac{1}{2}\Bigg[
(N-1)\big(\tau^ih_{l\.\pos{v}}\@+h_{l\.\pos{v}}\tau^i\big)
+(N-1)\Bigg(\!
1-\cos\!\bigg(\frac{1}{2}l\.\pos{v}\@\sum_{j=1}^3\@\bar{A}_l^{(j)}\pos{v}\@\bigg)
\!\Bigg)\@\tau^i
-\!\sum_{p=1}^{N-1}\sin\!\Bigg(\sum_{q=1}^p\@\alpha^i_q-\!\!\!\sum_{r=p+1}^N\!\!\!\alpha^i_r\Bigg)
\Bigg]\,.
\end{align}
Finally, the last term in the obtained expression is additionally expandable as follows,
\begin{align}
\label{h,F_linear_sinus}
\sum_{p=1}^{N-1}\sin\!\Bigg(\sum_{q=1}^p\@\alpha^i_q-\!\!\!\sum_{r=p+1}^N\!\!\!\alpha^i_r\Bigg)
=\frac{1}{2}\.\overline{\Delta l}\.\pos{v}\Big(\bar{A}^j_l\!\.(\bar{v}_{1}\!\.)-\bar{A}^j_l\!\.(\bar{v}_{N}\!\.)\Big)
+\mathcal{O}\Big(\@\big(\overline{\Delta l}\.\pos{v}\big)^{\@3}\Big)\,.
\end{align}

\ \\

	%%%%%%%%%	%%%%%%%%%	%%%%%%%%%	%%%%%%%%%

	%%%%%%%%%	%%%%%%%%%	%%%%%%%%%	%%%%%%%%%
%
\acknowledgments{
This work was partially supported by the National Natural Science Foundation of China grants Nos. 11675145 and 11975203.
}
%
	%%%%%%%%%	%%%%%%%%%	%%%%%%%%%	%%%%%%%%%

	%%%%%%%%%	%%%%%%%%%	%%%%%%%%%	%%%%%%%%%

\ \\


\begin{thebibliography}{999}

\newpage

%\cite{Bilski:2021_RCT_I}
\bibitem{Bilski:2021_RCT_I}
J.~Bilski,
``Relativistic classical theory I. General relativity,''
in preparation (2021).

%\cite{Einstein:1916vd}
\bibitem{Einstein:1916vd}
A.~Einstein,
%``The Foundation of the General Theory of Relativity,''
Annalen Phys. \textbf{49}, no.7, 769-822 (1916)
doi:10.1002/andp.200590044.

%\cite{Einstein:1907}
\bibitem{Einstein:1907} 
A.~Einstein,
%``\"Uber das Relativit\"atsprinzip und die aus demselben gezogenen Folgerungen,"
%``On the relativity principle and the conclusions drawn from it,"
Jahrb Radioaktivit\"{a}t Elektronik {\bf 4}, 411-462 (1907).

%\cite{Einstein:1911vc}
\bibitem{Einstein:1911vc}
A.~Einstein,
%``On The influence of gravitation on the propagation of light,''
Annalen Phys. \textbf{35}, 898-908 (1911)
doi:10.1002/andp.200590033.

%\cite{Einstein:1907iag}
\bibitem{Einstein:1907iag}
A.~Einstein,
%``\"Uber die vom Relativit\"atsprinzip geforderte Tr\"agheit der Energie,''
%``\On the inertia of energy required by the relativity principle,''
Annalen Phys. \textbf{23}, no.7, 371-384 (1907)
doi:10.1002/andp.19073280713.

%\cite{Hamilton:1834}
\bibitem{Hamilton:1834}
W.~R.~Hamilton,
%``On a general method in dynamics; by which the study of the motions of all free systems of attracting or repelling points is reduced to the search and differentiation of one central relation, or characteristic function,''
Philosophical transactions of the Royal Society of London \textbf{124}, 247-308 (1834).

%\cite{Hamilton:1835}
\bibitem{Hamilton:1835}
W.~R.~Hamilton,
%``Second essay on a general method in dynamics,''
Philosophical transactions of the Royal Society of London \textbf{125}, 95-144 (1835).

%\cite{Dirac:1950pj}
\bibitem{Dirac:1950pj} 
P.~A.~M.~Dirac,
%``Generalized Hamiltonian dynamics,''
Can.\ J.\ Math.\  {\bf 2}, 129 (1950)
doi:10.4153/CJM-1950-012-1.

%\cite{Caratheodory:1909}
\bibitem{Caratheodory:1909}
C.~Carath\'eodory,
%``Untersuchungen \"uber die Grundlagen der Thermodynamik,''
Mathematische Annalen {\bf 67}, 3, 355-386 (1909).

%\cite{Planck:1926}
\bibitem{Planck:1926}
M.~Planck,
%``\"Uber die Begründung des zweiten Hauptsatzes der Thermodynamik,''
Naturwissenschaften {\bf 14}, 13 249-261 (1926).

%\cite{De_Donder:1930}
\bibitem{De_Donder:1930}
T.~De Donder,
%``Th\'eorie invariantive du calcul des variations,''
Gaulthier-Villars \& Cie., Paris, (1930).

%\cite{Weyl:1935}
\bibitem{Weyl:1935}
H.~Weyl,
%``Geodesic fields in the calculus of variations,''
Ann Math {\bf 36}, 607-629 (1935)
doi:10.2307/1968645.

%\cite{Ivanenko:1984vf}
\bibitem{Ivanenko:1984vf}
D.~Ivanenko and G.~Sardanashvily,
%``The Gauge Treatment of Gravity,''
Phys. Rept. \textbf{94}, 1-45 (1983)
doi:10.1016/0370-1573(83)90046-7.

%\cite{Arnowitt:1959ah}
\bibitem{Arnowitt:1959ah}
R.~L.~Arnowitt, S.~Deser and C.~W.~Misner,
%``Dynamical Structure and Definition of Energy in General Relativity,''
Phys.\ Rev.\  \textbf{116}, 1322-1330 (1959)
doi:10.1103/PhysRev.116.1322.

%\cite{Arnowitt:1960es}
\bibitem{Arnowitt:1960es}
R.~L.~Arnowitt, S.~Deser and C.~W.~Misner,
%``Canonical variables for general relativity,''
Phys. Rev. \textbf{117}, 1595-1602 (1960)
doi:10.1103/PhysRev.117.1595.

%\cite{Arnowitt:1962hi}
\bibitem{Arnowitt:1962hi}
R.~L.~Arnowitt, S.~Deser and C.~W.~Misner,
%``The Dynamics of general relativity,''
Gen. Rel. Grav. \textbf{40}, 1997-2027 (2008)
doi:10.1007/s10714-008-0661-1
[arXiv:gr-qc/0405109 [gr-qc]].

%\cite{Thiemann:1996ay}
\bibitem{Thiemann:1996ay}
T.~Thiemann,
%``Anomaly - free formulation of nonperturbative, four-dimensional Lorentzian quantum gravity,''
Phys. Lett. B \textbf{380}, 257-264 (1996)
doi:10.1016/0370-2693(96)00532-1
[arXiv:gr-qc/9606088 [gr-qc]].

%\cite{Thiemann:1996aw}
\bibitem{Thiemann:1996aw}
T.~Thiemann,
%``Quantum spin dynamics (QSD),''
Class. Quant. Grav. \textbf{15}, 839-873 (1998)
doi:10.1088/0264-9381/15/4/011
[arXiv:gr-qc/9606089 [gr-qc]].

%\cite{Thiemann:1997rt}
\bibitem{Thiemann:1997rt}
T.~Thiemann,
%``QSD 5: Quantum gravity as the natural regulator of matter quantum field theories,''
Class. Quant. Grav. \textbf{15}, 1281-1314 (1998)
doi:10.1088/0264-9381/15/5/012
[arXiv:gr-qc/9705019 [gr-qc]].

%\cite{Ashtekar:2004eh}
\bibitem{Ashtekar:2004eh}
A.~Ashtekar and J.~Lewandowski,
%``Background independent quantum gravity: A Status report,''
Class. Quant. Grav. \textbf{21}, R53 (2004)
doi:10.1088/0264-9381/21/15/R01
[arXiv:gr-qc/0404018 [gr-qc]].

%\cite{Thiemann:2007zz}
\bibitem{Thiemann:2007zz}
T.~Thiemann,
%``Modern canonical quantum general relativity,''
Cambridge, UK: Cambridge Univ. Pr. (2007)
doi:10.1017/CBO9780511755682
[arXiv:gr-qc/0110034 [gr-qc]].

%\cite{Bilski:2020poi}
\bibitem{Bilski:2020poi}
J.~Bilski,
%``Implementation of the holonomy representation of the Ashtekar connection in loop quantum gravity,''
[arXiv:2012.14441 [gr-qc]].

%\cite{Henneaux:1992ig}
\bibitem{Henneaux:1992ig}
M.~Henneaux and C.~Teitelboim,
%``Quantization of gauge systems,''
Princeton, USA: Univ. Pr. (1992),
doi:10.2307/j.ctv10crg0r.

%\cite{Ashtekar:2003hd}
\bibitem{Ashtekar:2003hd}
A.~Ashtekar, M.~Bojowald and J.~Lewandowski,
%``Mathematical structure of loop quantum cosmology,''
Adv. Theor. Math. Phys. \textbf{7}, no.2, 233-268 (2003)
doi:10.4310/ATMP.2003.v7.n2.a2
[arXiv:gr-qc/0304074 [gr-qc]].

%\cite{Bojowald:2008zzb}
\bibitem{Bojowald:2008zzb}
M.~Bojowald,
%``Loop quantum cosmology,''
Living Rev. Rel. \textbf{11}, 4 (2008).

%\cite{Ashtekar:2009vc}
\bibitem{Ashtekar:2009vc}
A.~Ashtekar and E.~Wilson-Ewing,
%``Loop quantum cosmology of Bianchi I models,''
Phys. Rev. D \textbf{79}, 083535 (2009)
doi:10.1103/PhysRevD.79.083535
[arXiv:0903.3397 [gr-qc]].

%\cite{Ashtekar:2011ni}
\bibitem{Ashtekar:2011ni}
A.~Ashtekar and P.~Singh,
%``Loop Quantum Cosmology: A Status Report,''
Class. Quant. Grav. \textbf{28}, 213001 (2011)
doi:10.1088/0264-9381/28/21/213001
[arXiv:1108.0893 [gr-qc]].

%\cite{Bilski:2021fki}
\bibitem{Bilski:2021fki}
J.~Bilski and A.~Wang,
%``Lattice classical cosmology,''
[arXiv:2101.02223 [gr-qc]].

%\cite{Bilski:2019tji}
\bibitem{Bilski:2019tji}
J.~Bilski and A.~Marcian\`o,
%``Critical Insight into the Cosmological Sector of Loop Quantum Gravity,''
Phys. Rev. D \textbf{101}, no.6, 066026 (2020)
doi:10.1103/PhysRevD.101.066026
[arXiv:1905.00001 [gr-qc]].

%\cite{Bilski:2020xfq}
\bibitem{Bilski:2020xfq}
J.~Bilski,
%``Lie algebra of Ashtekar-Barbero connections,''
[arXiv:2012.10465 [gr-qc]].

%\cite{Kibble:1961ba}
\bibitem{Kibble:1961ba}
T.~W.~B.~Kibble,
%``Lorentz invariance and the gravitational field,''
J. Math. Phys. \textbf{2}, 212-221 (1961)
doi:10.1063/1.1703702.

%\cite{Cartan:1922}
\bibitem{Cartan:1922}
E.~Cartan,
%``Sur une g\'en\'eralisation de la notion de courbure de Riemann et les espaces \`a torsion,''
Comptes rendus de l'Acad\'emie des Sciences de Paris \textbf{174}, 593-595 (1922).

%\cite{Cartan:1923zea}
\bibitem{Cartan:1923zea}
E.~Cartan,
%``Sur les vari\'et\'es \`a connexion affine et la th\'eorie de la relativit\'e g\'en\'eralis\'ee. (premi\`ere partie),''
Annales Sci. Ecole Norm. Sup. \textbf{40}, 325-412 (1923).

%\cite{Cartan:1924yea}
\bibitem{Cartan:1924yea}
E.~Cartan,
%``Sur les vari\'et\'es \`a connexion affine et la th\'eorie de la relativit\'e g\'en\'eralis\'ee. (premi\`ere partie) (Suite),''
Annales Sci. Ecole Norm. Sup. \textbf{41}, 1-25 (1924).

%\cite{Cartan:1925}
\bibitem{Cartan:1925}
E.~Cartan,
%``Sur les vari\'et\'es \`a connexion affine et la th\'eorie de la relativit\'e g\'en\'eralis\'ee. (deuxi\`eme partie),''
Annales Sci. Ecole Norm. Sup. \textbf{42}, 17-88 (1925).

%\cite{Hehl:1976kj}
\bibitem{Hehl:1976kj}
F.~W.~Hehl, P.~Von Der Heyde, G.~D.~Kerlick and J.~M.~Nester,
%``General Relativity with Spin and Torsion: Foundations and Prospects,''
Rev. Mod. Phys. \textbf{48}, 393-416 (1976)
doi:10.1103/RevModPhys.48.393.

%\cite{Hehl:1994ue}
\bibitem{Hehl:1994ue}
F.~W.~Hehl, J.~D.~McCrea, E.~W.~Mielke and Y.~Ne'eman,
%``Metric affine gauge theory of gravity: Field equations, Noether identities, world spinors, and breaking of dilation invariance,''
Phys. Rept. \textbf{258}, 1-171 (1995)
doi:10.1016/0370-1573(94)00111-F
[arXiv:gr-qc/9402012 [gr-qc]].

%\cite{Ashtekar:1986yd}
\bibitem{Ashtekar:1986yd}
A.~Ashtekar,
%``New Variables for Classical and Quantum Gravity,''
Phys. Rev. Lett. \textbf{57}, 2244-2247 (1986)
doi:10.1103/PhysRevLett.57.2244.

%\cite{Ashtekar:1987gu}
\bibitem{Ashtekar:1987gu}
A.~Ashtekar,
%``New Hamiltonian Formulation of General Relativity,''
Phys. Rev. D \textbf{36}, 1587-1602 (1987)
doi:10.1103/PhysRevD.36.1587.

%\cite{Barbero:1994ap}
\bibitem{Barbero:1994ap}
J.~F.~Barbero G.,
%``Real Ashtekar variables for Lorentzian signature space times,''
Phys. Rev. D \textbf{51}, 5507-5510 (1995)
doi:10.1103/PhysRevD.51.5507
[arXiv:gr-qc/9410014 [gr-qc]].

%\cite{Mercuri:2006um}
\bibitem{Mercuri:2006um}
S.~Mercuri,
%``Fermions in Ashtekar-Barbero connections formalism for arbitrary values of the Immirzi parameter,''
Phys. Rev. D \textbf{73}, 084016 (2006)
doi:10.1103/PhysRevD.73.084016
[arXiv:gr-qc/0601013 [gr-qc]].

%\cite{Bojowald:2007nu}
\bibitem{Bojowald:2007nu}
M.~Bojowald and R.~Das,
%``Canonical gravity with fermions,''
Phys. Rev. D \textbf{78}, 064009 (2008)
doi:10.1103/PhysRevD.78.064009
[arXiv:0710.5722 [gr-qc]].

%\cite{Bilski:2021ysc}
\bibitem{Bilski:2021ysc}
J.~Bilski,
%``Continuously distributed holonomy-flux algebra,''
[arXiv:2101.05295 [gr-qc]].

%\cite{Bohr:1920}
\bibitem{Bohr:1920}
N.~Bohr,
%``\"Uber die Serienspektra der Elemente,''
Z. Physik {\bf 2}, 423-469 (1920)
doi:10.1007/BF01329978.

%\cite{Wilson:1974sk}
\bibitem{Wilson:1974sk}
K.~G.~Wilson,
%``Confinement of Quarks,''
Phys.\ Rev.\ D {\bf 10}, 2445 (1974)
doi:10.1103/PhysRevD.10.2445.

%\cite{Giles:1981ej}
\bibitem{Giles:1981ej}
R.~Giles,
%``The Reconstruction of Gauge Potentials From Wilson Loops,''
Phys. Rev. D \textbf{24}, 2160 (1981)
doi:10.1103/PhysRevD.24.2160.

%\cite{Mandelstam:1968hz}
\bibitem{Mandelstam:1968hz}
S.~Mandelstam,
%``Feynman rules for electromagnetic and Yang-Mills fields from the gauge independent field theoretic formalism,''
Phys. Rev. \textbf{175}, 1580-1623 (1968)
doi:10.1103/PhysRev.175.1580.

%\cite{Mandelstam:1978ed}
\bibitem{Mandelstam:1978ed}
S.~Mandelstam,
%``Charge - Monopole Duality and the Phases of Nonabelian Gauge Theories,''
Phys. Rev. D \textbf{19}, 2391 (1979)
doi:10.1103/PhysRevD.19.2391.

%\cite{Mandelstam:1962mi}
\bibitem{Mandelstam:1962mi}
S.~Mandelstam,
%``Quantum electrodynamics without potentials,''
Annals Phys. \textbf{19}, 1-24 (1962)
doi:10.1016/0003-4916(62)90232-4.

%\cite{Makeenko:1979pb}
\bibitem{Makeenko:1979pb}
Y.~M.~Makeenko and A.~A.~Migdal,
%``Exact Equation for the Loop Average in Multicolor QCD,''
Phys. Lett. B \textbf{88}, 135 (1979)
[erratum: Phys. Lett. B \textbf{89}, 437 (1980)]
doi:10.1016/0370-2693(79)90131-X.

%\cite{Makeenko:1980vm}
\bibitem{Makeenko:1980vm}
Y.~Makeenko and A.~A.~Migdal,
%``Quantum Chromodynamics as Dynamics of Loops,''
Sov. J. Nucl. Phys. \textbf{32}, 431 (1980)
doi:10.1016/0550-3213(81)90258-3.

%\cite{Gambini:1982bg}
\bibitem{Gambini:1982bg}
R.~Gambini and A.~Trias,
%``Chiral Formulation of \{Yang-Mills\} Equations: A Geometric Approach,''
Phys. Rev. D \textbf{27}, 2935 (1983)
doi:10.1103/PhysRevD.27.2935.

%\cite{Gambini:1980wm}
\bibitem{Gambini:1980wm}
R.~Gambini and A.~Trias,
%``Second Quantization of the Free Electromagnetic Field as Quantum Mechanics in the Loop Space,''
Phys. Rev. D \textbf{22}, 1380 (1980)
doi:10.1103/PhysRevD.22.1380.

%\cite{diBartolo:1983pt}
\bibitem{diBartolo:1983pt}
C.~di Bartolo, F.~Nori, R.~Gambini and A.~Trias,
%``Loop Space Quantum Formulation of Free Electromagnetism,''
Lett. Nuovo Cim. \textbf{38}, 497 (1983)
doi:10.1007/BF02787033.

%\cite{Gambini:1986ew}
\bibitem{Gambini:1986ew}
R.~Gambini and A.~Trias,
%``Gauge Dynamics in the C Representation,''
Nucl. Phys. B \textbf{278}, 436-448 (1986)
doi:10.1016/0550-3213(86)90221-X.

%\cite{Gambini:1996ik}
\bibitem{Gambini:1996ik}
R.~Gambini and J.~Pullin,
%``Loops, knots, gauge theories and quantum gravity,''
doi:10.1017/CBO9780511524431.

%\cite{Bilski:2021_RCT_III}
\bibitem{Bilski:2021_RCT_III}
J.~Bilski,
``Relativistic classical theory III. Gravitational field on a lattice,''
in preparation (2021/22).

%\cite{Ashtekar:1994mh}
\bibitem{Ashtekar:1994mh} 
A.~Ashtekar and J.~Lewandowski,
%``Projective techniques and functional integration for gauge theories,''
J.\ Math.\ Phys.\  {\bf 36}, 2170 (1995)
doi:10.1063/1.531037
[gr-qc/9411046].

%\cite{Ashtekar:1994wa}
\bibitem{Ashtekar:1994wa} 
A.~Ashtekar and J.~Lewandowski,
%``Differential geometry on the space of connections via graphs and projective limits,''
J.\ Geom.\ Phys.\  {\bf 17}, 191 (1995)
doi:10.1016/0393-0440(95)00028-G
[hep-th/9412073].

%\cite{Ashtekar:1995zh}
\bibitem{Ashtekar:1995zh}
A.~Ashtekar, J.~Lewandowski, D.~Marolf, J.~Mourao and T.~Thiemann,
%``Quantization of diffeomorphism invariant theories of connections with local degrees of freedom,''
J. Math. Phys. \textbf{36}, 6456-6493 (1995)
doi:10.1063/1.531252
[arXiv:gr-qc/9504018 [gr-qc]].

%\cite{Thiemann:1997rv}
\bibitem{Thiemann:1997rv}
T.~Thiemann,
%``QSD 3: Quantum constraint algebra and physical scalar product in quantum general relativity,''
Class. Quant. Grav. \textbf{15}, 1207-1247 (1998)
doi:10.1088/0264-9381/15/5/010
[arXiv:gr-qc/9705017 [gr-qc]].

%\cite{Thiemann:1996av}
\bibitem{Thiemann:1996av}
T.~Thiemann,
%``Quantum spin dynamics (qsd). 2.,''
Class. Quant. Grav. \textbf{15}, 875-905 (1998)
doi:10.1088/0264-9381/15/4/012
[arXiv:gr-qc/9606090 [gr-qc]].

%\cite{Fleischhack:2004jc}
\bibitem{Fleischhack:2004jc}
C.~Fleischhack,
%``Representations of the Weyl algebra in quantum geometry,''
Commun. Math. Phys. \textbf{285}, 67-140 (2009)
doi:10.1007/s00220-008-0593-3
[arXiv:math-ph/0407006 [math-ph]].

%\cite{Lewandowski:2005jk}
\bibitem{Lewandowski:2005jk}
J.~Lewandowski, A.~Okolow, H.~Sahlmann and T.~Thiemann,
%``Uniqueness of diffeomorphism invariant states on holonomy-flux algebras,''
Commun. Math. Phys. \textbf{267}, 703-733 (2006)
doi:10.1007/s00220-006-0100-7
[arXiv:gr-qc/0504147 [gr-qc]].

%\cite{Palatini:1919}
\bibitem{Palatini:1919}
A.~Palatini,
%``\"Deduzione invariantiva delle equazioni gravitazionali dal principio di Hamilton,''
Rend. Circ. Matem. Palermo {\bf 43}, 203-212 (1919)
10.1007/BF03014670.

%\cite{Proca:1900nv}
\bibitem{Proca:1900nv}
A.~Proca,
%``Sur la theorie ondulatoire des electrons positifs et negatifs,''
J. Phys. Radium \textbf{7}, 347-353 (1936)
doi:10.1051/jphysrad:0193600708034700

%\cite{Yang:1954ek}
\bibitem{Yang:1954ek}
C.~N.~Yang and R.~L.~Mills,
%``Conservation of Isotopic Spin and Isotopic Gauge Invariance,''
Phys. Rev. \textbf{96}, 191-195 (1954)
doi:10.1103/PhysRev.96.191

%\cite{Immirzi:1996dr}
\bibitem{Immirzi:1996dr}
G.~Immirzi,
%``Quantum gravity and Regge calculus,''
Nucl. Phys. B Proc. Suppl. \textbf{57}, 65-72 (1997)
doi:10.1016/S0920-5632(97)00354-X
[arXiv:gr-qc/9701052 [gr-qc]].

%\cite{Rovelli:1997na}
\bibitem{Rovelli:1997na}
C.~Rovelli and T.~Thiemann,
%``The Immirzi parameter in quantum general relativity,''
Phys. Rev. D \textbf{57}, 1009-1014 (1998)
doi:10.1103/PhysRevD.57.1009
[arXiv:gr-qc/9705059 [gr-qc]].

%\cite{Thiemann:1997rq}
\bibitem{Thiemann:1997rq}
T.~Thiemann,
%``Kinematical Hilbert spaces for Fermionic and Higgs quantum field theories,''
Class. Quant. Grav. \textbf{15}, 1487-1512 (1998)
doi:10.1088/0264-9381/15/6/006
[arXiv:gr-qc/9705021 [gr-qc]].

%\cite{Bojowald:2010qpa}
\bibitem{Bojowald:2010qpa}
M.~Bojowald,
%``Canonical Gravity and ApplicationsCosmology, Black Holes, and Quantum Gravity,''
Cambridge, UK: Cambridge Univ. Pr. (2011)

%\cite{Holst:1995pc}
\bibitem{Holst:1995pc}
S.~Holst,
%``Barbero's Hamiltonian derived from a generalized Hilbert-Palatini action,''
Phys. Rev. D \textbf{53}, 5966-5969 (1996)
doi:10.1103/PhysRevD.53.5966
[arXiv:gr-qc/9511026 [gr-qc]].

%\cite{Ashtekar:1993wf}
\bibitem{Ashtekar:1993wf}
A.~Ashtekar and J.~Lewandowski,
%``Representation theory of analytic holonomy C* algebras,''
[arXiv:gr-qc/9311010 [gr-qc]].

%\cite{Marolf:1994cj}
\bibitem{Marolf:1994cj}
D.~Marolf and J.~M.~Mourao,
%``On the support of the Ashtekar-Lewandowski measure,''
Commun. Math. Phys. \textbf{170}, 583-606 (1995)
doi:10.1007/BF02099150
[arXiv:hep-th/9403112 [hep-th]].

%\cite{Zapata:1997db}
\bibitem{Zapata:1997db}
J.~A.~Zapata,
%``A Combinatorial approach to diffeomorphism invariant quantum gauge theories,''
J. Math. Phys. \textbf{38}, 5663-5681 (1997)
doi:10.1063/1.532159
[arXiv:gr-qc/9703037 [gr-qc]].

%\cite{Zapata:1997da}
\bibitem{Zapata:1997da}
J.~A.~Zapata,
%``Combinatorial space from loop quantum gravity,''
Gen. Rel. Grav. \textbf{30}, 1229-1245 (1998)
doi:10.1023/A:1026699012787
[arXiv:gr-qc/9703038 [gr-qc]].

%\cite{Regge:1961px}
\bibitem{Regge:1961px}
T.~Regge,
%``GENERAL RELATIVITY WITHOUT COORDINATES,''
Nuovo Cim. \textbf{19}, 558-571 (1961)
doi:10.1007/BF02733251

%\cite{Ambjorn:2012jv}
\bibitem{Ambjorn:2012jv}
J.~Ambjorn, A.~Goerlich, J.~Jurkiewicz and R.~Loll,
%``Nonperturbative Quantum Gravity,''
Phys. Rept. \textbf{519}, 127-210 (2012)
doi:10.1016/j.physrep.2012.03.007
[arXiv:1203.3591 [hep-th]].

%\cite{Whitehead:1940}
\bibitem{Whitehead:1940}
J.~H.~C.~Whitehead,
%``On C1-Complexes,''
Annals Math., Second Series \textbf{41} (4), 809-824 (1940)
doi:10.2307/1968861

%\cite{Thurston97}
\bibitem{Thurston97}
W.~P.~Thurston, S. Levy (ed.),
%``Three-dimensional geometry and topology. Vol. 1,''
Princeton Mathematical Series, 35, Princeton Univ. Pr. (1997).

%\cite{Moise:1977}
\bibitem{Moise:1977}
E.~Moise,
%``Geometric Topology in Dimensions 2 and 3,''
Springer-Verlag (1977),
ISBN 0-387-90220-1

%\cite{Munkres:1960}
\bibitem{Munkres:1960}
J.~Munkres,
%``Obstructions to the smoothing of piecewise-differentiable homeomorphisms,''
Annals Math. \textbf{72} (3), 521-554 (1960)
doi:10.2307/1970228

%\cite{Whitehead:1961}
\bibitem{Whitehead:1961}
J.~H.~C.~Whitehead,
%``Manifolds with Transverse Fields in Euclidean Space,''
Annals Math. \textbf{73} (1), 154-212 (1961)
doi:10.2307/1970286

%\cite{Dirac:1925jy}
\bibitem{Dirac:1925jy}
P.~A.~M.~Dirac,
%``The fundamental equations of quantum mechanics,''
Proc. Roy. Soc. Lond. A \textbf{109}, 642-653 (1925)
doi:10.1098/rspa.1925.0150

%\cite{Heisenberg:1929xj}
\bibitem{Heisenberg:1929xj}
W.~Heisenberg and W.~Pauli,
%``On Quantum Field Theory. (In German),''
Z. Phys. \textbf{56}, 1-61 (1929)
doi:10.1007/BF01340129

%\cite{DeWitt:1967yk}
\bibitem{DeWitt:1967yk}
B.~S.~DeWitt,
%``Quantum Theory of Gravity. 1. The Canonical Theory,''
Phys. Rev. \textbf{160}, 1113-1148 (1967)
doi:10.1103/PhysRev.160.1113

%\cite{Dittrich:2014ala}
\bibitem{Dittrich:2014ala}
B.~Dittrich,
%``The continuum limit of loop quantum gravity - a framework for solving the theory,''
doi:10.1142/9789813220003\_0006
[arXiv:1409.1450 [gr-qc]].

%\cite{Bilski:2020fmn}
\bibitem{Bilski:2020fmn}
J.~Bilski,
%``Regularization of the cosmological sector of loop quantum gravity with bosonic matter and the related problems with the general covariance of quantum corrections,''
[arXiv:2001.04491 [gr-qc]].

%\cite{Vines:2014uoa}
\bibitem{Vines:2014uoa}
J.~Vines and D.~A.~Nichols,
%``Properties of an affine transport equation and its holonomy,''
Gen. Rel. Grav. \textbf{48}, no.10, 127 (2016)
doi:10.1007/s10714-016-2118-2
[arXiv:1412.4077 [gr-qc]].

%\cite{Wigner:1931}
\bibitem{Wigner:1931}
E.~Wigner,
%``Gruppentheorie und ihre Anwendung auf die Quantenmechanik der Atomspektren,''
Vieweg+Teubner Verlag, Wiesbaden (1931)
doi:10.1007/978-3-663-02555-9.

%\cite{Wigner:1939cj}
\bibitem{Wigner:1939cj}
E.~P.~Wigner,
%``On Unitary Representations of the Inhomogeneous Lorentz Group,''
Annals Math. \textbf{40}, 149-204 (1939)
doi:10.2307/1968551

%\cite{Weinberg:1995mt}
\bibitem{Weinberg:1995mt}
S.~Weinberg,
%``The Quantum theory of fields. Vol. 1: Foundations,''
Cambridge, UK: Cambridge Univ. Pr. (1995)
doi:10.1017/CBO9781139644167.

%\cite{Marolf:1994wh}
\bibitem{Marolf:1994wh}
D.~Marolf,
%``Quantum observables and recollapsing dynamics,''
Class. Quant. Grav. \textbf{12}, 1199-1220 (1995)
doi:10.1088/0264-9381/12/5/011
[arXiv:gr-qc/9404053 [gr-qc]].

%\cite{Wigner:1957ep}
\bibitem{Wigner:1957ep}
E.~P.~Wigner,
%``Relativistic Invariance and Quantum Phenomena,''
Rev. Mod. Phys. \textbf{29}, 255-268 (1957)
doi:10.1103/RevModPhys.29.255

%\cite{Salecker:1957be}
\bibitem{Salecker:1957be}
H.~Salecker and E.~P.~Wigner,
%``Quantum limitations of the measurement of space-time distances,''
Phys. Rev. \textbf{109}, 571-577 (1958)
doi:10.1103/PhysRev.109.571

%\cite{Charles:2010}
\bibitem{Charles:2010}
L.~Charles,
%``On the quantization of polygon spaces,''
Asian J. Math. \textbf{14}, 1 (2010)
doi:10.4310/AJM.2010.v14.n1.a6
[arXiv:0806.1585 [math.SG]].

%\cite{Bianchi:2010gc}
\bibitem{Bianchi:2010gc}
E.~Bianchi, P.~Dona and S.~Speziale,
%``Polyhedra in loop quantum gravity,''
Phys. Rev. D \textbf{83}, 044035 (2011)
doi:10.1103/PhysRevD.83.044035
[arXiv:1009.3402 [gr-qc]].

%\cite{Heisenberg:1925zz}
\bibitem{Heisenberg:1925zz}
W.~Heisenberg,
%``A quantum-theoretical reinterpretation of kinematic and mechanical relations,''
Z. Phys. \textbf{33}, 879-893 (1925)
doi:10.1007/BF01328377

%\cite{Lewandowski:1993zq}
\bibitem{Lewandowski:1993zq}
J.~Lewandowski, E.~T.~Newman and C.~Rovelli,
%``Variations of the parallel propagator and holonomy operator and the Gauss law constraint,''
J. Math. Phys. \textbf{34}, 4646-4654 (1993)
doi:10.1063/1.530362

%\cite{Dona:2010hm}
\bibitem{Dona:2010hm}
P.~Dona and S.~Speziale,
%``Introductory lectures to loop quantum gravity,''
[arXiv:1007.0402 [gr-qc]].

%\cite{Dittrich:2007th}
\bibitem{Dittrich:2007th}
B.~Dittrich and T.~Thiemann,
%``Are the spectra of geometrical operators in Loop Quantum Gravity really discrete?,''
J. Math. Phys. \textbf{50}, 012503 (2009)
doi:10.1063/1.3054277
[arXiv:0708.1721 [gr-qc]].

%\cite{Lewandowski:1996gk}
\bibitem{Lewandowski:1996gk}
J.~Lewandowski,
%``Volume and quantizations,''
Class. Quant. Grav. \textbf{14}, 71-76 (1997)
doi:10.1088/0264-9381/14/1/010
[arXiv:gr-qc/9602035 [gr-qc]].

%\cite{Ashtekar:1997fb}
\bibitem{Ashtekar:1997fb}
A.~Ashtekar and J.~Lewandowski,
%``Quantum theory of geometry. 2. Volume operators,''
Adv. Theor. Math. Phys. \textbf{1}, 388-429 (1998)
doi:10.4310/ATMP.1997.v1.n2.a8
[arXiv:gr-qc/9711031 [gr-qc]].

%\cite{Thiemann:1996au}
\bibitem{Thiemann:1996au}
T.~Thiemann,
%``Closed formula for the matrix elements of the volume operator in canonical quantum gravity,''
J. Math. Phys. \textbf{39}, 3347-3371 (1998)
doi:10.1063/1.532259
[arXiv:gr-qc/9606091 [gr-qc]].

%\cite{Giesel:2005bk}
\bibitem{Giesel:2005bk}
K.~Giesel and T.~Thiemann,
%``Consistency check on volume and triad operator quantisation in loop quantum gravity. I.,''
Class. Quant. Grav. \textbf{23}, 5667-5692 (2006)
doi:10.1088/0264-9381/23/18/011
[arXiv:gr-qc/0507036 [gr-qc]].

%\cite{Baez:1995md}
\bibitem{Baez:1995md}
J.~C.~Baez,
%``Spin networks in nonperturbative quantum gravity,''
[arXiv:gr-qc/9504036 [gr-qc]].

%\cite{Thiemann:1996hw}
\bibitem{Thiemann:1996hw}
T.~Thiemann,
%``The Inverse loop transform,''
J. Math. Phys. \textbf{39}, 1236-1248 (1998)
doi:10.1063/1.532344
[arXiv:hep-th/9601105 [hep-th]].


	%%%%%%%%%	%%%%%%%%%	%%%%%%%%%	%%%%%%%%%

	%%%%%%%%%	%%%%%%%%%	%%%%%%%%%	%%%%%%%%%


\end{thebibliography}
\end{document}